\documentclass[
 reprint,
 amsmath,amssymb,
 aps,nofootinbib,preprintnumbers,longbibliography,superscriptaddress
]{revtex4-1}

\usepackage[utf8]{inputenc}
\usepackage{graphicx}
\usepackage{dcolumn}
\usepackage{bm}
\usepackage{hyperref}
\usepackage{lipsum}
\usepackage{xcolor}
\usepackage{calc}
\usepackage{accents}
\usepackage{comment}
\usepackage{float}
\usepackage{diagbox}
\usepackage{soul}
\usepackage{braket}
\usepackage[normalem]{ulem}

\font\BF=cmmib10
\def\k{{\hbox{\BF k}}}

\def\hinvMpc{h\,{\rm Mpc}^{-1}}
\def\Mpcinvh{{\rm Mpc}/h}
\newcommand{\code}[1]{\texttt{#1}}
\makeatletter
\newcommand\footnoteref[1]{\protected@xdef\@thefnmark{\ref{#1}}\@footnotemark}
\makeatother

\newcommand{\Planck}{{\it Planck}}

\begin{document}


\title{Consistency of effective field theory analyses of the BOSS power spectrum}
\author{Th\'eo Simon}
\email{theo.simon@umontpellier.fr}
\affiliation{Laboratoire Univers \& Particules de Montpellier (LUPM), CNRS \& Universit\'e de Montpellier (UMR-5299), Place Eug\`ene Bataillon, F-34095 Montpellier Cedex 05, France}
\author{Pierre Zhang}
\email{pierrexyz@protonmail.com}
\affiliation{Department of Astronomy, School of Physical Sciences,
University of Science and Technology of China, Hefei, Anhui 230026, China \\
CAS Key Laboratory for Research in Galaxies and Cosmology,
University of Science and Technology of China, Hefei, Anhui 230026, China \\
School of Astronomy and Space Science, \\
University of Science and Technology of China, Hefei, Anhui 230026, China}
\author{Vivian Poulin}
\affiliation{Laboratoire Univers \& Particules de Montpellier (LUPM), CNRS \& Universit\'e de Montpellier (UMR-5299), Place Eug\`ene Bataillon, F-34095 Montpellier Cedex 05, France}
\author{Tristan L.~Smith}
\affiliation{Department of Physics and Astronomy, Swarthmore College, Swarthmore, PA 19081, USA}

\begin{abstract}

We assess the robustness of $\Lambda$CDM results from the full-shape analysis of BOSS power spectrum using the one-loop prediction of the effective field theory of large-scale structure (EFTofLSS). 
In former EFT analyses, the two public likelihoods \texttt{PyBird} and \texttt{CLASS-PT} lead to results in agreement only at the $1\sigma$ level, which may appear unsatisfactory given that they are derived from the same BOSS dataset and the same theory model. 
To identify the origin of the difference, we perform a thorough comparison of the various analyses choices made between the two pipelines. 
We find that most of the difference can be attributed to the choices of prior on the EFT parameters, dubbed ``West-coast'' (WC) prior and ``East-coast'' (EC) prior, respectively associated to \texttt{PyBird} and \texttt{CLASS-PT}.
In particular, because posteriors are non-Gaussian, projection effects from the marginalization over the EFT parameters shift the posterior mean of the cosmological parameters with respect to the maximum a posteriori up to $\sim 1\sigma$ in the WC prior and up to $\sim 2\sigma$ in the EC prior. 
Additionally, we quantify that maximum a posteriori cosmological parameters extracted from BOSS given the two prior choices are consistent at $\lesssim 1\sigma$. The consistency improves to $\lesssim 0.4\sigma$ when doubling the width of both priors. 
While this reveals that current EFT analyses are subject to prior effects, we show that cosmological results obtained in combination with CMB, or from forthcoming large-volume data, are less sensitive to those effects. 
In addition, we evaluate the impact on the cosmological constraints from various BOSS power-spectrum measurements. While we find broad agreements across all pre-reconstructed measurements considered ($<0.6\sigma$), we show that the two available BOSS post-reconstructed measurements in Fourier space, once combined with the EFT full-shape analysis, lead to discrepant Hubble parameter $H_0$ at $\sim 0.9\sigma$. Finally, given the various effects we discuss, we argue that the clustering amplitude $\sigma_8$ measured with BOSS is not in statistical tension with that inferred from \textit{Planck} under $\Lambda$CDM.

\end{abstract}

\maketitle

\section{\label{sec:Intro}Introduction}

In recent years, developments of the one-loop prediction of the galaxy power spectrum in redshift space from the effective field theory of large-scale structures (EFTofLSS)\footnote{See also the introduction footnote in, \emph{e.g.},~\cite{DAmico:2022osl} for relevant related works on the EFTofLSS.}~\cite{Baumann:2010tm,Carrasco:2012cv,Senatore:2014via,Senatore:2014eva,Senatore:2014vja,Perko:2016puo} have made possible the determination of the $\Lambda$CDM parameters from the full-shape analysis of SDSS/BOSS data~\cite{BOSS:2016wmc} at precision higher than that from conventional BAO and RSD analyses, and even comparable to that of CMB experiments. 
This provides an important consistency test for the $\Lambda$CDM model, while providing competitive constraints on models beyond $\Lambda$CDM (see, \emph{e.g.}, Refs.~\cite{DAmico:2019fhj,Ivanov:2019pdj,Colas:2019ret,DAmico:2020kxu,DAmico:2020tty,Chen:2021wdi,Zhang:2021yna,Zhang:2021uyp,Philcox:2021kcw,Simon:2022csv,Chudaykin:2022nru,Smith:2022iax,Simon:2022ftd,Kumar:2022vee,Nunes:2022bhn,Niedermann:2020qbw,Lague:2021frh,Carrilho:2022mon,Simon:2022adh,Schoneberg:2023rnx,Allali:2023zbi}).
These  analyses were also recently extended to the inclusion of the BOSS bispectrum analyzed at the one-loop level~\cite{DAmico:2022osl,DAmico:2022gki}.

In this paper, we perform a thorough comparison of the cosmological constraints derived from the full-shape analysis of BOSS power spectrum from the EFTofLSS, in order to assess the consistency of the various analyses presented in the literature. 
Indeed, a proper comparison between these various analyses is still lacking, and the implication for the robustness of the constraints has yet to be established. 
The EFT implementation and BOSS data we will focus on in this study are packaged in the \code{PyBird} likelihood, based on the EFT prediction and likelihood from~\code{PyBird}~\footnote{\url{https://github.com/pierrexyz/pybird}}~\cite{DAmico:2020kxu} and the \code{CLASS-PT} likelihood,  based on the EFT prediction from~\code{CLASS-PT}~\footnote{\url{https://github.com/michalychforever/CLASS-PT}}~\cite{Chudaykin:2020aoj} and likelihood from Ref.~\cite{Philcox:2021kcw}.~\footnote{\url{https://github.com/oliverphilcox/full_shape_likelihoods}}
Cosmological constraints in $\Lambda$CDM obtained from these two likelihoods for the BOSS full-shape analysis with a big-bang nucleosynthesis (BBN) prior were originally presented in Ref.~\cite{Zhang:2021yna} and Ref.~\cite{Philcox:2021kcw}. 
While results are in broad agreement, differences occur at the $1\sigma$ level between the two approaches, in particular on the primordial power spectrum amplitude $A_s$ and the cold dark matter density $\omega_{\rm cdm}$, that can have an impact on the variance of matter fluctuations on a $8 h^{-1}$Mpc scale,   $\sigma_8$. 
As a result, the level of the tension on the $S_8\equiv \sigma_8(\Omega_m/0.3)^{0.5}$ parameter compared to the CMB prediction can vary between these analyses, from mild to insignificant. This is particularly relevant to understand the scale-dependence of the growing ``$S_8$ tension'' \cite{Heymans:2020gsg,DES:2021wwk,Lange:2020mnl,Amon:2022ycy,Abdalla:2022yfr}. Moreover, it casts some doubts on the robustness (and potentially on the validity) of the constraints derived on $\Lambda$CDM (and extensions) from the EFTofLSS applied to BOSS data.

In this work, we aim at understanding what drives the differences seen at the level of the posteriors of the cosmological parameters. 
There are several analyses choices that differ between the two pipelines, from the choice of prior on the EFT parameters, for which several prescriptions have been suggested in the literature, to the BOSS measurements themselves. 
Specifically, we ask: (i) How sensitive are cosmological constraints derived from the full-shape analysis of BOSS power spectrum to those effects?;  (ii) How do the various BOSS data measurements used in previous full-shape analysis, that are obtained with different estimators, split in different redshift bins, or combined with various post-reconstructed measurements, impact the cosmological results?

To answer those questions, we perform a series of analyses of the BOSS full-shape data, varying  one-by-one (in order of importance) the prior choices, the BOSS measurements used (full-shape and post-reconstructed BAO parameters), the scale cuts and the number of multipoles~\footnote{By multipoles, we refer to the Legendre polynomial $\mathcal{L}_\ell$ decomposition in multipoles $P_\ell(k)$ of the 3D power spectrum $P(k,\mu)$, \emph{i.e.}, $P(k,\mu)=\sum_{\ell} \mathcal{L}_\ell(\mu) P_\ell(k)$, where $k$ is the norm of the mode $\k$ and $\mu$ is the cosine of its angle with the line-of-sight. In this work we consider multipoles restricted to the first even ones, namely $\ell = \lbrace 0,2 \rbrace $ (the monopole and the quadrupole), or $\ell = \lbrace 0,2,4 \rbrace $ (including also the hexadecapole).} included. 
Importantly, we find that cosmological constraints are sensitive to the choice of prior on the EFT parameter space, and the two different choices of prior used in the \code{PyBird} and \code{CLASS-PT} analyses drive most of the differences in the results. 
On the other hand, the different BOSS full-shape measurements leads to at most $0.6\sigma$ difference among all cosmological parameters, while the different post-reconstructed BAO measurements can affect constraints by up to $0.9\sigma$.
Yet, when the choices of prior and data are the same, we show that the two pipelines agree at better than $0.2\sigma$, which consists in an important validation check of the two public likelihoods available.

For all analyses in this paper, we work within $\Lambda$CDM.\footnote{In a companion paper \cite{Simon:2022adh}, we explore the impact within a popular extension of $\Lambda$CDM suggested to resolve the Hubble tension \cite{Schoneberg:2021qvd,Riess:2021jrx,Abdalla:2022yfr}, namely early dark energy \cite{Poulin:2018cxd}.}
Except when combined with \Planck{}~\cite{Planck:2018lbu}, we impose a Gaussian prior on $\omega_b \sim \mathcal{N}(0.02268, 0.00038)$.~\footnote{This prior is inspired from BBN experiments~\cite{Schoneberg_2019}, based on the theoretical prediction of \cite{Consiglio_2018}, the experimental Deuterium fraction of~\cite{Cooke_2018} and the experimental Helium fraction of~\cite{Aver_2015}.} 
We scan over the physical dark matter density $\omega_{cdm}$, the reduced Hubble constant $h$, the log-amplitude of the primordial fluctuations $\ln(10^{10}A_s)$, and the spectral tilt $n_s$, with large flat prior. 
We fix the total neutrino mass to minimal following \Planck{} prescription~\cite{Planck:2018lbu}. 
We sample our posteriors using the Metropolis-Hasting algorithm in~\code{MontePython}~\cite{Brinckmann:2018cvx} with convergence given by the Gelman-Rubin criterion $R-1 < 0.01$. 
Finally, we extract the maximum a posteriori (MAP) parameters from the procedure highlighted in appendix of Ref.~\cite{Schoneberg:2021qvd}, and triangle plots are produced using \code{GetDist}~\cite{Lewis:2019xzd}.

Our paper is organized as follows. 
In Sec.~\ref{sec:prior}, we review the two prior choices on the EFT parameters used in previous analyses with the two aforementioned likelihoods, and discuss the various prior effects at play in the determination of the cosmological parameters from the Bayesian analysis. 
In Sec.~\ref{sec:impactprior}, we assess the impact from those prior choices on the cosmological constraints from the EFT analysis of BOSS power spectrum. 
We scrutiny the impacts given various BOSS data measurements of the pre-reconstructed two-point functions in Sec.~\ref{sec:measurements}, and of the post-reconstructed ones in Sec.~\ref{sec:reconBAO}.
Finally, we summarize our findings and conclude in Sec.~\ref{sec:conclusions}. 
In App.~\ref{app:scalecut}, we quantify the (minor) differences introduced due to choices of scale cuts and number of multipoles included in the analyses. 
For completeness, we provide a comparison of the two likelihoods in their respective baseline configurations in App.~\ref{app:direct}. 

\begin{figure*}
\centering
{\large \emph{WC vs EC prior: $P_\ell \ (\ell=0,2)$}}\\[0.2cm]
\includegraphics[width=1.87\columnwidth]{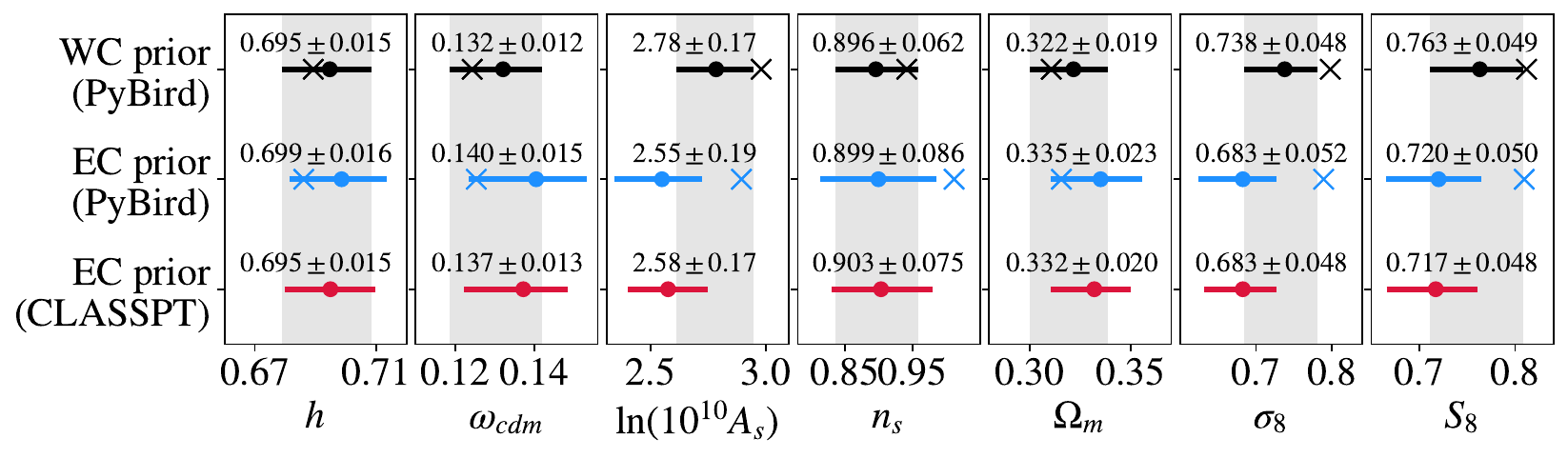}
\includegraphics[width=1.55\columnwidth]{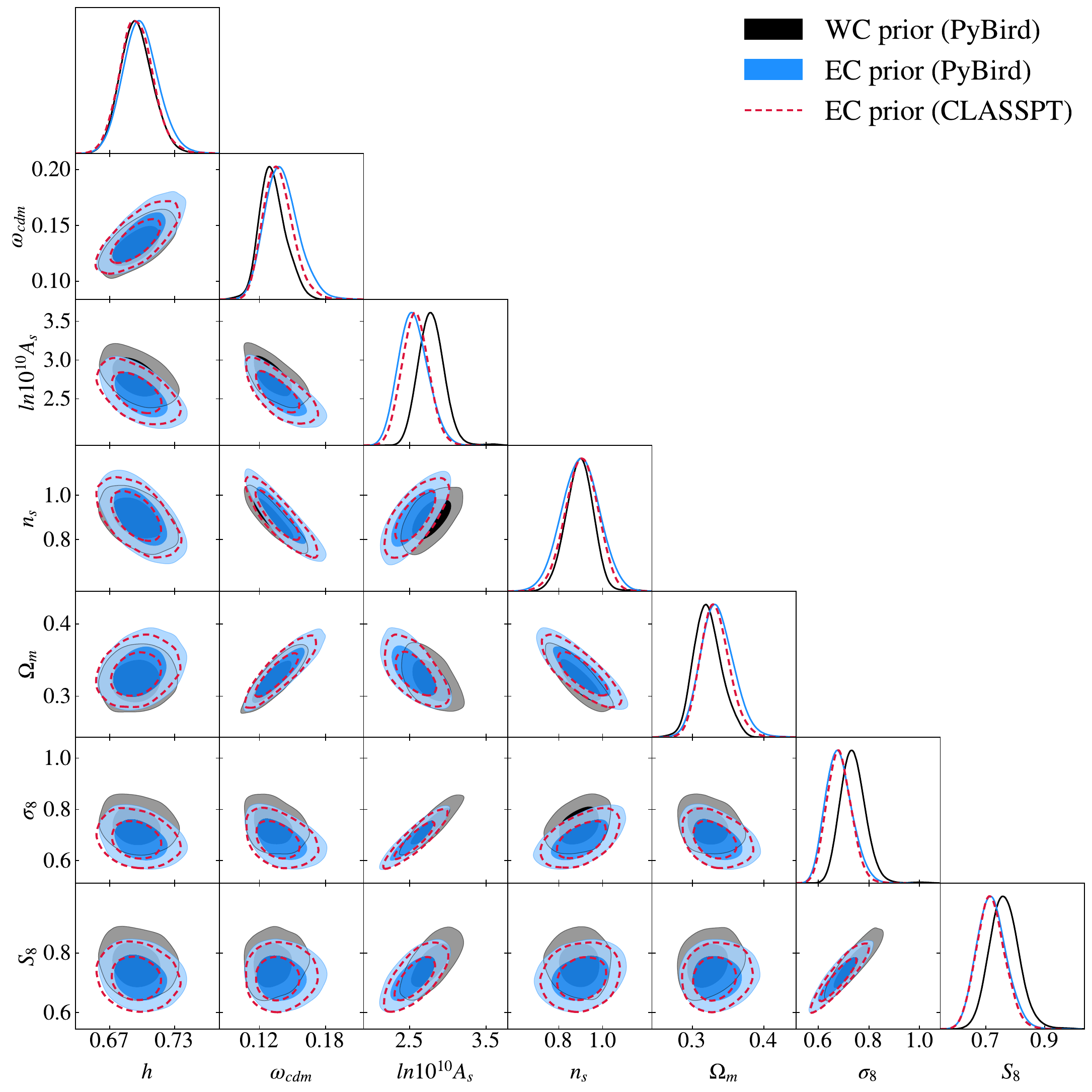}
\caption{
Comparison of $\Lambda$CDM results (1D and 2D credible intervals) from the full-shape analyses of BOSS power spectrum using the \code{PyBird} likelihood or the \code{CLASS-PT} likelihood. 
Here we use the same data measurements, $P_\textsc{quad}^{z_1/z_3}$ as specified in Tab.~\ref{tab:twopoint_summary}, and same analysis configuration:
we fit two multipoles, $\ell=0,2$, and use $k_{\rm max}=0.20/0.25\hinvMpc$ for the $z_1/z_3$ redshift bins.
Given the same prior choice, the EC prior, we reproduce from the \code{PyBird} likelihood the results from the \code{CLASS-PT} likelihood to very good agreement (see blue and red posteriors): we obtain shifts $\lesssim 0.2\sigma$ on the means and the errors bars similar at $\lesssim 15\%$. 
Given that the two pipelines have been developed independently, this comparison provides a validation check of their implementation.
In contrast, the WC and the EC prior choices lead to substantial differences on the 1D marginalized posteriors (see black and blue posteriors). The gray bands on the 1D posteriors are centered on the results obtained with the WC priors. 
The MAP (depicted by the crosses) are however in better agreement.
}
\label{fig:prior}
\end{figure*}

\section{The role of EFT priors}\label{sec:prior}

The one-loop prediction to the galaxy power spectrum in redshift space depends on a number of EFT parameters. 
Those are marginalized over in order to obtain constraints on the cosmological parameters. 
There are various ways that the EFT prediction can be parametrized, but all are equivalent at the one-loop order, in the sense that they are simply changes of basis (\emph{i.e.}, linear transformations) of each others. 
However, differences can appear at the level of the posteriors, as soon as one needs to impose priors on the EFT parameters. 
There are two effects that can arise from the choice of priors.
Let us give a precise definition for a given parameter $\Omega$ of interest (a cosmological parameter) and one nuisance ``EFT'' parameter $c$. 
The generalization to more parameters is straightforward. 
Considering a Gaussian prior $e^{-\tfrac{1}{2}(c/\sigma)^2}$ on $c$, we identify the following effects on the 1D posterior of $\Omega$:
\begin{itemize}
    \item \emph{The prior weight effect}: this refers to how much the prior is weighting in the likelihood given that the true value of $c$ will be different than the central value of our prior: $e^{-\tfrac{1}{2}((c-\hat c)/\sigma)^2}$, with $\hat c$ the true value. 
    This can lead to a shift of the most-likely value of $\Omega$ away from its true value. 
    \item \emph{The prior volume projection effect}: 
    this refers to the marginalization integral over $c$ given its prior: $\int dc \, e^{-\tfrac{1}{2}((c-\hat c)/\sigma)^2} \dots$ 
    As the likelihood will be a function of $\Omega$ and $c$, that usually enter in the model not just linearly but also as $\Omega \times c$, etc., the posterior of $\Omega$ will be non-Gaussian.
    The effect is a shift of the mean of $\Omega$ away from its most-likely value.
\end{itemize}

Here, we quantify the impact  on the inferred cosmological parameters that different choices in the prior of the EFT parameters can have upon marginalization.

\subsection{The two EFT priors} 
There has been several prescriptions for the EFT parameter priors that have been suggested in the literature. 
Generally, one would like to keep EFT parameters within physical range, such that the one-loop contributions cannot be larger than the tree-level part given the perturbative nature of the EFTofLSS. 
The simplest way to implement this requirement is to ask the EFT parameters controlling the loop contributions to be $\sim \mathcal{O}(b_1)$, where $b_1$ is the linear bias. 
 
We here compare two choices of prior on the EFT parameters made in the original analyses with the \code{PyBird} likelihood and the \code{CLASS-PT} likelihood. 
Following Ref.~\cite{Nishimichi:2020tvu}, we dub those prior choices ``West coast'' (WC) prior and ``East coast'' (EC) prior, respectively. 
\\

\paragraph{WC prior:} 
The WC prior is designed to encompass the region physically-allowed by the EFTofLSS~\cite{Senatore:2014eva}. 
For each sky-cut, we assign one set of EFT parameters, and impose the following priors to keep them within physical range~\cite{DAmico:2020tty}: 
\begin{itemize}
    \item $b_1 \sim$ flat $[0,4]$,
    \item $c_2 = (b_2+b_4)/\sqrt{2} \sim$ flat $[-4 ,4]$,
    \item $\lbrace b_3, c_{\rm ct}, 2 c_{r,1}, c_{e,0}, c_{e,2} \rbrace \sim \mathcal{N}(0, 2)$,
    \item $\lbrace c_4 = (b_2-b_4)/\sqrt{2}, c_{r,2}, c_{e,1} \rbrace \sim 0$,
\end{itemize}
where $\mathcal{N}(m, \sigma)$ is a Gaussian prior centered on $m$ with a standard deviation $\sigma$.  
Here $b_1$ is the linear bias and $b_2, b_3, b_4$ are the nonlinear EFTofLSS biases~\cite{Senatore:2014eva,Angulo:2015eqa,Fujita:2016dne}.
$c_{\rm ct}$ is dark-matter / higher-derivative counterterm coefficient appearing in front of $\sim k^2 / k_{\rm M}^2 P_{\rm lin}(k)$~\cite{Carrasco:2012cv,Senatore:2014eva}. 
$c_{r,1}, c_{r,2}$ are the counterterm coefficients renormalizing products of velocity operators appearing the expansion of the density field in redshift space~\cite{Senatore:2014vja,Perko:2016puo,DAmico:2021ymi}, that are appearing in front of $\sim k^2/k_{\rm R}^2 P_{\rm lin}(k)$. 
$c_{e,0}, c_{e,1}, c_{e,2}$ are the stochastic term coefficients~\cite{Perko:2016puo}, respectively of the shot noise $\bar n^{-1}$, monopole $\sim k^2/k_{\rm M}^2$ and quadrupole $\sim k^2/k_{\rm M}^2$. 
The renormalization scales are measured to be $k_{\rm NL} = k_{\rm M} = 0.7 \hinvMpc$ and $k_{\rm R} = 0.35 \hinvMpc$~\cite{DAmico:2021ymi}, and the mean galaxy density is set to $\bar{n} = 4 \cdot 10^{-4} (\Mpcinvh)^3$. 
The EFT parameters set to $0$ have too low signal-to-noise ratio to be measured from BOSS two-point function (namely, $c_4$ and $c_{\epsilon,1}$), or are degenerate with already present EFT parameters when using only two multipoles (namely $c_{r,2}$)~\cite{DAmico:2019fhj}.~\footnote{Notice than when we perform checks adding the hexadecapole, we then free $c_{r,2}$ with a prior $\sim \mathcal{N}(0,2)$ as the degeneracy is broken. 

}
In total, the WC prior consists of $9$ EFT parameters per sky-cut when fitting two multipoles, and $10$ when fitting three multipoles. 
We also perform checks freeing $c_4$ and $c_{e,1}$, as well as adding the next-to-next-leading order redshift-space counterterm $\tilde c$ (defined in the following). In this case, both priors have the same number of EFT parameters and an equivalent set of associated theoretical predictions. \\

\paragraph{EC prior:} 
The EC prior is motivated by the coevolution model and simulations~\cite{Ivanov:2021kcd} (and see Refs. therein). 
The basis of galaxy biases $\lbrace \tilde b_1, \tilde b_2, b_{\mathcal{G}_2}, b_{\Gamma_3} \rbrace$ developed in Ref.~\cite{Mirbabayi:2014zca} is related to the EFTofLSS basis as (see, \emph{e.g.},~\cite{Fujita:2020xtd}):
\begin{align}
    b_1 &= \tilde b_1, \quad b_2 = \tilde b_1 + \frac{7}{2}b_{\mathcal{G}_2}, \quad \nonumber \\
    b_3 &= \tilde b_1 + 15 b_{\mathcal{G}_2} + 6  b_{\Gamma_3}, \quad b_4 = \frac{1}{2}\tilde b_2 - \frac{7}{2}b_{\mathcal{G}_2}\, . \label{eq:basis}
\end{align}
As for the counterterms and the stochastic terms, although almost all scaling functions are present in the two likelihoods, there are differences in their definition, leading to differences in their prior. 
In particular, in the EC prior, $k_{\rm M}$ or $k_{\rm R}$ are absorbed in the definition of the counterterm coefficients $c_0$, $c_2$, $c_4$, while $k_{\rm NL}^{-0/2} \bar n^{-1}$ appears explicitly in front of their $k^0 / k^2$ stochastic terms, with choice $k_{\rm NL} = 0.45 \hinvMpc$ and $\bar n \simeq 3 \cdot 10^{-4} (\Mpcinvh)^3$. 
Furthermore, the EC prior also includes in their baseline a next-to-next-leading order term $\sim \tilde c \, k^4 P_{\rm lin}(k) $. 
The EC prior on the EFT parameters consists of~\cite{Philcox:2021kcw}:
\begin{itemize}
    \item $\tilde b_1 \sim$ flat $[0,4]$,
    \item $\lbrace \tilde b_2, b_{\mathcal{G}_2} \rbrace \sim \mathcal{N}(0,1), b_{\Gamma_3} \sim \mathcal{N}(\tfrac{23}{42}(b_1-1),1)$, 
    \item $c_0 / [\Mpcinvh]^2 \sim \mathcal{N}(0,30)$, $c_2 / [\Mpcinvh]^2 \sim \mathcal{N}(30,30)$, $c_4 / [\Mpcinvh]^2 \sim \mathcal{N}(0,30)$, 
    \item $\lbrace c_{e,0}, c_{e,1}, c_{e,2} \rbrace \sim \mathcal{N}(0,2)$,

    \item $\tilde c / [\Mpcinvh]^4 \sim \mathcal{N}(500,500)$.
\end{itemize}
In total, the EC prior consists of $11$ EFT parameters per sky-cut when fitting two multipoles, and an extra one, $c_4$, when fitting three multipoles.  

\subsection{Prior weight and volume projection effects} 

As mentioned above, the two basis are merely linear combinations of the other ones. 
However, we stress that the two prior choices are not equivalent, for two reasons. 

First, given the definition above, the allowed ranges of variation are not equivalent. 
As a result, they can lead to different \emph{prior weight effect} (on the likelihood of the cosmological parameters of interest) as defined previously. This raises two important questions regarding the prior choice and the prior weight effect: Is one prior choice more restrictive (\emph{i.e.}, more informative) than the other one?  How significantly does the prior choice disfavor physically-allowed region, and lead to potential bias in the measured cosmological parameters? 

Second, the \emph{metric} on the parameter space is different: although one can go from one basis to the other through linear transformations, we do not keep the jacobians of the transformations, \emph{i.e.}, the integral measures that enter in the marginalization. 
If the posteriors are Gaussian, \emph{e.g.}, in the limit where the parameters are well determined, this is not so much an issue. 
However, in our case, the posteriors are non-Gaussian. 
This is obvious in the case of the cosmological parameters, but it is also the case for EFT ones, as for example $b_1$ enters quadratically in the prediction. 
In fact, even the EFT parameters that enter at most linearly in the prediction, and thus quadratically in the likelihood, do not lead to Gaussian posteriors as they often (if not always) correlate with other parameters, such as $b_1$, $A_s$, etc. 
Given the relatively large number of EFT parameters to marginalize over, this might lead to a rather large \emph{prior volume projection effect} that affects the marginalized posteriors, as defined previously. 
Given the non-Gaussianity of the posteriors, a natural question to ask is therefore: do the differences in the parametrization, producing effectively different integral measures upon marginalization, lead to discrepancies on the measured value of the cosmological parameters? 

In the following, we perform a detailed analysis to address those issues.

\subsection{Pipeline validation check} 
Before comparing the results from the two prior choices, let us first present an important check. 
To test the validity of the two pipelines, we implement in the \code{PyBird} likelihood the EC prior. 
On the same data and at same configuration (same number of multipoles and same $k_{\rm max}$), we obtain the posteriors shown in Fig.~\ref{fig:prior} (see also Fig.~\ref{fig:prior_3mult} of App.~\ref{app:scalecut} for the equivalent analyses with three multipoles). 
The residual differences are $\lesssim 0.2\sigma$ on the 1D posteriors of the cosmological parameters. 
Beyond serving as validation check of those two pipelines built independently, this also means that the different IR-resummation schemes, that differ at the two-loop level, are indeed not leading to appreciable shifts in the posteriors, as expected from the size of theory error (compared to BOSS error bars) at the scales we analyze.~\footnote{\code{PyBird} implements the original IR-resummation scheme proposed in Ref.~\cite{Senatore:2014via}, generalized to redshift space in Ref.~\cite{Lewandowski:2015ziq}, and made numerically practical in Ref.~\cite{DAmico:2020kxu}. 
In this approach, the bulk displacements are resummed directly on the full shape, and higher-order terms that are neglected are proven to be small at each order in perturbations~\cite{Senatore:2014via} (see also~\cite{Senatore:2017pbn}). 
\code{CLASS-PT} implements instead the IR-resummation scheme proposed in Ref.~\cite{Blas:2016sfa}, and generalized to redshift space in Ref.~\cite{Ivanov:2018gjr}. 
This alternative scheme has been shown to be an approximation of the former one in Ref.~\cite{Lewandowski:2018ywf}, where one consider only the resummation of the bulk displacements around the BAO peak, $r_{\rm BAO} \sim 110 \Mpcinvh$.
For this scheme to be made practical, one further relies on a wiggle-no-wiggle split procedure to isolate the BAO part. 
These approximations were shown to be smaller than the two-loop contribution in Ref.~\cite{Chudaykin:2020aoj}}.

\begin{figure*}
\centering
{\large \emph{WC vs EC prior $(\times 2)$: $P_\ell \ (\ell=0,2)$}}\\[0.2cm]
\includegraphics[width=1.87\columnwidth]{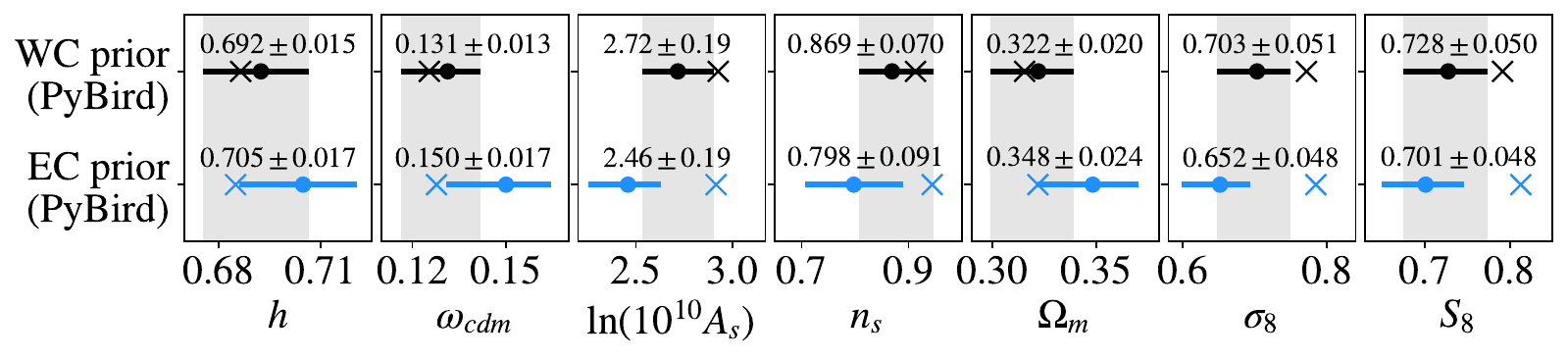}
\caption{
Same figure as the top panel of Fig.~\ref{fig:prior}, but this time increasing the allowed prior width for the EFT parameters by a factor of two. The shifts between the mean and the MAP are $\lesssim 1.4\sigma$ for the WC prior and $\lesssim 2.7\sigma$ for the EC prior. This should be compared with the shifts we obtain for the usual EFT prior width, namely $\lesssim 1.2\sigma$ for the WC prior and $\lesssim 2.0\sigma$ for the EC prior.
}
\label{fig:priorx2}
\end{figure*}

\begin{table}
\begin{tabular}{|l|c|c|c|c|}
 \hline

Parameter & WC pr.  & EC pr. & WC pr. $\times 2$ & EC pr. $\times 2$ \\
\hline
$h$ & 0.6893 & 0.6861                       & 0.6865 & 0.6850 \\
$\omega_{\rm cdm}$ & 0.1243 & 0.1253        & 0.1254 & 0.1277 \\
$\text{ln}(10^{10} A_s)$  & 2.980 & 2.894   & 2.926 & 2.915 \\
$n_s$  & 0.941 & 1.011                      & 0.913 & 0.944 \\
$\Omega_{\rm m}$  & 0.3107   & 0.3158       & 0.3155  & 0.3219 \\
$\sigma_8$ & 0.7979 & 0.7891                & 0.7718 & 0.7848 \\
$S_8$ & 0.8120 & 0.8096                     & 0.7915 & 0.8129\\
\hline
$b_{1}$ & 1.977& $-$                        & 1.962 & $-$ \\
$c_{2}$ & 0.4058 & $-$                      & -0.0478 & $-$ \\
$c_{4}$ & $-$  & $-$                        & 3.999 & $-$ \\
$b_{3}$ & 0.7003 & $-$                      & -0.1567 & $-$ \\
$c_{ct}$ & -0.2901 & $-$                    & 0.0927 & $-$ \\
$c_{r,1}$ & -0.6575 & $-$                   & 1.246 & $-$ \\
$c_{e,0}$ & 1.706 & $-$                    & 2.131 & $-$ \\
$c_{e,1}$ & $-$ & $-$                       & 3.919 & $-$ \\
$c_{e,2}$ & -0.3780 & $-$                   & 0.1944 & $-$ \\
$\tilde c/[{\rm Mpc}/h]^4$ & $-$ & $-$    & 134.3 & $-$ \\
\hline
$\tilde b_{1}$ & $-$ & 2.181                & $-$ & 2.038 \\
$\tilde b_{2}$ & $-$ & -1.382               & $-$ & -2.725 \\
$b_{\mathcal{G}_2}$ & $-$ & 0.0977          & $-$ & -0.2013 \\
$b_{\Gamma_3}$ & $-$ & 0.0571               & $-$ & -0.3848 \\
$c_{0} / [{\rm Mpc}/h]^2$ & $-$ & 19.06     & $-$ & 23.27 \\
$c_{2}/[{\rm Mpc}/h]^2$ & $-$ & 43.88       & $-$ & 36.07 \\
$c_{e,0}$ & $-$ & 0.3509                    & $-$ & 0.5684 \\
$c_{e,1}$ & $-$ & -0.0440                   & $-$ & 0.4738 \\
$c_{e,2}$ & $-$ & 0.6255                    & $-$ & 0.4041 \\
$\tilde c/[{\rm Mpc}/h]^4$ & $-$ & 160.3    & $-$ & 111.5 \\
\hline
$\chi^2_{\rm min}$ & 352.6 & 343.7        & 336.0 & 336.4 \\
\hline
$N_{\rm data}$ &\multicolumn{4}{|c|}{344}\\
\hline
\end{tabular}
\caption{MAP of the cosmological parameters and EFT parameters corresponding to the analyses of Fig.~\ref{fig:prior}, obtained either with the WC or the EC prior. 
For clarity, we only show the EFT parameters associated to the NGC $z_3$ sky-cut. 
We also report the associated effective $\chi^2$ values.
Here we quote the MAP, as defined in the main text, which is \emph{not} the values obtained maximizing the likelihood where the EFT parameters that enter the model linearly are marginalized over analytically. The MAP can be obtained with such likelihood~\cite{DAmico:2020kxu} (see also Ref.~\cite{DAmico:2022osl}), but it is not sufficient to simply maximize this likelihood.
}
\label{tab:bestfit}
\end{table}

\begin{table}
\begin{tabular}{|l|c|c|}
 \hline

Parameter & WC ( $\to$ Pr. $\times 2$) & EC ($\to$ Pr. $\times 2$)\\
\hline
$h$ & 0.4 $\sigma$ ($\to 0.4 \sigma$) &  0.8 $\sigma$ ($\to 1.1 \sigma$) \\
$\omega_{\rm cdm}$ & 0.6 $\sigma$ ($\to 0.4 \sigma$) &  1.0 $\sigma$ ($\to 1.3 \sigma$) \\
$\text{ln}(10^{10} A_s)$ & -1.2 $\sigma$ ($\to -1.1\sigma$) & -1.3 $\sigma$ ($\to -2.4 \sigma$) \\
$n_s$  & -0.7 $\sigma$ ($\to -0.6 \sigma$) & -1.3 $\sigma$ ($\to  -1.6 \sigma$) \\
$\Omega_{\rm m}$  & 0.5 $\sigma$ ($\to$ 0.3 $\sigma$) & 0.8 $\sigma$  ($\to 1.1 \sigma$) \\
$\sigma_8$ & -1.2 $\sigma$ ($\to -1.3 \sigma$) & -2.0 $\sigma$  ($\to$ -2.7 $\sigma$) \\
$S_8$ & -1.0 $\sigma$ ($\to  -1.3 \sigma$) & -1.8 $\sigma$ ($\to$-2.3 $\sigma$) \\
\hline
\end{tabular}
\caption{
A summary of prior volume projection effects on the posterior mean: distance of the mean from the MAP.  $\sigma$ is taken as the $68\%$ C.L. error bars. The number in parenthesis give the distance when multiplying the prior width by two.
}
\label{tab:priorproj}
\end{table}

\begin{table}
\begin{tabular}{|l|c|c|c|c|}
 \hline

Parameter $X$ &  $\Delta X$(MAP) & $\to$ Pr. $\times 2$\\

\hline
$h$ & 0.2 $\sigma$ & $\to$ 0.1 $\sigma$ \\
$\omega_{\rm cdm}$ & -0.1 $\sigma$ & $\to$ -0.2 $\sigma$  \\
$\text{ln}(10^{10} A_s)$ & 0.5 $\sigma$ & $\to$ 0.1 $\sigma$ \\
$n_s$  & 0.9 $\sigma$ & $\to$ -0.4 $\sigma$  \\
$\Omega_{\rm m}$  & -0.2 $\sigma$ & $\to$ -0.3 $\sigma$  \\
$\sigma_8$ & 0.2 $\sigma$ & $\to$ -0.3 $\sigma$ \\
$S_8$ & 0.1 $\sigma$ & $\to$ -0.4 $\sigma$ \\
\hline
\end{tabular}
\caption{A summary of prior weight effects on the MAP: distance $(X^{\rm WC}-X^{\rm EC})$ between the MAP obtained with the WC and EC prior in units of $\sigma$, the average of the $68\%$-CL error bars derived from the two priors.    The number in the right column give the distance when multiplying the prior width by two.
}
\label{tab:priorweight}
\end{table}

\section{Impact of EFT priors in $\Lambda$CDM}\label{sec:impactprior}

\subsection{Highlighting the role of the priors}
To illustrate the impact of the prior choice, we compare the marginalized posteriors of the cosmological parameters within $\Lambda$CDM obtained with one or another prior choice (WC or EC), using the exact same data measurements, at the exact same scale cut and number of multipoles. In Fig.~\ref{fig:prior}, we show the results when analyzing  $P_\textsc{quad}^{z_1/z_3}$ as specified in Tab.~\ref{tab:twopoint_summary}, with the same analysis configuration, namely we fit two multipoles, $\ell=0,2$, and use $k_{\rm max}=0.20/0.25\hinvMpc$ for the $z_1/z_3$ redshift bins. 
Additional comparisons with different data configurations are provided in App.~\ref{app:scalecut}, Figs.~\ref{fig:prior_3mult} and~\ref{fig:CLASSPT_vs_PyBird}.

Let us quote the largest shifts for two analysis configurations: 
\begin{itemize}
    \item[\textbullet] Fitting $\ell=0,2$ at $k_{\rm max} = 0.25 \hinvMpc$ in $z_3$ (\emph{i.e.}, the \code{PyBird} native configuration), we find differences $<0.5 \sigma$ on all cosmological parameters between the two likelihoods, except larger ones on $\ln(10^{10}A_s)$, $\sigma_8$, and $S_8$, of $1.2\sigma$, $1.1\sigma$ and $0.9\sigma$.
   \item[\textbullet] Fitting $\ell=0,2,4$ at $k_{\rm max} = 0.20 \hinvMpc$ in $z_3$ (\emph{i.e.}, the \code{CLASS-PT} native configuration), we find differences $<0.5 \sigma$ on all cosmological parameters between the two likelihoods, except large ones on $\ln(10^{10}A_s)$, $\Omega_m$, $\sigma_8$ and $S_8$, of $1.2 \sigma$, $0.7\sigma$, $1\sigma$, and $0.7\sigma$.
 \end{itemize}

This shows that the choice of prior on the EFT parameters can lead to differences in the posteriors.
These can arise either from prior weights, in the sense that the allowed ranges are informing (potentially disfavoring) the ``true'' value that the EFT parameters want to take; or the prior volume lead to important projection effects, given the large number of EFT parameters that we marginalize over. \\

\paragraph{Prior volume projection effects.}
One way to estimate the prior volume projection effects is to compare the MAP values in Tab.~\ref{tab:bestfit} to the $68\%$-credible intervals in Fig.~\ref{fig:prior}. 
We summarize those shifts in Tab.~\ref{tab:priorproj}. 
In particular, one can compute the shifts of the mean to the MAP, where the MAP is (by definition) \emph{not} affected by prior volume projection effects. 
Here we refer to the ``MAP'' as the most likely value obtained by maximizing the likelihood of the data together with a conditional probability distribution given by the prior chosen for the EFT (nuisance) parameters. 
We stress that to obtained such MAP, the nuisance parameters are not marginalized over, {\it i.e.}, they are not integrated over given their prior probability distribution.

With the EC prior, we find for some cosmological parameters that the MAP values are not lying within the $68\%$-credible intervals: 
for example, we find shifts of $\sim 2\sigma$ on $\ln(10^{10}A_s)$, $\sigma_8$, or $S_8$. 
With the WC prior, we find that the MAP and the mean are consistent at $\lesssim 1.2\sigma$ for all cosmological parameters.
These shifts are particularly relevant when assessing the level of tension with the $\sigma_8$ and $S_8$ measurements from \Planck{}. While it might appear that $\sigma_8$ measured from EFTBOSS data are systematically lower than those deduced from \Planck{} under $\Lambda$CDM, we find here that a large part of the apparent tension comes from a projection effect that shift the $\sigma_8$ value by $1.2\sigma$ and $2\sigma$ for the WC and EC prior respectively compared to the MAP (and by a similar amount for $S_8$). In fact, the MAP we derived for both priors (see Tab.~\ref{tab:bestfit}) is in very good agreement with the reconstructed value from \Planck{} TTTEEE+lowE+lensing under $\Lambda$CDM, $\sigma_8=0.8111\pm0.0060$ \cite{Planck:2018vyg}. 

Finding smaller prior volume projection effects with the WC prior than with the EC prior is consistent with the fact that the prior widths for the EFT parameters are, in general, slightly more restrictive in the WC prior than in the EC prior (see discussion in Sec.~\ref{sec:prior}). 
To further demonstrate the prior volume effect, we increasing the prior widths for the EFT parameters by a factor of two. 
One can see from Fig.~\ref{fig:priorx2} and Tab.~\ref{tab:priorproj} that the prior volume projection effects grow as expected: the mean-to-MAP distances are now up to $\sim 1.3\sigma$ with the WC prior and up to $\sim 2.7\sigma$ with the EC prior, with $\sigma_8$ suffering again from the largest projection effect. 
A similar analysis was recently performed in Ref.~\cite{Carrilho:2022mon} in the context of $\Lambda$CDM and a model of dark energy with a free-to-vary equation of state $w$ and interaction rate with dark matter. Working with the EC priors defined above, they show that broadening the width of the priors can strongly affect posteriors distributions of cosmological parameters, in good agreement with our findings.
A more complete diagnosis would be to look at the profile likelihoods, that are however computationally challenging to obtain. 
We discuss this frequentist approach in Sec.~\ref{sec:beatprior}. \\

\paragraph{Prior weight effects.}
One simple way to quantify the effect due to the prior weight is to consider the MAP from the two prior choices, given in Tab.~\ref{tab:bestfit}. Indeed, these are not affected by the projection effects discussed above, which only occur when performing the marginalization integrals over the EFT parameters (within their priors), and therefore are mostly biased by the prior weight effect (barring computational errors / inaccuracies).

In Tab.~\ref{tab:priorweight}, we quantify the consistency between the most-likely values of all cosmological parameters $X$ derived with the two prior choices (EC or WC), by computing the distance $(X^{\rm WC}-X^{\rm EC})/\sigma$, where $\sigma$ is taken as the average of the $68\%$-CL error bars derived from the two priors.~\footnote{In principle, it would be more accurate to estimate the consistency between the best-fits via a profile likelihood. We take the  $68\%$-credible intervals  obtained from the posterior distribution as a simple proxy, although these are potentially affected by the projection effects mentioned above.} One can see that they are different at $\lesssim 1\sigma$, with the largest difference being for $n_s$. 
It is also informative to compare the $\min \chi^2$ values, in order to check whether the fit is acceptable for both priors.
From Tab.~\ref{tab:bestfit}, we see that with the EC prior, the $\Lambda$CDM model leads to a slight better $\min \chi^2$ at $\Delta\chi^2 \sim 9$ than with the WC prior, but also introduces $2$ extra free parameters per sky-cut.
Assuming all data points and parameters to be uncorrelated, we estimate that both prior choices lead to a comparable goodness-of-fit, with a $p{\rm -value} \simeq5\%$. 

Finally, to further demonstrate the role of the prior in informing the determination of the cosmological parameters, we enlarge the allowed range for the EFT parameters in both prior choices by a factor of two.~\footnote{For the WC prior, we also free $c_4$ and $c_{\epsilon,1}$ with range $4$, and add the next-to-next leading redshift-space counterterm $\tilde c$ as in the EC prior, such that the two priors have equivalent sets of associated theoretical predictions.}
We now find that the $\min \chi^2$ values are comparable: $336.0$ and $336.4$ from the WC and the EC prior, respectively, with corresponding $p$-values $\simeq 7\%$.
More importantly, the most-likely values of the cosmological parameters are now compatible at $\lesssim 0.4\sigma$ (compared to $\lesssim 1\sigma$ before enlargement).\\

\paragraph{Summary:}
On the one hand, we have shown that prior volume projection effects lead to shifts up to $\sim 1\sigma$ and $\sim 2\sigma$ on the posterior means from the WC and EC prior, respectively (see Tab.~\ref{tab:priorproj}). 
This effect is particularly noticeable in shifting downward the mean value of $\sigma_8$, which lead to an apparent small tension with \Planck{} under $\Lambda$CDM (at $1.5\sigma$ and $2.5\sigma$ for the WC and EC prior respectively), compared to the MAP that is in good agreement with \Planck{} at $\lesssim 0.5\sigma$ for all prior choices.
The prior weight effects, on the other hand, are responsible for differences in the most-likely values up to $\sim 1\sigma$ between the two prior choices.
Additionally, the $\Lambda$CDM model provides an acceptable description of the data regardless of the prior.
Let us stress that the effects from the prior that we have found here are sizeable (with respect to the error bars) only because current data are of relatively small volume (and therefore larger statistical errors). 
In the following, we argue that those effects becomes less relevant as soon as more data are added in the cosmological analysis. 

Before moving on, we make the following comment. 
One may wonder if the present study allows us to draw lesson on how to choose appropriately priors on the EFT parameters. 
We have demonstrated that the two EFT priors allow for the same maximal likelihood point once enlarged enough. 
This is expected since we stress again that the two parametrizations are equivalent, as they are 
related by a change of basis to each other: as such, once the prior is large enough, the prior weight becomes negligible with respect to the likelihood of the data, and the maximal likelihood point is recovered. 
Therefore, one possible criteria to choose the prior is to require that the size of the one-loop contribution stays smaller than the tree-level, such that the perturbative nature of the theory is preserved. 
Progress in this direction are ongoing. 
Nevertheless, we anticipate than none of the choice for the EFT parameters satisfying such criteria will be immune to the prior volume projection effects given BOSS data volume. 
We therefore now move on to look at the situation given larger data volume.

\begin{figure*}
\centering
{\large \emph{Prior effects in current and forthcoming data}}\\[0.2cm]
\includegraphics[width=1.5\columnwidth]{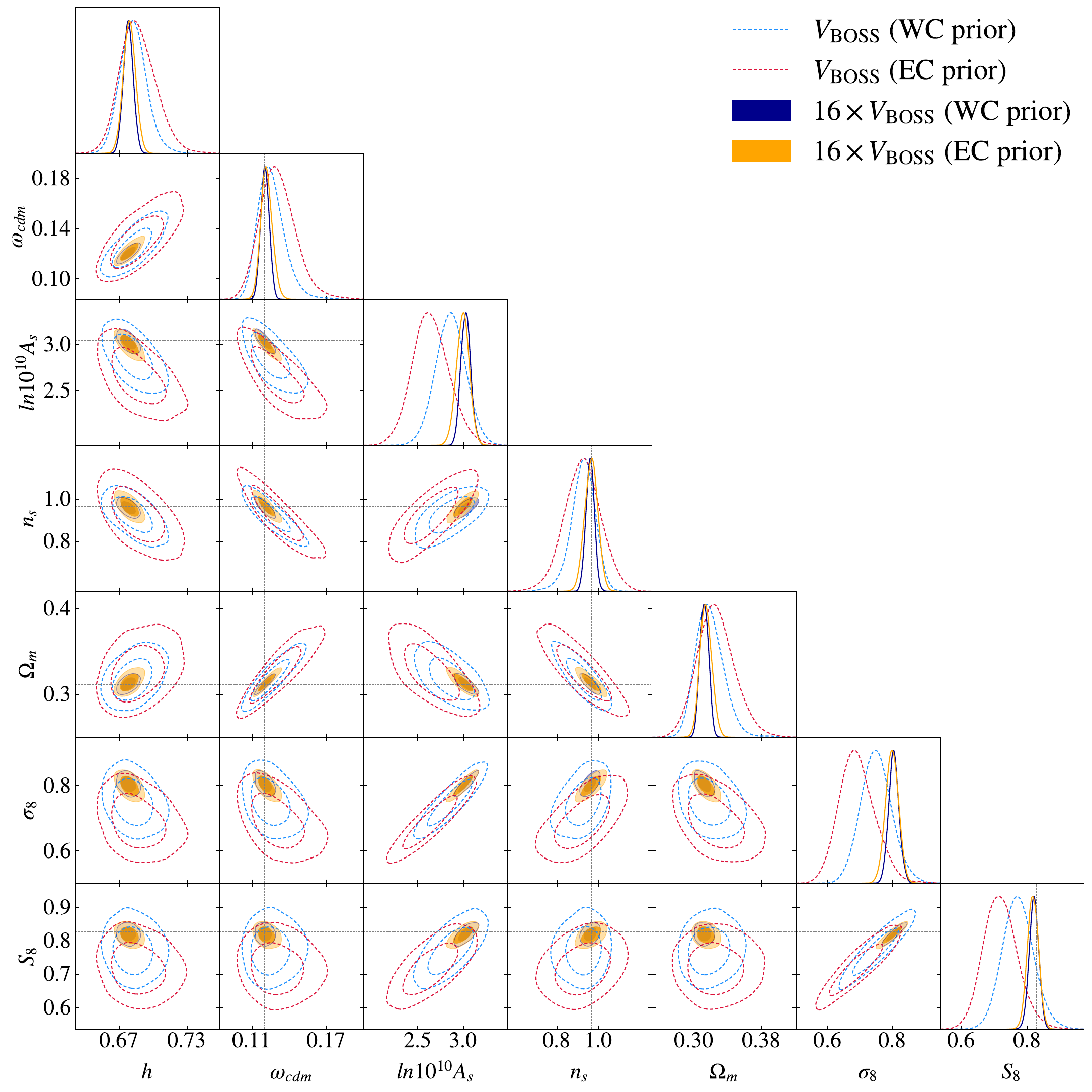}
\caption{
$\Lambda$CDM results (1D and 2D credible intervals) from the same likelihood as Fig.~\ref{fig:prior}, \code{PyBird}, but on noiseless synthetic data generated with the EFT prediction close to the MAP of BOSS. 
In particular, we use the same covariance as for the BOSS analysis, represented by $V_{\rm BOSS}$. 
We perform this analysis either with the WC prior or the EC prior.
The vertical lines represent the truth. 
For BOSS data volume, $V_{\rm BOSS}$, we observe shifts in the 1D posteriors from the prior effects up to $\sim 1.2\sigma$  for the WC prior, and up to  $\sim 2.0\sigma$ for the EC prior. 
For forthcoming survey-like data volume, $\sim 16\times V_{\rm BOSS}$, we see that the cosmological parameters are instead recovered  at $\lesssim 0.5\sigma$ with the WC prior and $\lesssim 0.7\sigma$ with the EC prior.
}
\label{fig:BOSS_Synth}
\end{figure*}

\begin{figure*}
\centering
{\large \emph{WC vs EC prior in combination with {\it Planck} data}}\\[0.2cm]
\includegraphics[trim=0.32cm 0cm 0cm 0cm, clip=true, width=1.5\columnwidth]{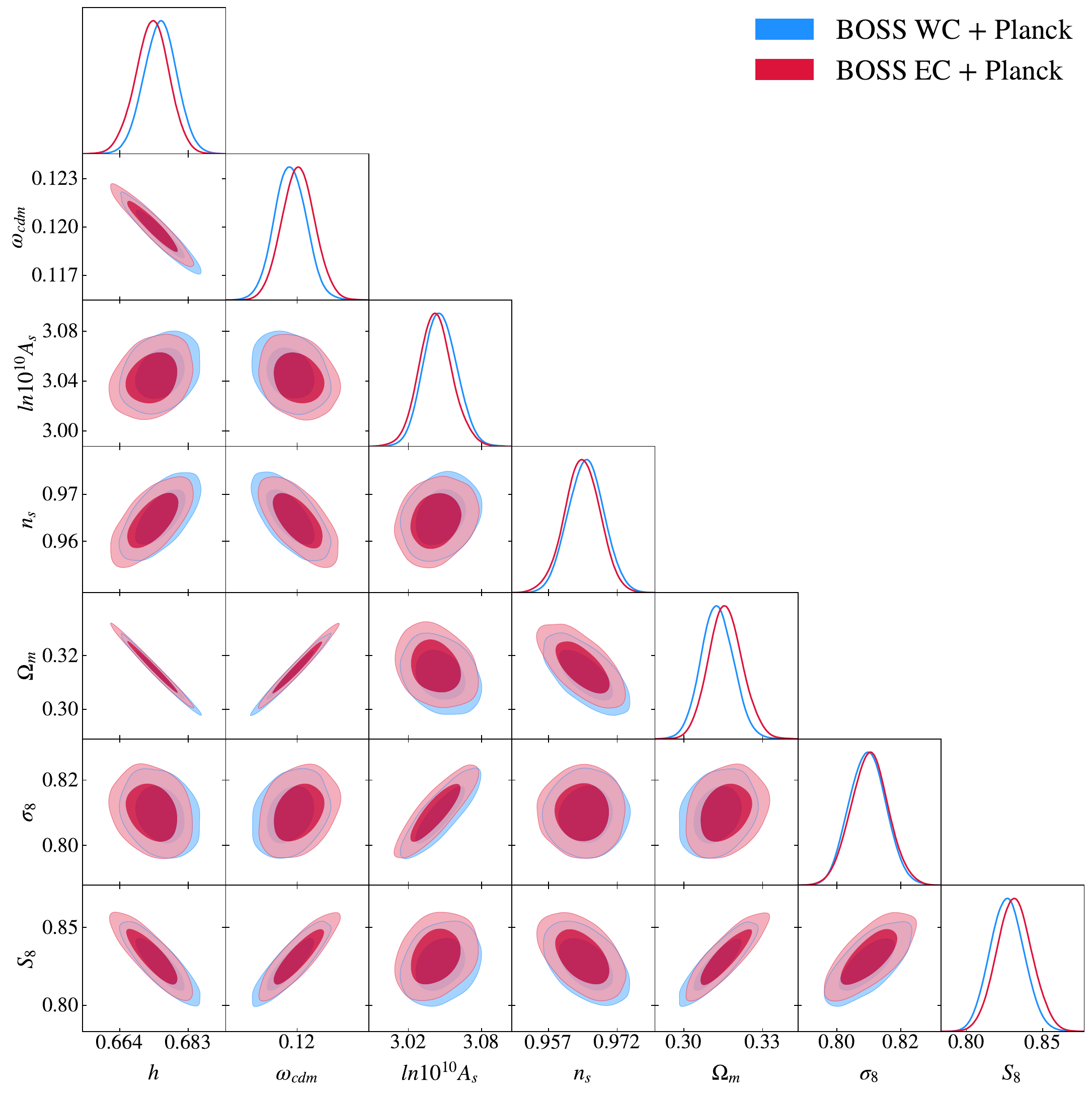}
\caption{
Once combined with {\it Planck} TT,TE,EE+lowE+lensing~\cite{Planck:2018vyg}, the full-shape analysis of BOSS with the EC and WC prior choices lead to similar results within $\lesssim 0.5\sigma$ on all cosmological parameters. 
}
\label{fig:BOSS_LCDM}
\end{figure*}

\subsection{How to beat the prior weights and volume effects}\label{sec:beatprior}

\paragraph{Forthcoming surveys data:}
While we have established that effects from the prior are of utmost importance for BOSS, one may ask whether these will still be important given a larger data volume, \emph{e.g.}, from forthcoming surveys such as DESI \cite{Aghamousa:2016zmz} or Euclid \cite{Amendola:2016saw}.
To answer this question, following Ref.~\cite{DAmico:2022osl}, we measure effect from the prior by fitting synthetic data generated with our prediction on the MAP of the data. 
The results are presented in Fig.~\ref{fig:BOSS_Synth} for the WC prior as well as for the EC prior. 
For the volume of BOSS, one can see as expected that the prior effects are important, as the posterior means are far away from the truth, namely a $\lesssim 1.2\sigma$ shift for the WC prior and a $\lesssim 2.0\sigma$ shift for the EC prior.
However, by re-scaling the covariance of BOSS by $16$, which corresponds roughly to the volume of the forthcoming galaxy surveys, one can see that the prior effects are less important: 
the shifts of the mean to the truth are now $\lesssim 0.5\sigma$ for the WC prior and $\lesssim 0.7\sigma$ for the EC prior.
There are several caveats to this simple exercise. 
First, here we have simply rescaled the covariance of BOSS, and used the synthetic data generated from the MAP to BOSS data. 
These are far from the specifications of forthcoming surveys in terms of targeting, shot noise, redshifts, etc., although we anticipate that this should not change the conclusions. 
Maybe more importantly, keeping in mind that the $k_{\rm max}$ is determined as the highest scale at which the theory error remains under control with respect to the statistical error, the $k_{\rm max}$ will presumably not be as high for larger data volume. 
Therefore, the size of the error bars seen in Fig.~\ref{fig:BOSS_Synth} are likely underestimated. 
This in principle can allow for more effects from the prior, which remain to be precisely quantified.
We refer to App.~C of Ref.~\cite{Nishimichi:2020tvu} as well as Ref.~\cite{DAmico:2021ymi} for more realistic prospects of the EFT analysis on a DESI-like surveys with the WC prior. 
Finally, the forthcoming data will be cut into very different redshift bins than the ones of BOSS. 
If one assigns one set of EFT parameters per redshift bin in the analysis, the thinner is the slicing, the bigger the prior volume will get. 
If this becomes an issue, one can imagine to be more informative, for example add a correlation on the EFT parameters from one redshift bin to another, given that one expects them to not be so different. 
This effectively reduces the number of EFT parameters to marginalize over, \emph{i.e.}, reduces the prior volume and the associated projection effects. 
We refer to Ref.~\cite{DAmico:2022osl} for a practical implementation of such correlated prior in an EFT analysis of BOSS data. \\

\paragraph{Combining with CMB:}
In Fig.~\ref{fig:BOSS_LCDM}, we show the combination of the EFT analysis of BOSS power spectrum, using either the WC or the EC prior, with {\it Planck} TTTEEE+lowE+lensing data \cite{Planck:2018vyg}. 
The inclusion of {\it Planck} data brings the two analyses into good agreement: we observe at most shifts $\lesssim 0.5\sigma$ on the means, and the errors bars are similar at $\lesssim 5\%$.
The {\it Planck} data represents a considerable data volume with respect to BOSS, such that it is not surprising that the cosmological constraints are dominated by {\it Planck}. 
As such, all prior effects observed earlier are then less prone to bias the cosmological results. 
\\

\paragraph{Profile Likelihood:}
Although we have shown that increasing the data volume, either from the survey or by combining with CMB experiments, help to mitigate prior effects, the question of how to extract reliable cosmological summary statistics from smaller data volume remains. 
One possibility is to go back to the frequentist approach: instead of sampling the likelihood to obtain posteriors that we then marginalize to get credible intervals, we can simply look at the profile likelihoods and read the confidence intervals.  In the context of \Planck~CMB data, Ref.~\cite{Planck:2013nga} showed that the frequentist analysis yields similar distribution as the Bayesian analysis within $\Lambda$CDM. However, it as already been pointed out that this is not necessarily the case for beyond-$\Lambda$CDM model, such as early dark energy \cite{Herold:2021ksg,Reeves:2022aoi,Gomez-Valent:2022hkb}. 
As we have illustrated, this can have several advantages over the Bayesian approach: one is free to choose very agnostic prior, \emph{i.e.}, broad prior ranges, thus avoiding potential bias from the prior weight, without paying the price of being subject to large prior volume projection effects, as the confidence intervals are not derived upon marginalization. Some efforts in this direction are in progress.

\begin{table*}
\begin{tabular}{|l|l|l|l|l|l|}
\hline
\multicolumn{6}{|c|}{\bf Pre-reconstructed measurements}\\
\hline
        & Ref.       & Estimator       & Code                    & Redshift split     & Window             \\ \hline
$\mathcal{P}_\textsc{fkp}^\textsc{lz/cm}$              & \cite{Gil-Marin:2015sqa} & FKP  & \code{Rustico}\footnote{\url{https://github.com/hectorgil/Rustico}}\cite{Gil-Marin:2015sqa}                 & LOWZ / CMASS       & Inconsistent norm. \\
$P_\textsc{fkp}^\textsc{lz/cm}$             & \cite{Zhang:2021yna} & FKP  & \code{PowSpec}\footnote{\url{https://github.com/cheng-zhao/powspec}}~\cite{Zhao:2020bib} / \code{nbodykit}\footnote{\url{https://github.com/bccp/nbodykit}}~\cite{Hand:2017pqn} & LOWZ / CMASS       & Consistent norm.   \\
$\xi^\textsc{lz/cm}$            & \cite{Zhang:2021yna} & Landy \& Slazay & \code{FCFC}\footnote{\url{https://github.com/cheng-zhao/FCFC}}~\cite{Zhao:2023iwb}                    & LOWZ / CMASS       & Window-free                 \\
$P_\textsc{fkp}^{z_1/z_3}$            & \cite{Beutler:2021eqq}~\footnote{\url{https://fbeutler.github.io/hub/deconv_paper.html}}  & FKP  & --                       &  $z_1$ / $z_3$ & Consistent norm.   \\
$P_\textsc{quad}^{z_1/z_3}$            & \cite{Philcox:2021kcw} & Quadratic       & \code{Spectra without Windows}\footnote{\url{https://github.com/oliverphilcox/Spectra-Without-Windows}}~\cite{Philcox:2020vbm} &  $z_1$ / $z_3$ & Window-free                 \\ \hline  
\multicolumn{6}{c}{}\\[-0.5em]
\hline
\multicolumn{6}{|c|}{\bf Post-reconstructed measurements}\\
\hline
& Ref. & -- & -- & Redshift split & Method \\ \hline
$\alpha^\textsc{lz/cm}_{rec}$           & \cite{Gil-Marin:2015nqa} & --               & --                       & LOWZ / CMASS       & \cite{DAmico:2020kxu}                 \\
$\alpha^{z_1/z_3}_{rec}$           & \cite{BOSS:2016hvq} & --               & --                       &  $z_1$ / $z_3$ &  \cite{DAmico:2020kxu}               \\
$\beta^{z_1/z_3}_{rec}$            & \cite{BOSS:2016hvq} & --               & --                       &  $z_1$ / $z_3$ & \cite{Philcox:2020vvt}                  \\ \hline
\end{tabular}
\caption{Comparison of pre-reconstructed and post-reconstructed BOSS two-point function measurements: reference, estimator, code of the measurements, redshift split [LOWZ: $0.2<z<0.43 \  (z_{\rm eff}=0.32)$, CMASS: $0.43<z<0.7  \ (z_{\rm eff}=0.57)$; $z_1$: $0.2<z<0.5 \  (z_{\rm eff}=0.38)$,  $z_3$: $0.5<z<0.7  \ (z_{\rm eff}=0.61)$], and window function treatment. 
For the post-reconstructed measurements, while we instead provide under ``Method'' the references presenting the algorithm used to extract the reconstructed BAO parameters and how the cross-correlation with the pre-reconstructed measurements is performed, ``Ref.'' now refers to the public post-reconstructed measurements used.
The SDSS-III BOSS DR12 galaxy sample data are described in Refs.~\cite{BOSS:2016wmc,Kitaura:2015uqa}. 
The pre-reconstructed measurements are from BOSS catalogs DR12 (v5) combined CMASS-LOWZ~\footnote{\url{https://data.sdss.org/sas/dr12/boss/lss/}}~\cite{Reid:2015gra}. 
}
\label{tab:twopoint_summary}
\end{table*}

\section{Comparison of BOSS measurements}\label{sec:measurements}

On top of various EFT prior choices, there are various BOSS two-point function measurements (that can be) used in full-shape analyses. 
Here, we present a detailed comparison on the posteriors obtained from the EFT analysis given various BOSS measurements. 
In particular, we ask what are the differences that can occur given the various treatments of the window functions. 
The characteristics of each measurements are listed in Tab.~\ref{tab:twopoint_summary}, while a more in-depth description is available in Sec.~\ref{sec:contenders}.
All analyses in this section are performed using the same pipeline: same prior choice on the EFT parameters, same scale cuts, and same number of multipoles, to ensure that we are only sensitive to differences due to the various measurements under scrutiny.

\subsection{Contenders} 
\label{sec:contenders}
Here we compare four pre-reconstructed and two post-reconstructed two-point function measurements from the BOSS sample, summarized in Tab.~\ref{tab:twopoint_summary}:
\begin{itemize}
    \item $P_\textsc{fkp}^\textsc{lz/cm}$: pre-reconstructed power spectrum measured for the full-shape analysis (abbreviated ``FS'' analysis in the following) presented in Ref.~\cite{Zhang:2021yna}. 
    The corresponding window functions were consistently normalized with $Q_0(s\rightarrow 0) \sim 0.9$ at vanishing separation, matching the measurements normalization (see App.~A of~\cite{Simon:2022adh}). 
    \item $\xi^\textsc{lz/cm}$: pre-reconstructed correlation function measured for the FS analysis presented in Ref.~\cite{Zhang:2021yna}. 
    The correlation function estimator is free from window function effects. 
    \item $P_\textsc{fkp}^{z_1/z_3}$: pre-reconstructed power spectrum measured in Ref.~\cite{Beutler:2021eqq}. 
    The corresponding window functions were consistently normalized matching the corresponding measurements normalization. 
    We analyze $P_\textsc{fkp}^{z_1/z_3}$ by deconvolving the window functions from the theory prediction by redefinition of the data vector and covariances at the level of the likelihood, as described in Ref.~\cite{Beutler:2021eqq}. 
    The window functions furthermore include the integral constraints~\cite{deMattia:2019vdg}. 
    \item $P_\textsc{quad}^{z_1/z_3}$: pre-reconstructed power spectrum measured using the quadratic ``window-free'' estimator of~\cite{Philcox:2020vbm}.
    \item $\alpha^\textsc{lz/cm}_{rec}$: BAO transverse and parallel parameters measured in Ref.~\cite{DAmico:2020kxu} from post-reconstructed power spectrum measured in Ref.~\cite{Gil-Marin:2015nqa}.
    \item $\alpha^{z_1/z_3}_{rec}$: BAO transverse and parallel parameters measured in this work (following methodology described, \emph{e.g.}, in Ref.~\cite{BOSS:2016hvq}) from post-reconstructed power spectrum measured in Ref.~\cite{BOSS:2016hvq}.
\end{itemize}

$P_\textsc{fkp}^\textsc{lz/cm}$, $\xi^\textsc{lz/cm}$, and $\alpha^\textsc{lz/cm}_{rec}$ are cut into LOWZ and CMASS redshift bins, $0.2<z<0.43 \  (z_{\rm eff}=0.32)$, $0.43<z<0.7  \ (z_{\rm eff}=0.57)$, respectively. 
$P_\textsc{fkp}^{z_1/z_3}$, $P_\textsc{quad}^{z_1/z_3}$ and $\alpha^{z_1/z_3}_{rec}$ are cut into  $z_1$ and $z_3$ redshift bins, $0.2<z<0.5 \  (z_{\rm eff}=0.38)$ and $0.5<z<0.7  \ (z_{\rm eff}=0.61)$, respectively. 
The scale cut for BOSS FS analysis has been determined on large-volume high-fidelity HOD simulations in Refs.~\cite{Colas:2019ret,Nishimichi:2020tvu,DAmico:2020kxu,Zhang:2021yna} and from a theory-error estimate in Ref.~\cite{Zhang:2021yna} for LOWZ / CMASS split to $(k_{\rm min}, k_{\rm max}) = (0.01, 0.20/0.23) \hinvMpc$ in Fourier space and $(s_{\rm min}, s_{\rm max}) = (25/20, 200) \Mpcinvh$ in configuration space.
When the data are split into  $z_1$ and $z_3$ instead, we rescale $k_{\rm max}$, using Eq.~(40) of~\cite{DAmico:2019fhj}, in order to have an equivalence with the LOWZ / CMASS separation. Especially, since  $z_3$ is effectively slightly higher redshift and with less data volume than CMASS, we re-scale the associated $k_{\rm max}$ to $k_{\rm max}^{z_3}=0.25 \hinvMpc$, while we keep $k_{\rm max}^{z_1}=0.20 \hinvMpc$.
Finally, we precise that the reconstructed BAO parameters are always combined with a FS analysis of pre-reconstructed measurements. 
$\alpha^\textsc{lz/cm}_{rec}$ and $\alpha^{z_1/z_3}_{rec}$ listed above for completeness will be compared in the next section. 

\begin{figure*}
    \centering
    {\large \emph{Comparison of BOSS pre-reconstructed measurements}}\\[0.2cm]
    \includegraphics[width=.90\textwidth]{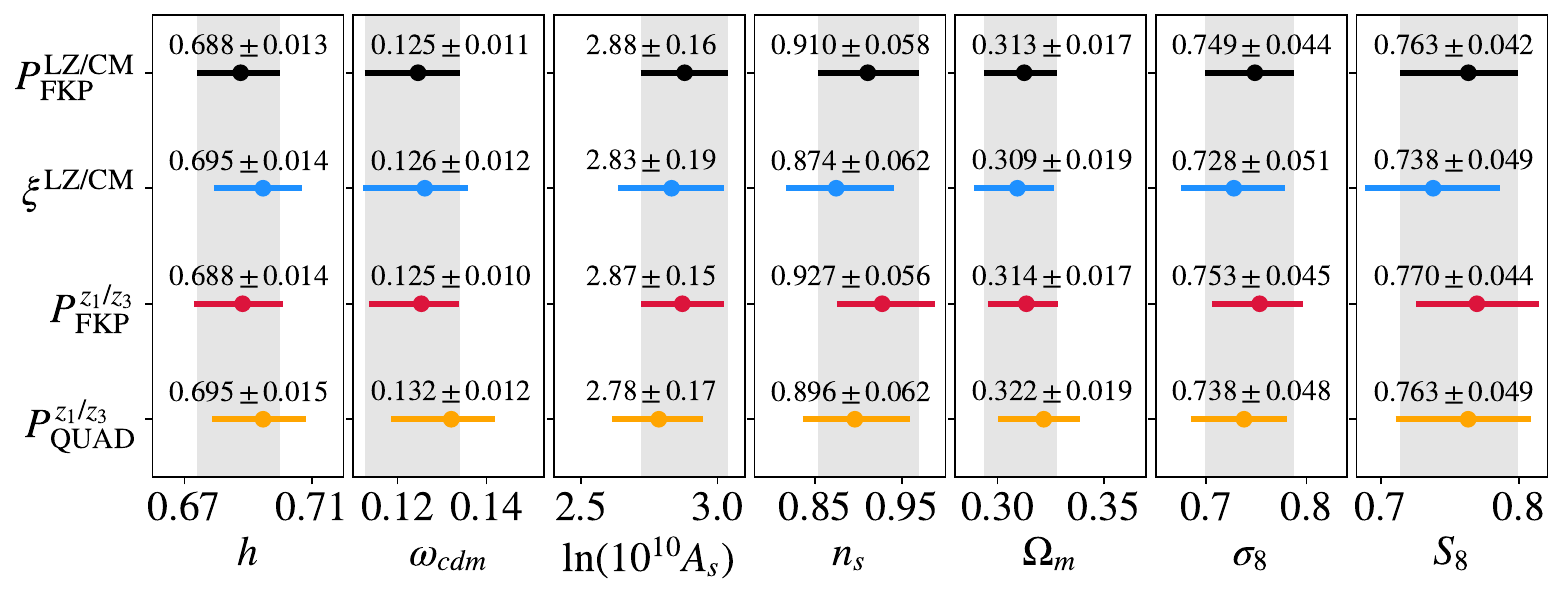}
\includegraphics[width=1.7\columnwidth]{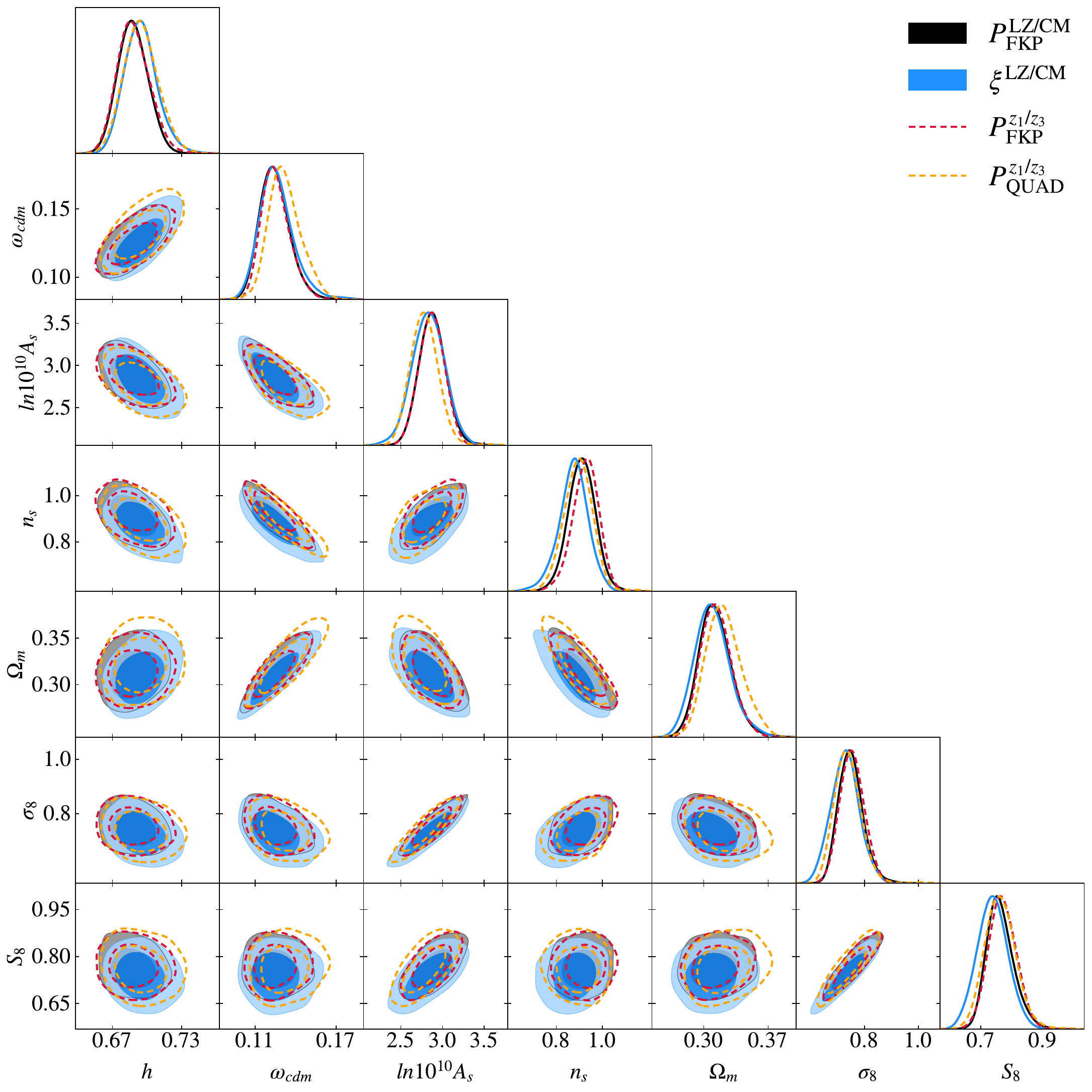}
\caption{
Comparison of $\Lambda$CDM results (1D and 2D credible intervals) from BOSS full-shape analyses of various pre-reconstructed two-point function measurements ($P_\textsc{fkp}^\textsc{lz/cm}, \xi^\textsc{lz/cm}, P_\textsc{fkp}^{z_1/z_3}, P_\textsc{quad}^{z_1/z_3}$). Details on the naming convention and relevant information are summarized in Tab.~\ref{tab:twopoint_summary}. 
The gray bands on the 1D posteriors are centered on the results obtained with $P_\textsc{fkp}^\textsc{lz/cm}$.
}
\label{fig:measurements}
\end{figure*}
 
\subsection{The matchups} 
\label{sec:matchups}

We now compare the cosmological results from a FS analysis within $\Lambda$CDM of the various BOSS data presented previously.
Summary of the cosmological results are given in Fig.~\ref{fig:measurements}.
 \\

We divide the contenders into the following matchups:\\ 

\begin{figure*}
\centering
    {\large \emph{Comparison of BOSS post-reconstructed measurements}}\\[0.2cm]
\includegraphics[width=0.90\textwidth]{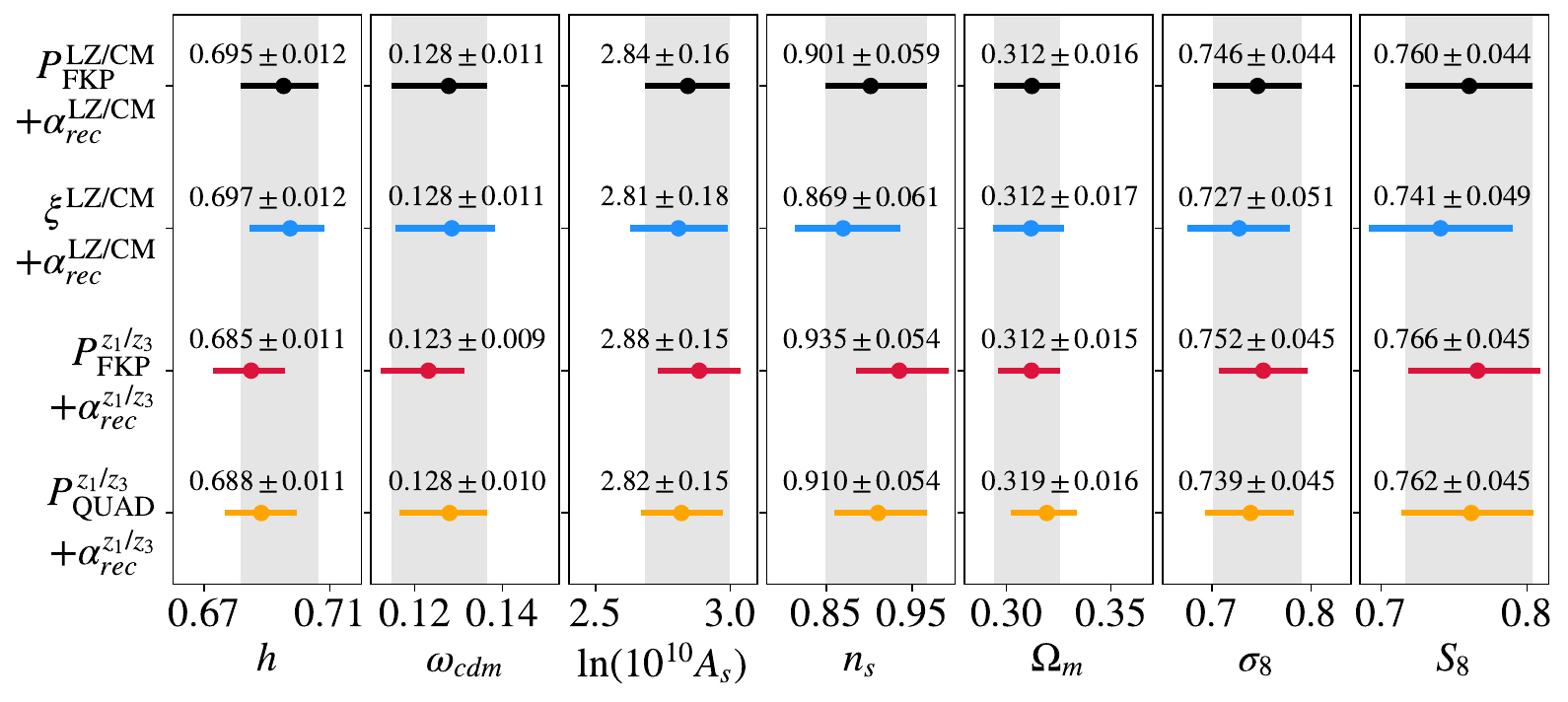}
\includegraphics[width=1.5\columnwidth]{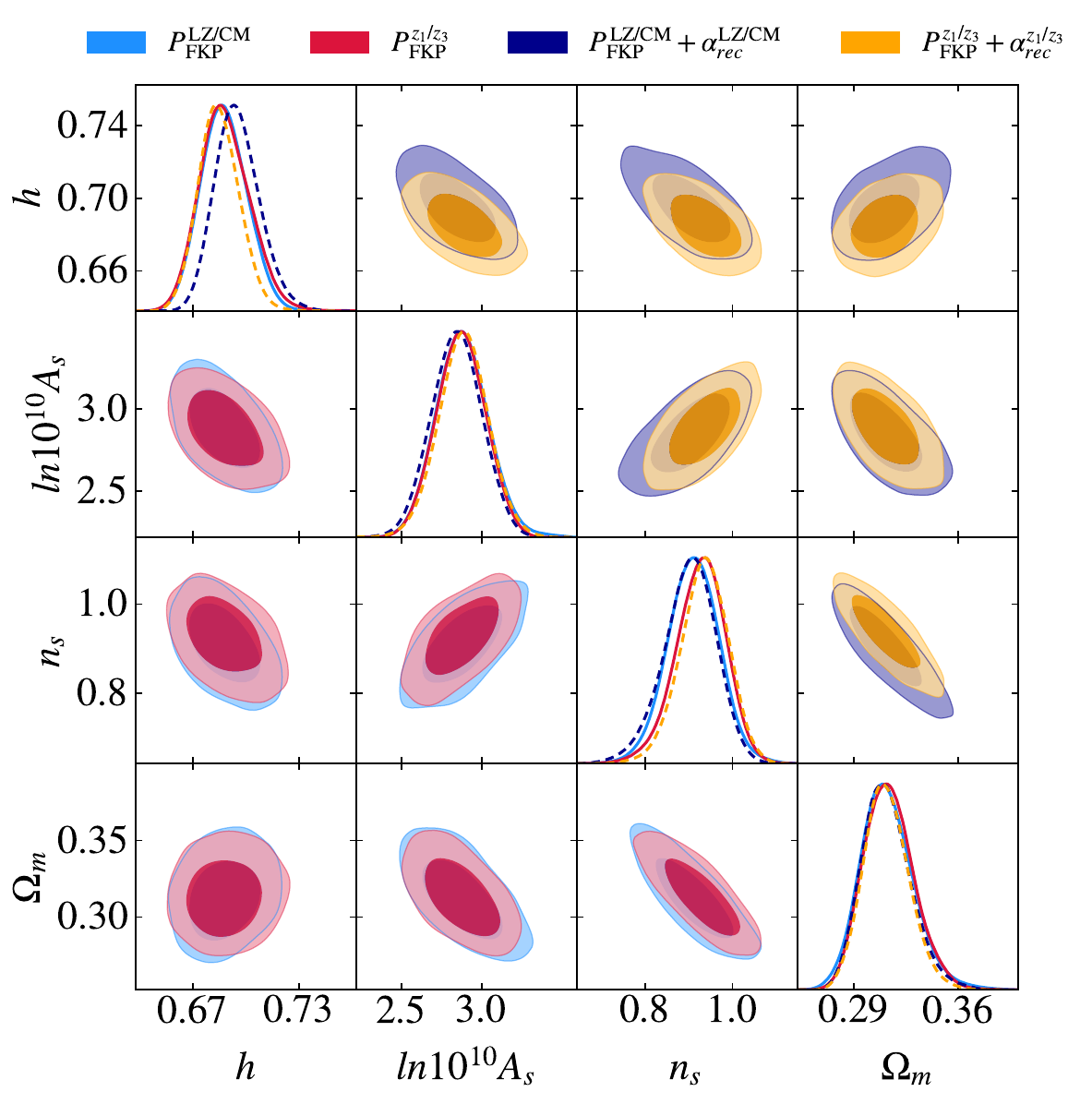}
\caption{{\it Upper panel:} same as the 1D posterior distributions of Fig.~\ref{fig:measurements} but combined with various post-reconstructed BAO parameters: $\alpha^\textsc{lz/cm}_{rec}$, $\alpha^{z_1/z_3}_{rec}$. 
The gray bands are centered on the results obtained with $P_\textsc{fkp}^\textsc{lz/cm} + \alpha^\textsc{lz/cm}_{rec}$. 
{\it Lower panel:} 2D posteriors from the full-shape analyses of BOSS power spectrum with two choices of redshift splits: $P_\textsc{fkp}^\textsc{lz/cm}$, $P_\textsc{fkp}^{z_1/z_3}$. 
We also show their combinations with $\alpha^\textsc{lz/cm}_{rec}$ and $\alpha^{z_1/z_3}_{rec}$, respectively.  
Details on the naming convention and relevant information are summarized in Tab.~\ref{tab:twopoint_summary}. 
While the two choices of redshift split lead to consistent results at $\lesssim 0.1\sigma$ on $h$, the addition of the BAO parameters, that extracted from the two available BOSS post-reconstructed measurements in Fourier space, lead to differences on $h$ of $\sim 0.9\sigma$. }
\label{fig:bao}
\end{figure*}

\paragraph{$P_\textsc{fkp}^\textsc{lz/cm}$ vs.~$\xi^\textsc{lz/cm}$ (\emph{i.e.}, the Fourier vs. configuration space matchup).}
Such matchup was already presented in Ref.~\cite{Zhang:2021yna} but with varying neutrino masses. 
Here we re-do the same comparison with one massive neutrino fixed to minimal mass, finding similar conclusions: the difference in the 1D posteriors is about 
$\lesssim 0.6 \sigma$ for all cosmological parameters. 
Importantly, as seen in Fig.~\ref{fig:bao}, the consistency is brought to better agreement when the same reconstructed BAO parameters $\alpha^\textsc{lz/cm}_{rec}$ is added to both: $\lesssim 0.2 \sigma$ for all cosmological parameter, except on $\sigma_8$, $S_8$, and $n_s$, which are consistent at about $0.3 - 0.5\sigma$. 
Contrary to the other comparisons made here, the cosmological information between the two compared statistics is effectively quite different due to two reasons. First, the BAO signal is fully analyzed in configuration space, as it shows up as a peak around $110\Mpcinvh$, while the BAO wiggles in Fourier space above the scale cut are not analyzed. Second, the scale cuts are effectively different (see more discussions in Ref.~\cite{Zhang:2021yna}). 
Therefore, the addition of the same reconstructed BAO parameters effectively bring closer the BAO information content between the Fourier and configuration space analysis. 
However, we still expect some level of differences on the posteriors as the information content is not equivalent in the two analyses. 
In particular, as the correlation function is free from the window functions effect, such match between the two analyses tells us that the effect from the window function (normalization) is under relatively good control. 
$P_\textsc{fkp}^\textsc{lz/cm}$ and $\xi^\textsc{lz/cm}$ are thus declared both consistent. \\

\paragraph{$P_\textsc{fkp}^\textsc{lz/cm}$ vs. $P_\textsc{fkp}^{z_1/z_3}$ (\emph{i.e.}, the LOWZ / CMASS vs $z_1$ / $z_3$ redshift split matchup).}
We find that $P_\textsc{fkp}^\textsc{lz/cm}$ and $P_\textsc{fkp}^{z_1/z_3}$ and their respective window functions (consistently normalized), measured independently, are rather consistent ($\lesssim 0.3\sigma$). 
Here $P_\textsc{fkp}^{z_1/z_3}$ is analyzed by deconvolving the window from the theory predictions at the level of the likelihood as described in Ref.~\cite{Beutler:2021eqq}. 
Furthermore, \cite{Beutler:2021eqq} adds to the window of $P_\textsc{fkp}^{z_1/z_3}$ the integral constraints~\cite{deMattia:2019vdg}. 
Therefore, finding consistency between $P_\textsc{fkp}^\textsc{lz/cm}$ and $P_\textsc{fkp}^{z_1/z_3}$ gives us several important information: (i) it allows us to check the accuracy of the deconvolution procedure on BOSS data; (ii) it tells us that the integral constraints have minor effects on the cosmological results from BOSS; and (iii) that the LOWZ / CMASS and  $z_1$ / $z_3$ splits (and their respective scale cuts) lead to consistent cosmological measurements. 
$P_\textsc{fkp}^\textsc{lz/cm}$ vs. $P_\textsc{fkp}^{z_1/z_3}$ are thus declared both consistent. \\

\paragraph{$P_\textsc{fkp}^\textsc{lz/cm}$ vs. $P_\textsc{quad}^{z_1/z_3}$ (\emph{i.e.}, window vs. window-free matchup).} 
This comparison was initially performed in Ref.~\cite{Philcox:2020vbm} but using the \code{CLASS-PT} likelihood. 
Thanks to the \code{PyBird} likelihood, we find similar trend using the WC prior, with $P_\textsc{quad}^{z_1/z_3}$ leading to differences of about $0.5 - 0.6 \sigma$ on  $h$, $\omega_{\rm cdm}$,  $\ln(10^{10}A_s)$ and $\Omega_m$.
Similarly, $P_\textsc{fkp}^{z_1/z_3}$ and $P_\textsc{quad}^{z_1/z_3}$ are consistent at $\lesssim 0.6\sigma$ on all cosmological parameters. 
While Ref.~\cite{Philcox:2020vbm} argues that the $P_\textsc{quad}^{z_1/z_3}$ analysis is ``formally equivalent'' to the $P_\textsc{fkp}^{z_1/z_3}$ window-deconvolved analysis, we observe that the inverse covariance (schematically $W^{T} \cdot C^{-1} \cdot W$, where $W$ is the window function matrix, see again Ref.~\cite{Beutler:2021eqq}) in the deconvolved analysis is different than the inverse covariance built from measurements using the window-free quadratic estimator.
Another potential difference is the fact that $P_\textsc{fkp}^{z_1/z_3}$ is shot-noise subtracted while $P_\textsc{quad}^{z_1/z_3}$ is not. 
However, putting a prior centered on $1$ instead of $0$ (in unit of $\bar n^{-1}$) for the shot noise in the analysis $P_\textsc{quad}^{z_1/z_3}$ only shifts $\ln(10^{10}A_s)$ by $\sim 0.2\sigma$. 
Finally, we note that both $P_\textsc{fkp}^\textsc{lz/cm}$ and $P_\textsc{quad}^{z_1/z_3}$ are consistent with $\xi^\textsc{lz/cm}$ at $\lesssim 0.6\sigma$.

\subsection{Measurements comparison summary}

All in all, all BOSS pre-reconstructed full-shape measurements not affected by a window function normalization issue (see App.~A of \cite{Simon:2022adh} for a discussion about this issue and its impact on the cosmological parameters), namely $P_\textsc{fkp}^\textsc{lz/cm}$, $\xi^\textsc{lz/cm}$, $P_\textsc{fkp}^{z_1/z_3}$, and $P_\textsc{quad}^{z_1/z_3}$, measured from different estimators as figuring in Tab.~\ref{tab:twopoint_summary}, lead to broadly consistent results at $< 0.8\sigma$ on the 1D posteriors for all cosmological parameters, and with similar error bars within $\lesssim 10\%$ (see Fig.~\ref{fig:measurements}). 
To be more precise, taking $P_\textsc{fkp}^\textsc{lz/cm}$ as reference, the 1D posterior distribution of parameters reconstructed from $\xi^\textsc{lz/cm}$, $P_\textsc{fkp}^{z_1/z_3}$, and $P_\textsc{quad}^{z_1/z_3}$ are consistent at $\lesssim 0.6\sigma, \ 0.3\sigma$, and $0.6\sigma$, respectively. 
The addition of the same post-reconstructed BAO signal (by cross-correlation) to $P_\textsc{fkp}^\textsc{lz/cm}$ and $\xi^\textsc{lz/cm}$ brings them in consistency at $\lesssim 0.2\sigma$ for all cosmological parameters, with the exception of residual shifts of $\sim 0.3-0.5\sigma$ on $\sigma_8, \ S_8$, or $n_s$, as it can be seen on Figs.~\ref{fig:measurements}.

To summarize, we list the differences seen at the level of the posteriors (within $\Lambda$CDM), ordered from the most to the least important one, and the respective choices of measurements that they stem from: 
\begin{itemize}
    \item up to $0.6\sigma$ among all cosmological parameters from the choice of the power spectrum estimators ($P_\textsc{fkp}^\textsc{lz/cm}$ vs. $P_\textsc{quad}^{z_1/z_3}$);
    \item about $0.3 - 0.5\sigma$ on $\sigma_8, \ S_8$, or $n_s$, from the choice of Fourier-space analysis or configuration-space analysis ($P_\textsc{fkp}^\textsc{lz/cm}+\alpha^\textsc{lz/cm}_{rec}$ vs. $\xi^\textsc{lz/cm}+\alpha^\textsc{lz/cm}_{rec}$);
    \item $\lesssim 0.3\sigma$ on all cosmological parameters from the choice of the redshift bin split in either LOWZ and CMASS or  $z_1$ and $z_3$ ($P_\textsc{fkp}^\textsc{lz/cm}$ vs. $P_\textsc{fkp}^{z_1/z_3}$), as defined in Tab.~\ref{tab:twopoint_summary}. 
\end{itemize}
Besides the effects mentioned here, there are subleading ones affecting those comparisons that we have discussed above: the addition of the integral constraints in the analysis of FKP measurements or subtracting the shot noise in the power spectrum measurements lead to shifts of at most $\lesssim0.2\sigma$.
We now turn to the comparisons of reconstructed BAO parameters combined with the full-shape analysis.

\begin{figure*}[t!]
\centering
    {\large \emph{Comparison of BAO extraction methods}}\\[0.2cm]
\includegraphics[width=2.\columnwidth]{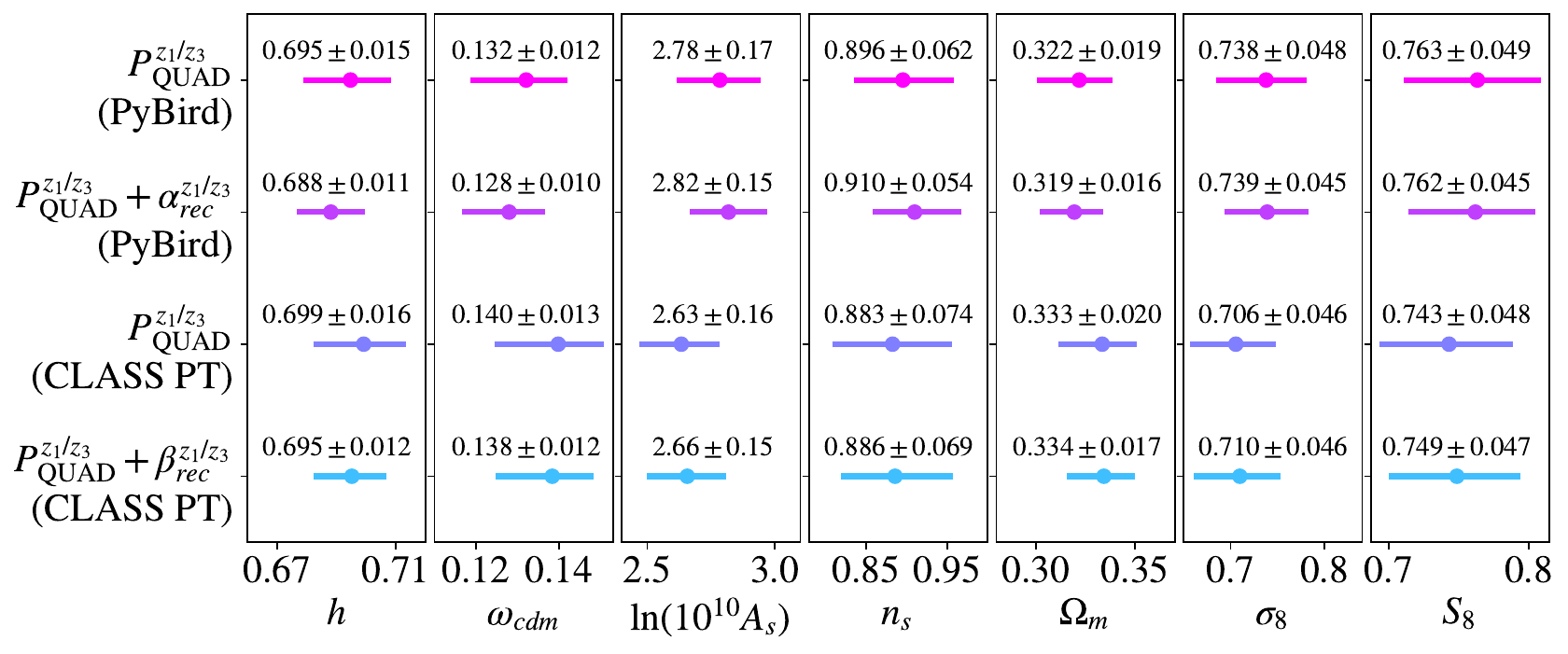}
\caption{
Comparison of $\Lambda$CDM results (1D credible intervals) from BOSS full-shape analyses using the \code{PyBird} likelihood or the \code{CLASS-PT} likelihood. 
The differences between the two likelihoods consist in the choices of prior on the EFT parameters, the number of multipoles analyzed and the value of $k_{\rm max}$. 
For the same pre-reconstructed measurements $P_\textsc{quad}^{z_1/z_3}$, although not analyzed with the same likelihood, one can see the differences from different BAO parameters, $\alpha^{z_1/z_3}_{rec}$ and $\beta^{z_1/z_3}_{rec}$, due to different extraction methods, since they are from the same post-reconstructed measurements. 
Relevant information regarding the measurements and their notations are summarized in Tab.~\ref{tab:twopoint_summary}. 
}
\label{fig:boss_bao_CLASSPT_pybird}
\end{figure*}

\section{Comparison of Reconstructed BAO}\label{sec:reconBAO}

\subsection{Inconsistency between post-reconstructed measurements} 

We here compare the two BOSS post-reconstructed measurements through the BAO parameters extracted with the same methods, as defined in previous section and in Tab.~\ref{tab:twopoint_summary}: $\alpha^\textsc{lz/cm}_{rec}$ vs. $\alpha^{z_1/z_3}_{rec}$. 
The results of this comparison are shown in Fig.~\ref{fig:bao}.
We find that adding the reconstructed signals $\alpha^\textsc{lz/cm}_{rec}$ and $\alpha^{z_1/z_3}_{rec}$ to $P_\textsc{fkp}^\textsc{lz/cm}$ and $P_\textsc{fkp}^{z_1/z_3}$, respectively, lead to substantial differences on the mean of $h$, at about $0.9\sigma$. 
This is to be contrasted with the consistency on $h$ that was better than $<0.1\sigma$ between $P_\textsc{fkp}^\textsc{lz/cm}$ and $P_\textsc{fkp}^{z_1/z_3}$ before the addition of the reconstructed BAO parameters. 
Indeed, the addition of the two reconstructed BAO measurements to the full-shape analysis shift $h$ in the opposite directions (see Fig.~\ref{fig:bao}).

Given that the reconstruction algorithm used for both reconstructed measurements is essentially the same, this is an unexpected result. 
Exploring the reconstruction algorithm is beyond the scope of this paper, and we leave a careful scrutiny of the reconstructed measurements to future work.
We observe that full-shape analyses combining the pre-reconstructed power spectrum either with reconstructed signal from configuration space~\cite{Chen:2021wdi}, or with the bispectrum analyzed at one loop up to $k_{\rm max} \sim 0.23 \hinvMpc$ (which comprises most of the additional information brought by the reconstructed signal)~\cite{DAmico:2022osl}, find shifts in $h$ in the same direction (and by a similar amount) as what we obtain when we add $\alpha^\textsc{lz/cm}_{rec}$, rather than $\alpha^{z_1/z_3}_{rec}$. Although the comparisons are far from straightforward given differences in the analysis setups, we take them as mild evidence that $\alpha^\textsc{lz/cm}_{rec}$ is more consistent than $\alpha^{z_1/z_3}_{rec}$ with what one should expect from the addition of the information from the reconstructed measurements. 
We nevertheless warn the reader that further studies are required to clarify this discrepancy. 
We note that the addition of the reconstructed BAO parameters also has an impact on $n_s$, since we have a shift of $0.6\sigma$ between 
$P_\textsc{fkp}^\textsc{lz/cm}+\alpha^\textsc{lz/cm}_{rec}$ and $P_\textsc{fkp}^{z_1/z_3}+\alpha^{z_1/z_3}_{rec}$, while the other parameters does not shift appreciably.\\

\subsection{Comparison of extraction methods of reconstructed BAO parameters}

After comparing the two available BOSS post-reconstructed measurements using the same BAO extraction methods,
$\alpha^\textsc{lz/cm}_{rec}$ and $\alpha^{z_1/z_3}_{rec}$, we now compare two sets of BAO parameters from the same post-reconstructed measurements, $\alpha^{z_1/z_3}_{rec}$ and $\beta^{z_1/z_3}_{rec}$, but extracted from two different methods as defined in the following.

The reconstructed BAO parameters are not obtained using the same methodology: in the \code{PyBird} likelihood, the BAO parameters ($\alpha^\textsc{lz/cm}_{rec}$ or $\alpha^{z_1/z_3}_{rec}$) are obtained following the standard method as described in, \emph{e.g.}, Ref.~\cite{BOSS:2016hvq}, while in the \code{CLASS-PT} likelihood, the BAO parameters ($\beta^{z_1/z_3}_{rec}$) are obtained following the method put forward in Ref.~\cite{Philcox:2020vvt}. 
The two methods are similar in spirit as they both focus on extracting the information from the reconstructed signal using only knowledge of ``the position of BAO peak'' through the Alcock-Paszinki parameters, as the broadband shape (and the BAO amplitude with respect to the broadband) is marginalized over. 
However, they differ slightly in their design.
In particular we note that in the \code{CLASS-PT} likelihood, some nuisance parameters such as the shot noise are not included in the model to fit the reconstructed power spectrum. 
Instead, an approximation for the theory error at high $k$ (where the shot noise contribution starts to be significant) is added to the data covariance to account for, among others, the shot noise contribution, which should be equivalent to the procedure in used by \code{PyBird} likelihood.~\footnote{There is an additional difference in the methodology, however, shown to be not relevant at the level of the constraints: $\beta^{z_1/z_3}_{rec}$ are obtained in Ref.~\cite{Philcox:2020vvt} by marginalizing over the damping of the BAO wiggles, while $\alpha^\textsc{lz/cm}_{rec}$ or $\alpha^{z_1/z_3}_{rec}$ are obtained following~\cite{BOSS:2016hvq} using a fixed damping amplitude parameter. As shown in Ref.~\cite{Philcox:2020vvt}, this does not lead to significant differences in the determination of the BAO parameters and their covariances.} 

In Fig.~\ref{fig:boss_bao_CLASSPT_pybird}, we can see the differences on the cosmological parameters arising from the two extraction methods. 
We compare $\alpha^{z_1/z_3}_{rec}$ with $\beta^{z_1/z_3}_{rec}$, that we remind that are from the same post-reconstructed measurements, combined with the same pre-reconstructed measurements $P_\textsc{quad}^{z_1/z_3}$, analyzed respectively with the \code{PyBird} or the \code{CLASS-PT} likelihood.
Here are the takeaways: 
\begin{itemize}
\item The addition of $\beta^{z_1/z_3}_{rec}$ to $P_\textsc{quad}^{z_1/z_3}$ in the \code{CLASS-PT} likelihood shifts $h$ in the same direction as the addition of $\alpha^{z_1/z_3}_{rec}$ to $P_\textsc{quad}^{z_1/z_3}$ in the \code{PyBird} likelihood, of about $1/3 \cdot \sigma$ and $1/2 \cdot \sigma$, respectively. 
This is expected, as the BAO parameters of $\beta^{z_1/z_3}_{rec}$ and $\alpha^{z_1/z_3}_{rec}$ are based on the same post-reconstructed measurements obtained in Ref.~\cite{BOSS:2016hvq} as seen in Tab.~\ref{tab:twopoint_summary}. 
\item The error bar reduction from the addition of the BAO parameters are quite comparable between the \code{PyBird} likelihood and the \code{CLASS-PT} likelihood. Indeed, taking the same pre-reconstructed measurements, $P_\textsc{quad}^{z_1/z_3}$, we find that the errors on $h, \ \ln(10^{10}A_s), n_s, \ \Omega_m$, and $\sigma_8$ are reduced respectively by $23\%, \ 13\%, \ 14\%, \ 18\%$, and $12\%$ in the \code{PyBird} likelihood when adding $\alpha^{z_1/z_3}_{rec}$, while they are reduced by $22\%, \ 3\%, \ 7\%, \ 16\%$, and $0\%$ in the \code{CLASS-PT} likelihood when adding $\beta^{z_1/z_3}_{rec}$. Therefore, keeping in mind that $\alpha^{z_1/z_3}_{rec}$ ans $\beta^{z_1/z_3}_{rec}$ are based on the same post-reconstructed measurements, we see that differences in the methods to extract and cross-correlate the BAO parameters lead to similar error bars within $\sim 10\%$. 
\end{itemize} 

To conclude, given the same post-reconstructed measurements, we do not find appreciable differences between the two extraction methods of reconstructed BAO parameters.

\section{Discussion and Conclusions}
\label{sec:conclusions}

The developments of the predictions for the galaxy clustering statistics from the EFTofLSS have made possible the study of BOSS data beyond the conventional analyses dedicated to extracting BAO and RSD information. 
The analyses available in the literature lead to differences on the reconstructed cosmological parameters that can be at the $1\sigma$ level. Given that they all come from the same BOSS data, this may be consider surprising and unsatisfactory. 
However, the analyses vary at a number of levels: the EFT parameters prior choices, the power spectrum estimator used for the measurements, the reconstructed BAO algorithm, the scale cut and the number of multipoles. 
In this paper, we have identified the analyses choices that can impact the cosmological constraints, and quantify the shifts in the full-shape analysis of BOSS power spectrum within $\Lambda$CDM. 
We summarize our findings below. \\

In Sec.~\ref{sec:prior}, we have looked  at two choices of prior used in previous BOSS full-shape analysis, the so-called ``West-coast'' (WC) and ``East-coast'' (EC) priors, that have been implemented in the \code{PyBird} and \code{CLASSPT} pipeline, respectively. 
Most importantly, we have identified that the prior assigned on the EFT parameters plays a non-negligible role in the determination of the cosmological parameters, for two reasons. 
\begin{itemize}
    \item First, in the Bayesian analysis, the marginalized constraints of the cosmological parameters are subject to prior volume projection effects from the marginalization over the EFT parameters, as the resulting posteriors are non-Gaussian.  We find that that prior volume projection effects shift the posterior mean from the MAP up to $\sim 1 \sigma$ with the WC prior and up to $\sim 2\sigma$ with the EC prior across all cosmological parameters.
    \item Second, from a frequentist perspective, we have found that the prior weight shift the MAP between the two analyses at $\lesssim 1\sigma$, with the $\min \chi^2$ different at $\Delta \chi^2 \sim 9$. 
    Once the prior range are enlarged by two, the MAP become consistent at $\lesssim 0.4\sigma$, and the $\min \chi^2$ are now comparable at $\Delta \chi^2 = 0.4$. 
    However, at the same time, the prior volume projection effect increases by up to $\sim 33\%$ depending on the prior choice and the cosmological parameters. 
    \item Nevertheless, we checked that when the pipelines follow the same prescription, results are in agreement at better than $0.2\sigma$.
    We conclude that the results between the two analyses are consistent, up to the various level of prior volume projection effects and prior weight effect, resulting from their respective choice of basis and more-or-less informative prior for the EFT parameters. 

\end{itemize}

We have then suggested several ways to mitigate the prior effects. 
\begin{itemize}
\item First, one can simply abandon the Bayesian view and come back to the frequentist one, for which the confidence intervals are not affected by prior volume projection effects as they are derived from profile likelihoods rather than from marginalized posteriors.
\item Setting aside the philosophical debate between Bayesian and frequentist, we have argued that for forthcoming larger data volume, all those prior effects will eventually become less important (with respect to the error bars).  In fact, the prior effects in the EFT analysis of BOSS have been realized only recently~\cite{DAmico:2022osl} because in the past, most of the validations, if not all, were performed with large-volume simulations.~\footnote{Note one exception in Ref.~\cite{Philcox:2021kcw}, where a large-volume simulation is analyzed with a covariance corresponding to BOSS total volume. 
Here the shift to the truth, that represents a sum of the theory error + prior effect, is find to be $\lesssim  0.4\sigma$ on all cosmological parameters. 
This is different than the shift we find in Fig.~\ref{fig:BOSS_Synth}, as in their case, there is only one sky, while in our case, we keep four skies as for the real analysis of BOSS data, with four independent sets of EFT parameters. When analyzing one-sky of synthetic data with covariance corresponding to BOSS total volume, we find $<1/5 \cdot \sigma$.}
 \item  Additionally, for the time being with BOSS, we have shown that when combined with \Planck{}, the results are less sensitive to those prior effects and the results are in good agreement. \\
\end{itemize}
For completeness, we have also scrutinized the impact on the cosmological constraints given various BOSS measurements. 
From the most significant to the least one, we have found: 
\begin{itemize}
    \item a difference of about $0.9\sigma$ on $h$ between the two public BOSS pre-reconstructed measurements in Fourier space ($\alpha^\textsc{lz/cm}_{rec}$ vs. $\alpha^{z_1/z_3}_{rec}$). This might constitute a warning that one should not use reconstructed measurements until this is clarified (see Sec.~\ref{sec:reconBAO} for more discussions);
    \item a difference of up to $0.6\sigma$ among all cosmological parameters between the FKP and quadratic estimators of the power spectrum ($P_\textsc{fkp}^\textsc{lz/cm}$ vs. $P_\textsc{quad}^{z_1/z_3}$);
    \item a difference of about $0.3 - 0.5\sigma$ on $\sigma_8$, $S_8$, and $n_s$, between the Fourier-space analysis and the configuration-space analysis ($P_\textsc{fkp}^\textsc{lz/cm}+\alpha^\textsc{lz/cm}_{rec}$ vs. $\xi^\textsc{lz/cm}+\alpha^\textsc{lz/cm}_{rec}$);
    \item Finally, a difference of $\lesssim 0.3\sigma$ on all cosmological parameters between the choices of redshift bin split in either LOWZ and CMASS or  $z_1$ and $z_3$ ($P_\textsc{fkp}^\textsc{lz/cm}$ vs. $P_\textsc{fkp}^{z_1/z_3}$). 
\end{itemize}

Besides the former EFT analyses of BOSS power spectrum using the \code{PyBird} likelihood or \code{CLASSPT} likelihood, let us also mention the work of Ref.~\cite{Chen:2021wdi} using another likelihood based on yet another public code developed independently, \code{Velocileptors}~\cite{Chen:2020fxs,Chen:2020zjt}. 
\code{Velocileptors} also implements predictions from a Lagrangian version of the EFTofLSS, which is equivalent, up to higher-order terms, to the Eulerian version of the EFTofLSS with IR-resummation~\cite{Senatore:2014via,Chen:2020zjt}. 
It would also be interesting to perform comparison with the \code{Velocileptors} likelihood with the prior choice used in the BOSS analysis of Ref.~\cite{Chen:2021wdi}. 
See some discussions in App.~\ref{app:direct}, and more importantly Ref.~\cite{Maus:2023rtr} that reaches similar conclusions as our current work on the prior volume projection effects in the EFT analysis within $\Lambda$CDM but with the \code{Velocileptors} pipeline. 
 Given that all analyses are equivalent in their parametrization (\emph{i.e.}, provide equivalent sets of fitting functions), all prior choices are equally motivated as long as they encompass the physically-allowed region of the EFT. For the current level of precision of the data, the various prior choices lead to various level of prior projection volume effect, but the results, \emph{i.e.}, MAP or multidimensional posteriors, are essentially the same.

We end the discussion with a closer look at $S_8$ and $\sigma_8$ in light of the BOSS full-shape analysis. 
At face-value, the $68\%$-credible interval on $S_8$ and $\sigma_8$ in this work are systematically lower than the value measured by \Planck{} under $\Lambda$CDM, with a statistical significance of $\sim 1.4\sigma$ ($2.2\sigma$) and $\sim 1.5\sigma$ ($2.5\sigma$) respectively for the WC (EC) priors. 
However, we have argued that part of this (small) discrepancy is due to a downwards shift compared to the MAP due to prior volume projection effect. These are more important for the EC prior ($\sim2\sigma$) than the WC prior ($\sim1.2\sigma$), and increase when doubling the widths of the EFT priors. 
In fact, the MAP values for  $S_8$ and $\sigma_8$ (Tab.~\ref{tab:bestfit}) measured with both priors are in good agreement with \Planck{} under $\Lambda$CDM, which infers $\sigma_8 =0.8111\pm0.006$ \cite{Planck:2018vyg} (see also~\cite{DAmico:2022osl,Amon:2022azi}).
Nevertheless, the values reconstructed from our analyses are also consistent with lower measurements of $S_8$ from lensing observations, see, \emph{e.g.},~\cite{Heymans:2012gg,Heymans:2020gsg,DES:2021wwk}. 
In fact, the analysis of BOSS data is done in the perturbative regime, \emph{i.e.}, we restrict the analysis at $k_{\rm max} \sim 0.2\hinvMpc$ where the EFTofLSS applies and in that sense, most of the cosmological information is from the large scales. 
In contrast, measurements of $S_8$ from lensing experiments rely on the modeling of small scales (way) beyond the nonlinear scales, where our EFT approach does not apply. 
Our reconstruction suggests that the deviation mostly occurs on scales smaller than those probed by our analysis (or at very low-$z<0.3$), although  because of large error bars, it is still compatible with a relatively low-$S_8$ on large scales, as hinted by the cross-correlation of DES and CMB lensing \cite{DES:2022urg}. 
For more discussion on the scale-dependence of the $S_8$ tension, we refer to Refs.~\cite{Lange:2020mnl,Amon:2022azi}. 

Finally, we mention that we have performed similar comparisons in the EDE model in a companion paper \cite{Simon:2022adh} to assess the level of robustness of the constraints on EDE.
Although the detailed comparisons we have performed in this series of papers helps quantifying at some level the systematic uncertainties associated with the measurements, we stress that we have not studied those related to BOSS galaxy catalog itself, which would require much more work. 
It will also be interesting to perform similar analysis for the bispectrum \cite{DAmico:2022osl,Philcox:2021kcw} as well as the recent eBOSS datasets \cite{eBOSS:2020yzd}, that can provide interesting additional constraining power on $\Lambda$CDM and extensions.

\begin{acknowledgements}
We thank Guido D'Amico, Marta Spinelli, and Oliver Philcox for useful discussions and insights. 
P.Z. would like to thank the organizers of the workshop \emph{LSS2022: Recent Developments in Theoretical Large-Scale Structure - IFPU} for hospitality in Trieste during the late stage of completion of this project. 
This work has been partly supported by the CNRS-IN2P3 grant Dark21. 
The authors acknowledge the use of computational resources from the Excellence Initiative of Aix-Marseille University (A*MIDEX) of the “Investissements d’Avenir” programme. These results have also been made possible thanks to LUPM's cloud computing infrastructure founded by Ocevu labex, and France-Grilles.
This project has received support from the European Union’s Horizon 2020 research and innovation program under the Marie Skodowska-Curie grant agreement No 860881-HIDDeN. This work used the Strelka Computing Cluster, which is run by Swarthmore College. T.L.S. is supported by NSF Grant No.~2009377, NASA Grant No.~80NSSC18K0728, and the Research Corporation.

\end{acknowledgements}

\appendix

\begin{figure*}
\centering
{\large \emph{WC vs EC prior: $P_\ell \ (\ell=0,2,4)$}}\\[0.2cm]
\includegraphics[width=1.87\columnwidth]{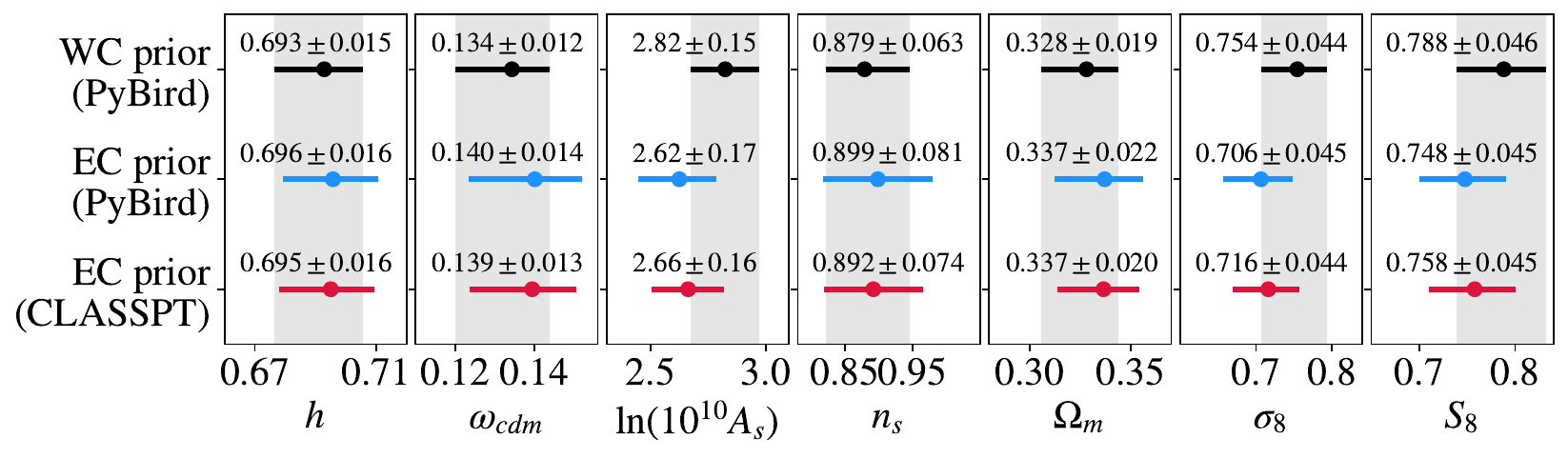}
\includegraphics[width=1.55\columnwidth]{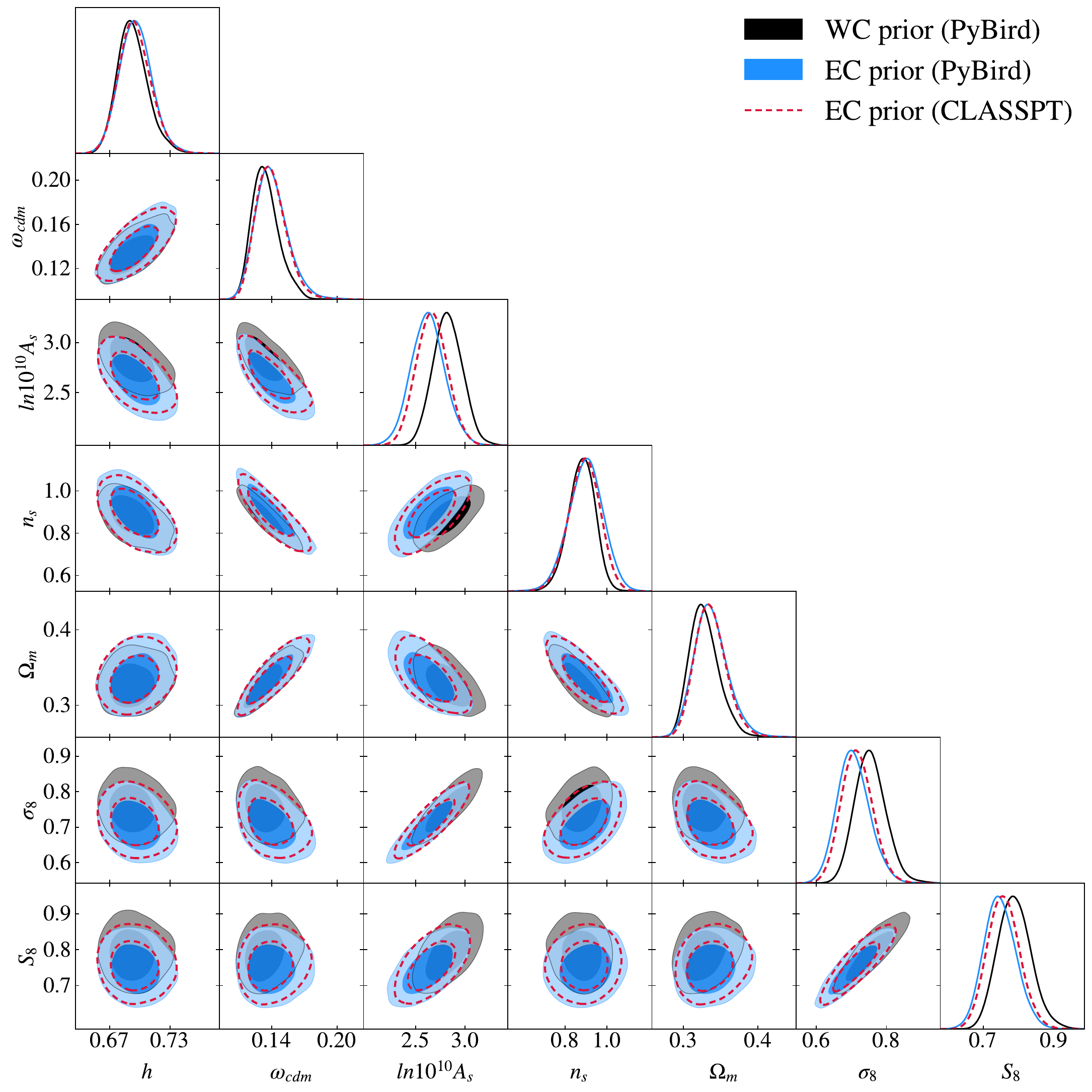}
\caption{
Comparison of $\Lambda$CDM results (1D and 2D credible intervals) from the full-shape analyses of BOSS power spectrum using the \code{PyBird} likelihood or the \code{CLASS-PT} likelihood. 
Here we use the same data measurements, $P_\textsc{quad}^{z_1/z_3}$ as specified in Tab.~\ref{tab:twopoint_summary}, and same analysis configuration:
we fit three multipoles, $\ell=0,2,4$, and use $k_{\rm max}=0.20/0.25\hinvMpc$ for the $z_1/z_3$ redshift bins.
Given the same prior choice, the EC prior, we reproduce from the \code{PyBird} likelihood the results from the \code{CLASS-PT} likelihood to very good agreement (see blue and red posteriors): we obtain shifts $\lesssim 0.2\sigma$ on the means and the errors bars similar at $\lesssim 10\%$. 
Given that the two pipelines have been developed independently, this comparison provides a validation check of their implementation.
In contrast, the WC and the EC prior choices lead to substantial differences on the 1D marginalized posteriors (see black and blue posteriors). The gray bands on the 1D posteriors are centered on the results obtained with the WC priors.
}
\label{fig:prior_3mult}
\end{figure*}

\section{Impact of scale cut and multipoles}\label{app:scalecut}

\begin{figure*}[t!]
    \centering
\includegraphics[width=0.90\textwidth]{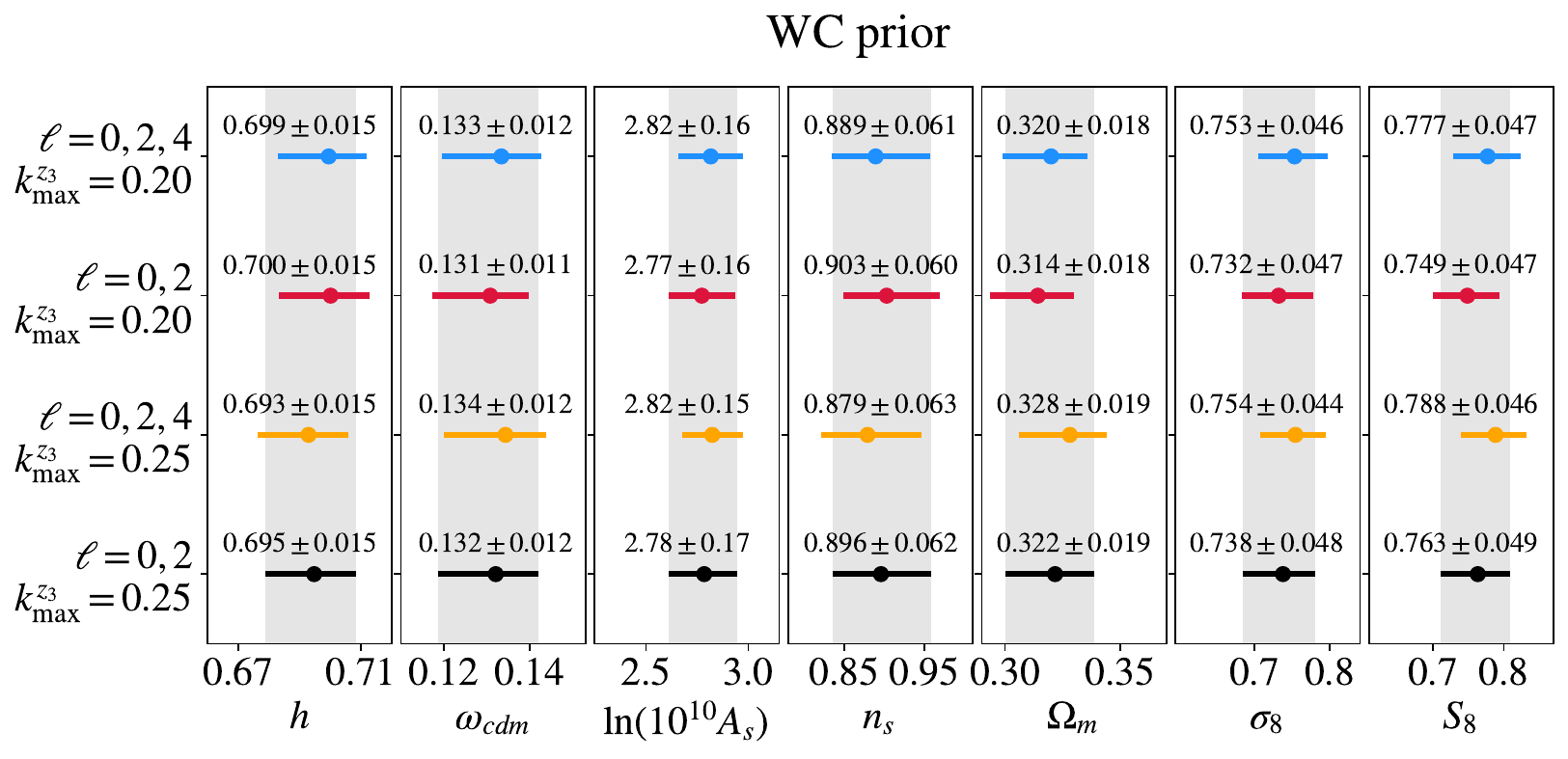}
\includegraphics[width=0.90\textwidth]{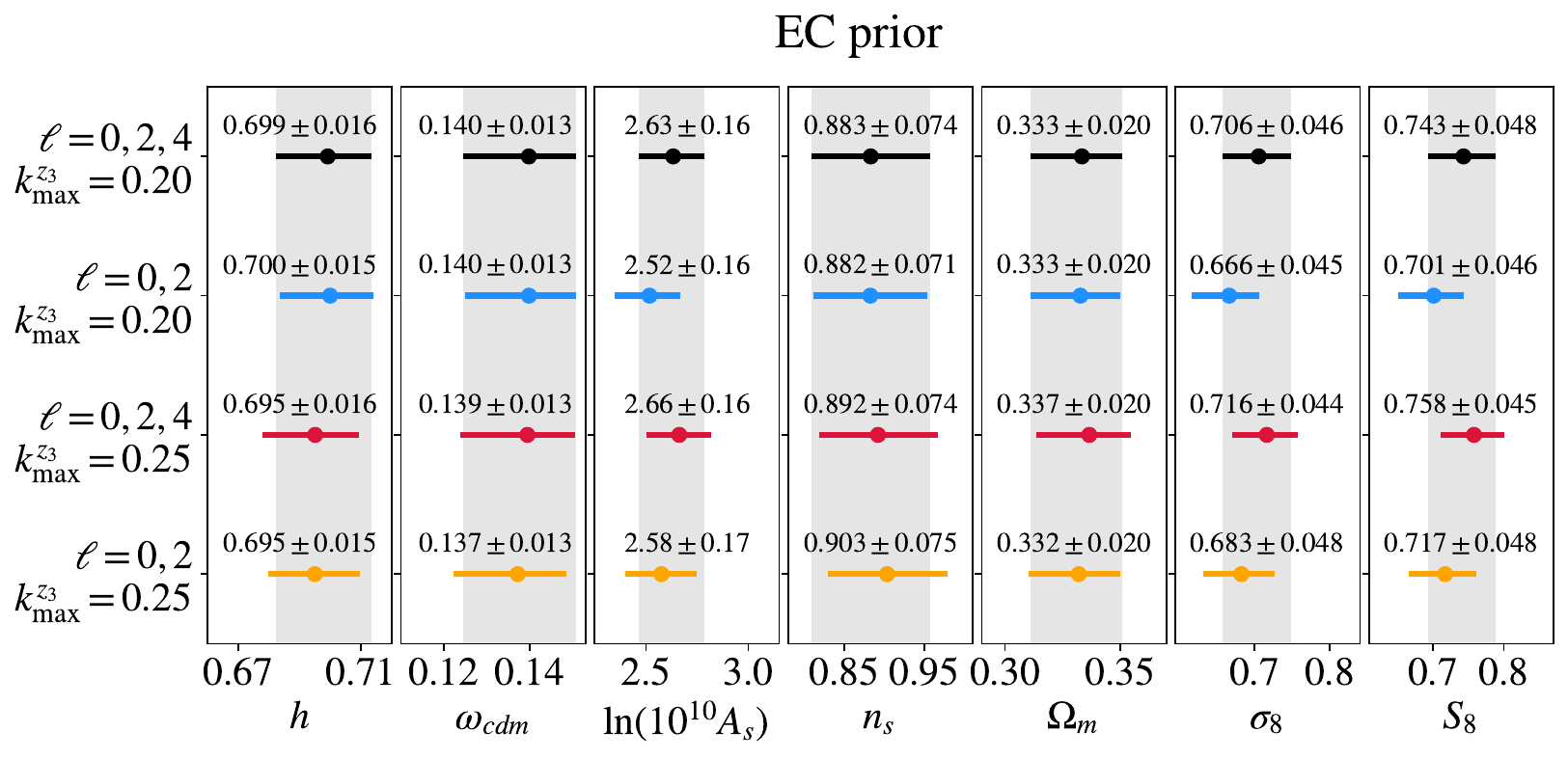}
\caption{
Comparison of $\Lambda$CDM results from BOSS full-shape analyses using the \code{PyBird} likelihood (WC prior) or the \code{CLASS-PT} likelihood (EC prior), for various analysis settings: number of multipoles analyzed ($\ell= 0,2$ or $\ell=0,2,4$), and $k_{\rm max}$ of $z_3$ ($k_{\rm max}^{z_3} = 0.20$ or $k_{\rm max}^{z_3} = 0.25$). 
$k_{\rm max}^{z_1} = 0.20$ for all analyses here, while all $k_{\rm max}$'s are given in $\hinvMpc$. 
Here we use the same data measurements,  $P_\textsc{quad}^{z_1/z_3}$, as specified in Tab.~\ref{tab:twopoint_summary}. 
The native baseline configurations used in previous BOSS full-shape analyses, highlighted by the gray bands, are $\ell=0,2$, $k_{\rm max}^{z_3} = 0.25$ for the \code{PyBird} likelihood, and $\ell=0,2,4$, $k_{\rm max}^{z_3} = 0.20$ for the \code{CLASS-PT} likelihood. 
}
\label{fig:CLASSPT_vs_PyBird}
\end{figure*}

In this appendix, we look at the differences when we change the scale cut and the number of multipoles analyzed. 
BOSS analyses using the \code{PyBird} likelihood usually include two multipoles, $\ell=0, 2$, with scale cut $k_{\rm max} = 0.25 (0.20) \hinvMpc$ for $z_3$ ($z_1$). 
The \code{CLASS-PT} likelihood include three multipoles, $\ell=0, 2, 4$, with scale cut $k_{\rm max} = 0.20 \hinvMpc$ for both $z_1$ and $z_3$.

In Fig.~\ref{fig:prior_3mult} we present a comparison between the WC and the EC prior, for the exact same data and configuration (same $k_{\rm max}$ and same number of multipoles), now considering three galaxy power-spectrum multipoles. 
This figure can be compared with Fig.~\ref{fig:prior}, where the same analysis was performed when considering two multipoles. 
One can see that, similar to Fig.~\ref{fig:prior}, the results of the \code{PyBird} and \code{CLASS-PT} likelihoods are in good agreement when using the same prior, also when the hexadecapole is included in the analysis.

Let us now look at how the results change when going from one choice of scale cut or multipoles to another one, for each prior choice. 
The results can be read from Fig.~\ref{fig:CLASSPT_vs_PyBird}, going from top to bottom, either looking in the upper panel or the lower panel. 
We find that: 
\begin{itemize}
    \item[\textbullet] with the WC prior, when either lowering the $k_{\rm max}$ from $0.25 \hinvMpc$ to $0.20 \hinvMpc$ in $z_3$, adding the hexadecapole, or changing both, we find at most a shift of $\lesssim 0.5 \sigma$ on the cosmological 1D posteriors. 
    \item[\textbullet] with EC prior, we find shifts up to about $0.3\sigma, \ 0.9 \sigma$, and $0.5\sigma$, respectively, when increasing the $k_{\rm max}$ from $0.20 \hinvMpc$ to $0.25 \hinvMpc$ in $z_3$, going from three to two multipoles, or changing both. 
\end{itemize} 

We stress that one does not expect the results between those various analysis settings to be the same, given that data are included (or removed). 
However, given that the signal-to-noise ratio of the hexadecapole is very low compared to the monopole and quadrupole, and that the data added between $k \in [ 0.20, 0.25] \hinvMpc$ are only a few bins, we expect to see only relatively small shifts in the posteriors. 
While this seems to be the case for the WC prior, the shifts are slightly larger for the EC prior when going from two to three multipoles.
As explained in previous section, the EC prior leads to larger prior volume projection effects, which can explain why we see larger differences in the current comparison.

\section{PyBird vs CLASS-PT: direct comparison}\label{app:direct}

\begin{figure*}[t!]
\centering
{\large \emph{\code{PyBird} likelihood vs \code{CLASS-PT} likelihood}}\\[0.2cm]
\includegraphics[width=1.82\columnwidth]{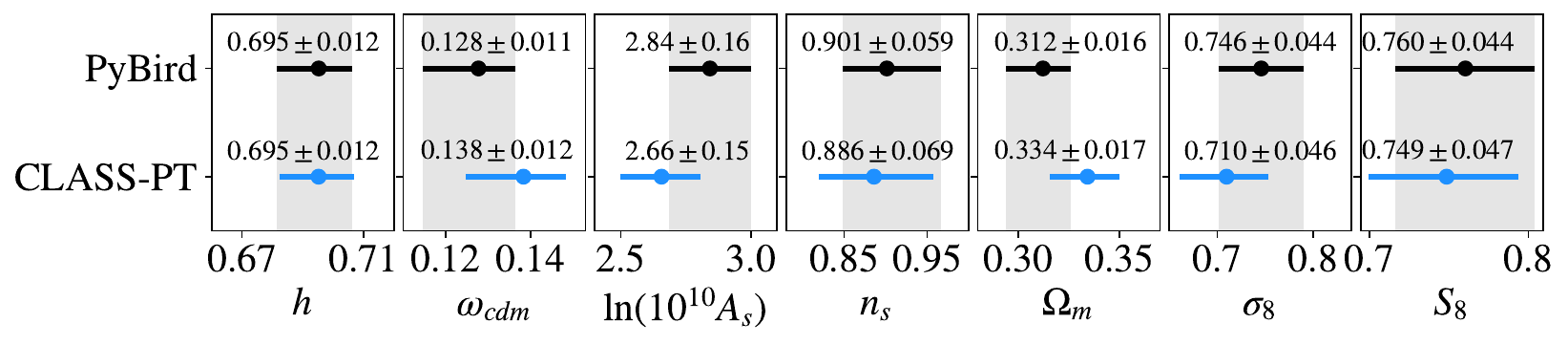}
\caption{
Comparison of $\Lambda$CDM results (1D credible intervals) from $P_\textsc{fkp}^\textsc{lz/cm}+\alpha^\textsc{lz/cm}_{rec}$ analyzed using the \code{PyBird} likelihood (\emph{i.e.}, the native data and configuration of \code{PyBird}), and $P_\textsc{quad}^{z_1/z_3}+\beta^{z_1/z_3}_{rec}$ analyzed using the \code{CLASS-PT} likelihood (\emph{i.e.}, the native data and configuration of \code{CLASS-PT}). 
Contrary to the analysis of Ref.~\cite{Zhang:2021yna} based on the \code{PyBird} likelihood, we fix the total neutrino mass to minimal, and, we do not use $Q_0$ or $B_0$ as Ref.~\cite{Philcox:2021kcw} in the \code{CLASS-PT} likelihood. 
The gray bands are centered on the results from \code{PyBird}. 
}
\label{fig:final_comparison}
\end{figure*}

For completeness, we provide now a comparison keeping all the analysis choices different in both likelihoods: the pre- and post-reconstructed measurements, scale cut, number of multipoles, and prior choices. This leads to the differences on the 1D posteriors that we see in Fig.~\ref{fig:final_comparison}. 
The differences between $P_\textsc{fkp}^\textsc{lz/cm}+\alpha^\textsc{lz/cm}_{rec}$ analyzed using the \code{PyBird} likelihood, with $P_\textsc{quad}^{z_1/z_3}+\beta^{z_1/z_3}_{rec}$ analyzed using the \code{CLASS-PT} likelihood, are about $0 \sigma, \ 0.9\sigma, \ 1.2\sigma, \ 0.2\sigma, \ 1.3\sigma, \ 0.8\sigma$ and $0.3\sigma$ on $h, \ \omega_{cdm}, \ \ln(10^{10}A_s), \ n_s, \ \Omega_m, \ \sigma_8$, and $S_8$, respectively.
As discussed in this paper, we have found that those differences are due to different prior choices, differences in the measurements used, and the full-shape analysis settings ($k_{\rm max}$ and number of multipoles). 
Therefore, if the shifts between the two base analyses do not seem to be that large in the end, $\lesssim 1.3\sigma$, we understand that there are cancellations arising from the different analysis choices.

As an intermediate result, we can compare these two likelihoods with the same dataset, i.e, by changing only the prior choices and the analysis settings ($k_{\rm max}$ and the number of multipoles). 
We find that the largest deviations between the \code{PyBird} likelihood and the \code{CLASS-PT} likelihood for $P_\textsc{quad}^{z_1/z_3}$ are on $\omega_{\rm cdm}$, $\ln(10^{10}A_s)$, $\Omega_m$,  and $\sigma_8$, as seen in Fig.~\ref{fig:boss_bao_CLASSPT_pybird}. 
Without reconstructed BAO, they are about $0.6\sigma$, $0.9\sigma$, $0.6\sigma$ and $0.7\sigma$, respectively.
With reconstructed BAO, the deviations tend to increase, since they become equal to $0.9\sigma$, $1.1\sigma$, $0.9\sigma$ and $0.6\sigma$, respectively.

To close this study, we mention a few other BOSS full-shape analyses using yet a different likelihood or measurements. 
First, Ref.~\cite{Chen:2021wdi}, that uses another prior choice (note in particular that they fix $\omega_b$ and $n_s$, other reconstructed measurements from configuration space~\cite{Vargas-Magana:2016imr}, and another methodology to analyze the reconstructed signal), finds $\Omega_m = 0.303 \pm 0.008, \ h=0.6923 \pm 0.0077, \ \ln(10^{10}A_s) = 2.81 \pm 0.12$, which overall is closer to $P_\textsc{fkp}^\textsc{lz/cm} + \alpha^\textsc{lz/cm}_{rec}$ than $P_\textsc{fkp}^{z_1/z_3} + \alpha^{z_1/z_3}_{rec}$, especially on $h$. 
It is actually also interesting to compare to their results without reconstructed signal, for which they obtain  $\Omega_m = 0.305 \pm 0.010, \ h=0.685 \pm 0.011, \ \ln(10^{10}A_s) = 2.84 \pm 0.13$. Here again, their results are closer to $P_\textsc{fkp}^\textsc{lz/cm}$ analyzed with WC prior, than, \emph{e.g.}, $P_\textsc{quad}^{z_1/z_3}$, analyzed either with the WC or EC prior. 
Second, Ref.~\cite{Brieden:2022lsd} put forward another approach, dubbed ShapeFit, that extends the traditional analysis BAO and redshift-space distortion measurements with one additional compressed parameter. 
They obtain on BOSS data (fixing $\omega_b$ and $n_s$): $\Omega_m = 0.300 \pm 0.006, h=0.6816 \pm 0.0067, \ln(10^{10}A_s) = 3.19 \pm 0.08$. 
Those results are also in better agreement with $P_\textsc{fkp}^\textsc{lz/cm}$ analyzed with the WC prior, than $P_\textsc{quad}^{z_1/z_3}$ analyzed either with the WC or EC prior. 
We warn that it is not straightforward to interpret those comparisons given that there are many differences in the analysis setup. 
In particular, we have checked that fixing $\omega_b$ instead of using a BBN prior, or fixing $n_s$, can shift the posteriors of the other cosmological parameters up to about $1\sigma$.

\newpage
\bibliography{biblio}

\begin{thebibliography}{86}%
\makeatletter
\providecommand \@ifxundefined [1]{%
 \@ifx{#1\undefined}
}%
\providecommand \@ifnum [1]{%
 \ifnum #1\expandafter \@firstoftwo
 \else \expandafter \@secondoftwo
 \fi
}%
\providecommand \@ifx [1]{%
 \ifx #1\expandafter \@firstoftwo
 \else \expandafter \@secondoftwo
 \fi
}%
\providecommand \natexlab [1]{#1}%
\providecommand \enquote  [1]{``#1''}%
\providecommand \bibnamefont  [1]{#1}%
\providecommand \bibfnamefont [1]{#1}%
\providecommand \citenamefont [1]{#1}%
\providecommand \href@noop [0]{\@secondoftwo}%
\providecommand \href [0]{\begingroup \@sanitize@url \@href}%
\providecommand \@href[1]{\@@startlink{#1}\@@href}%
\providecommand \@@href[1]{\endgroup#1\@@endlink}%
\providecommand \@sanitize@url [0]{\catcode `\\12\catcode `\$12\catcode
  `\&12\catcode `\#12\catcode `\^12\catcode `\_12\catcode `\%12\relax}%
\providecommand \@@startlink[1]{}%
\providecommand \@@endlink[0]{}%
\providecommand \url  [0]{\begingroup\@sanitize@url \@url }%
\providecommand \@url [1]{\endgroup\@href {#1}{\urlprefix }}%
\providecommand \urlprefix  [0]{URL }%
\providecommand \Eprint [0]{\href }%
\providecommand \doibase [0]{http://dx.doi.org/}%
\providecommand \selectlanguage [0]{\@gobble}%
\providecommand \bibinfo  [0]{\@secondoftwo}%
\providecommand \bibfield  [0]{\@secondoftwo}%
\providecommand \translation [1]{[#1]}%
\providecommand \BibitemOpen [0]{}%
\providecommand \bibitemStop [0]{}%
\providecommand \bibitemNoStop [0]{.\EOS\space}%
\providecommand \EOS [0]{\spacefactor3000\relax}%
\providecommand \BibitemShut  [1]{\csname bibitem#1\endcsname}%
\let\auto@bib@innerbib\@empty
\bibitem [{\citenamefont {D'Amico}\ \emph
  {et~al.}(2022{\natexlab{a}})\citenamefont {D'Amico}, \citenamefont {Donath},
  \citenamefont {Lewandowski}, \citenamefont {Senatore},\ and\ \citenamefont
  {Zhang}}]{DAmico:2022osl}%
  \BibitemOpen
  \bibfield  {author} {\bibinfo {author} {\bibfnamefont {Guido}\ \bibnamefont
  {D'Amico}}, \bibinfo {author} {\bibfnamefont {Yaniv}\ \bibnamefont {Donath}},
  \bibinfo {author} {\bibfnamefont {Matthew}\ \bibnamefont {Lewandowski}},
  \bibinfo {author} {\bibfnamefont {Leonardo}\ \bibnamefont {Senatore}}, \ and\
  \bibinfo {author} {\bibfnamefont {Pierre}\ \bibnamefont {Zhang}},\ }\bibfield
   {title} {\enquote {\bibinfo {title} {{The BOSS bispectrum analysis at one
  loop from the Effective Field Theory of Large-Scale Structure}},}\
  }\href@noop {} {\  (\bibinfo {year} {2022}{\natexlab{a}})},\ \Eprint
  {http://arxiv.org/abs/2206.08327} {arXiv:2206.08327 [astro-ph.CO]}
  \BibitemShut {NoStop}%
\bibitem [{\citenamefont {Baumann}\ \emph {et~al.}(2012)\citenamefont
  {Baumann}, \citenamefont {Nicolis}, \citenamefont {Senatore},\ and\
  \citenamefont {Zaldarriaga}}]{Baumann:2010tm}%
  \BibitemOpen
  \bibfield  {author} {\bibinfo {author} {\bibfnamefont {Daniel}\ \bibnamefont
  {Baumann}}, \bibinfo {author} {\bibfnamefont {Alberto}\ \bibnamefont
  {Nicolis}}, \bibinfo {author} {\bibfnamefont {Leonardo}\ \bibnamefont
  {Senatore}}, \ and\ \bibinfo {author} {\bibfnamefont {Matias}\ \bibnamefont
  {Zaldarriaga}},\ }\bibfield  {title} {\enquote {\bibinfo {title}
  {{Cosmological Non-Linearities as an Effective Fluid}},}\ }\href {\doibase
  10.1088/1475-7516/2012/07/051} {\bibfield  {journal} {\bibinfo  {journal}
  {JCAP}\ }\textbf {\bibinfo {volume} {07}},\ \bibinfo {pages} {051} (\bibinfo
  {year} {2012})},\ \Eprint {http://arxiv.org/abs/1004.2488} {arXiv:1004.2488
  [astro-ph.CO]} \BibitemShut {NoStop}%
\bibitem [{\citenamefont {Carrasco}\ \emph {et~al.}(2012)\citenamefont
  {Carrasco}, \citenamefont {Hertzberg},\ and\ \citenamefont
  {Senatore}}]{Carrasco:2012cv}%
  \BibitemOpen
  \bibfield  {author} {\bibinfo {author} {\bibfnamefont {John Joseph~M.}\
  \bibnamefont {Carrasco}}, \bibinfo {author} {\bibfnamefont {Mark~P.}\
  \bibnamefont {Hertzberg}}, \ and\ \bibinfo {author} {\bibfnamefont
  {Leonardo}\ \bibnamefont {Senatore}},\ }\bibfield  {title} {\enquote
  {\bibinfo {title} {{The Effective Field Theory of Cosmological Large Scale
  Structures}},}\ }\href {\doibase 10.1007/JHEP09(2012)082} {\bibfield
  {journal} {\bibinfo  {journal} {JHEP}\ }\textbf {\bibinfo {volume} {09}},\
  \bibinfo {pages} {082} (\bibinfo {year} {2012})},\ \Eprint
  {http://arxiv.org/abs/1206.2926} {arXiv:1206.2926 [astro-ph.CO]} \BibitemShut
  {NoStop}%
\bibitem [{\citenamefont {Senatore}\ and\ \citenamefont
  {Zaldarriaga}(2015)}]{Senatore:2014via}%
  \BibitemOpen
  \bibfield  {author} {\bibinfo {author} {\bibfnamefont {Leonardo}\
  \bibnamefont {Senatore}}\ and\ \bibinfo {author} {\bibfnamefont {Matias}\
  \bibnamefont {Zaldarriaga}},\ }\bibfield  {title} {\enquote {\bibinfo {title}
  {{The IR-resummed Effective Field Theory of Large Scale Structures}},}\
  }\href {\doibase 10.1088/1475-7516/2015/02/013} {\bibfield  {journal}
  {\bibinfo  {journal} {JCAP}\ }\textbf {\bibinfo {volume} {1502}},\ \bibinfo
  {pages} {013} (\bibinfo {year} {2015})},\ \Eprint
  {http://arxiv.org/abs/1404.5954} {arXiv:1404.5954 [astro-ph.CO]} \BibitemShut
  {NoStop}%
\bibitem [{\citenamefont {Senatore}(2015)}]{Senatore:2014eva}%
  \BibitemOpen
  \bibfield  {author} {\bibinfo {author} {\bibfnamefont {Leonardo}\
  \bibnamefont {Senatore}},\ }\bibfield  {title} {\enquote {\bibinfo {title}
  {{Bias in the Effective Field Theory of Large Scale Structures}},}\ }\href
  {\doibase 10.1088/1475-7516/2015/11/007} {\bibfield  {journal} {\bibinfo
  {journal} {JCAP}\ }\textbf {\bibinfo {volume} {1511}},\ \bibinfo {pages}
  {007} (\bibinfo {year} {2015})},\ \Eprint {http://arxiv.org/abs/1406.7843}
  {arXiv:1406.7843 [astro-ph.CO]} \BibitemShut {NoStop}%
\bibitem [{\citenamefont {Senatore}\ and\ \citenamefont
  {Zaldarriaga}(2014)}]{Senatore:2014vja}%
  \BibitemOpen
  \bibfield  {author} {\bibinfo {author} {\bibfnamefont {Leonardo}\
  \bibnamefont {Senatore}}\ and\ \bibinfo {author} {\bibfnamefont {Matias}\
  \bibnamefont {Zaldarriaga}},\ }\bibfield  {title} {\enquote {\bibinfo {title}
  {{Redshift Space Distortions in the Effective Field Theory of Large Scale
  Structures}},}\ }\href@noop {} {\  (\bibinfo {year} {2014})},\ \Eprint
  {http://arxiv.org/abs/1409.1225} {arXiv:1409.1225 [astro-ph.CO]} \BibitemShut
  {NoStop}%
\bibitem [{\citenamefont {Perko}\ \emph {et~al.}(2016)\citenamefont {Perko},
  \citenamefont {Senatore}, \citenamefont {Jennings},\ and\ \citenamefont
  {Wechsler}}]{Perko:2016puo}%
  \BibitemOpen
  \bibfield  {author} {\bibinfo {author} {\bibfnamefont {Ashley}\ \bibnamefont
  {Perko}}, \bibinfo {author} {\bibfnamefont {Leonardo}\ \bibnamefont
  {Senatore}}, \bibinfo {author} {\bibfnamefont {Elise}\ \bibnamefont
  {Jennings}}, \ and\ \bibinfo {author} {\bibfnamefont {Risa~H.}\ \bibnamefont
  {Wechsler}},\ }\bibfield  {title} {\enquote {\bibinfo {title} {{Biased
  Tracers in Redshift Space in the EFT of Large-Scale Structure}},}\
  }\href@noop {} {\  (\bibinfo {year} {2016})},\ \Eprint
  {http://arxiv.org/abs/1610.09321} {arXiv:1610.09321 [astro-ph.CO]}
  \BibitemShut {NoStop}%
\bibitem [{\citenamefont {Alam}\ \emph {et~al.}(2017)\citenamefont {Alam} \emph
  {et~al.}}]{BOSS:2016wmc}%
  \BibitemOpen
  \bibfield  {author} {\bibinfo {author} {\bibfnamefont {Shadab}\ \bibnamefont
  {Alam}} \emph {et~al.} (\bibinfo {collaboration} {BOSS}),\ }\bibfield
  {title} {\enquote {\bibinfo {title} {{The clustering of galaxies in the
  completed SDSS-III Baryon Oscillation Spectroscopic Survey: cosmological
  analysis of the DR12 galaxy sample}},}\ }\href {\doibase
  10.1093/mnras/stx721} {\bibfield  {journal} {\bibinfo  {journal} {Mon. Not.
  Roy. Astron. Soc.}\ }\textbf {\bibinfo {volume} {470}},\ \bibinfo {pages}
  {2617--2652} (\bibinfo {year} {2017})},\ \Eprint
  {http://arxiv.org/abs/1607.03155} {arXiv:1607.03155 [astro-ph.CO]}
  \BibitemShut {NoStop}%
\bibitem [{\citenamefont {D'Amico}\ \emph
  {et~al.}(2020{\natexlab{a}})\citenamefont {D'Amico}, \citenamefont {Gleyzes},
  \citenamefont {Kokron}, \citenamefont {Markovic}, \citenamefont {Senatore},
  \citenamefont {Zhang}, \citenamefont {Beutler},\ and\ \citenamefont
  {Gil-Marín}}]{DAmico:2019fhj}%
  \BibitemOpen
  \bibfield  {author} {\bibinfo {author} {\bibfnamefont {Guido}\ \bibnamefont
  {D'Amico}}, \bibinfo {author} {\bibfnamefont {Jérôme}\ \bibnamefont
  {Gleyzes}}, \bibinfo {author} {\bibfnamefont {Nickolas}\ \bibnamefont
  {Kokron}}, \bibinfo {author} {\bibfnamefont {Katarina}\ \bibnamefont
  {Markovic}}, \bibinfo {author} {\bibfnamefont {Leonardo}\ \bibnamefont
  {Senatore}}, \bibinfo {author} {\bibfnamefont {Pierre}\ \bibnamefont
  {Zhang}}, \bibinfo {author} {\bibfnamefont {Florian}\ \bibnamefont
  {Beutler}}, \ and\ \bibinfo {author} {\bibfnamefont {Héctor}\ \bibnamefont
  {Gil-Marín}},\ }\bibfield  {title} {\enquote {\bibinfo {title} {{The
  Cosmological Analysis of the SDSS/BOSS data from the Effective Field Theory
  of Large-Scale Structure}},}\ }\href {\doibase 10.1088/1475-7516/2020/05/005}
  {\bibfield  {journal} {\bibinfo  {journal} {JCAP}\ }\textbf {\bibinfo
  {volume} {05}},\ \bibinfo {pages} {005} (\bibinfo {year}
  {2020}{\natexlab{a}})},\ \Eprint {http://arxiv.org/abs/1909.05271}
  {arXiv:1909.05271 [astro-ph.CO]} \BibitemShut {NoStop}%
\bibitem [{\citenamefont {Ivanov}\ \emph {et~al.}(2019)\citenamefont {Ivanov},
  \citenamefont {Simonović},\ and\ \citenamefont
  {Zaldarriaga}}]{Ivanov:2019pdj}%
  \BibitemOpen
  \bibfield  {author} {\bibinfo {author} {\bibfnamefont {Mikhail~M.}\
  \bibnamefont {Ivanov}}, \bibinfo {author} {\bibfnamefont {Marko}\
  \bibnamefont {Simonović}}, \ and\ \bibinfo {author} {\bibfnamefont {Matias}\
  \bibnamefont {Zaldarriaga}},\ }\bibfield  {title} {\enquote {\bibinfo {title}
  {{Cosmological Parameters from the BOSS Galaxy Power Spectrum}},}\
  }\href@noop {} {\  (\bibinfo {year} {2019})},\ \Eprint
  {http://arxiv.org/abs/1909.05277} {arXiv:1909.05277 [astro-ph.CO]}
  \BibitemShut {NoStop}%
\bibitem [{\citenamefont {Colas}\ \emph {et~al.}(2020)\citenamefont {Colas},
  \citenamefont {D'amico}, \citenamefont {Senatore}, \citenamefont {Zhang},\
  and\ \citenamefont {Beutler}}]{Colas:2019ret}%
  \BibitemOpen
  \bibfield  {author} {\bibinfo {author} {\bibfnamefont {Thomas}\ \bibnamefont
  {Colas}}, \bibinfo {author} {\bibfnamefont {Guido}\ \bibnamefont {D'amico}},
  \bibinfo {author} {\bibfnamefont {Leonardo}\ \bibnamefont {Senatore}},
  \bibinfo {author} {\bibfnamefont {Pierre}\ \bibnamefont {Zhang}}, \ and\
  \bibinfo {author} {\bibfnamefont {Florian}\ \bibnamefont {Beutler}},\
  }\bibfield  {title} {\enquote {\bibinfo {title} {{Efficient Cosmological
  Analysis of the SDSS/BOSS data from the Effective Field Theory of Large-Scale
  Structure}},}\ }\href {\doibase 10.1088/1475-7516/2020/06/001} {\bibfield
  {journal} {\bibinfo  {journal} {JCAP}\ }\textbf {\bibinfo {volume} {06}},\
  \bibinfo {pages} {001} (\bibinfo {year} {2020})},\ \Eprint
  {http://arxiv.org/abs/1909.07951} {arXiv:1909.07951 [astro-ph.CO]}
  \BibitemShut {NoStop}%
\bibitem [{\citenamefont {D'Amico}\ \emph
  {et~al.}(2021{\natexlab{a}})\citenamefont {D'Amico}, \citenamefont
  {Senatore},\ and\ \citenamefont {Zhang}}]{DAmico:2020kxu}%
  \BibitemOpen
  \bibfield  {author} {\bibinfo {author} {\bibfnamefont {Guido}\ \bibnamefont
  {D'Amico}}, \bibinfo {author} {\bibfnamefont {Leonardo}\ \bibnamefont
  {Senatore}}, \ and\ \bibinfo {author} {\bibfnamefont {Pierre}\ \bibnamefont
  {Zhang}},\ }\bibfield  {title} {\enquote {\bibinfo {title} {{Limits on $w$CDM
  from the EFTofLSS with the PyBird code}},}\ }\href {\doibase
  10.1088/1475-7516/2021/01/006} {\bibfield  {journal} {\bibinfo  {journal}
  {JCAP}\ }\textbf {\bibinfo {volume} {01}},\ \bibinfo {pages} {006} (\bibinfo
  {year} {2021}{\natexlab{a}})},\ \Eprint {http://arxiv.org/abs/2003.07956}
  {arXiv:2003.07956 [astro-ph.CO]} \BibitemShut {NoStop}%
\bibitem [{\citenamefont {D'Amico}\ \emph
  {et~al.}(2020{\natexlab{b}})\citenamefont {D'Amico}, \citenamefont {Donath},
  \citenamefont {Senatore},\ and\ \citenamefont {Zhang}}]{DAmico:2020tty}%
  \BibitemOpen
  \bibfield  {author} {\bibinfo {author} {\bibfnamefont {Guido}\ \bibnamefont
  {D'Amico}}, \bibinfo {author} {\bibfnamefont {Yaniv}\ \bibnamefont {Donath}},
  \bibinfo {author} {\bibfnamefont {Leonardo}\ \bibnamefont {Senatore}}, \ and\
  \bibinfo {author} {\bibfnamefont {Pierre}\ \bibnamefont {Zhang}},\ }\bibfield
   {title} {\enquote {\bibinfo {title} {{Limits on Clustering and Smooth
  Quintessence from the EFTofLSS}},}\ }\href@noop {} {\  (\bibinfo {year}
  {2020}{\natexlab{b}})},\ \Eprint {http://arxiv.org/abs/2012.07554}
  {arXiv:2012.07554 [astro-ph.CO]} \BibitemShut {NoStop}%
\bibitem [{\citenamefont {Chen}\ \emph {et~al.}(2022)\citenamefont {Chen},
  \citenamefont {Vlah},\ and\ \citenamefont {White}}]{Chen:2021wdi}%
  \BibitemOpen
  \bibfield  {author} {\bibinfo {author} {\bibfnamefont {Shi-Fan}\ \bibnamefont
  {Chen}}, \bibinfo {author} {\bibfnamefont {Zvonimir}\ \bibnamefont {Vlah}}, \
  and\ \bibinfo {author} {\bibfnamefont {Martin}\ \bibnamefont {White}},\
  }\bibfield  {title} {\enquote {\bibinfo {title} {{A new analysis of galaxy
  2-point functions in the BOSS survey, including full-shape information and
  post-reconstruction BAO}},}\ }\href {\doibase 10.1088/1475-7516/2022/02/008}
  {\bibfield  {journal} {\bibinfo  {journal} {JCAP}\ }\textbf {\bibinfo
  {volume} {02}},\ \bibinfo {pages} {008} (\bibinfo {year} {2022})},\ \Eprint
  {http://arxiv.org/abs/2110.05530} {arXiv:2110.05530 [astro-ph.CO]}
  \BibitemShut {NoStop}%
\bibitem [{\citenamefont {Zhang}\ \emph {et~al.}(2022)\citenamefont {Zhang},
  \citenamefont {D'Amico}, \citenamefont {Senatore}, \citenamefont {Zhao},\
  and\ \citenamefont {Cai}}]{Zhang:2021yna}%
  \BibitemOpen
  \bibfield  {author} {\bibinfo {author} {\bibfnamefont {Pierre}\ \bibnamefont
  {Zhang}}, \bibinfo {author} {\bibfnamefont {Guido}\ \bibnamefont {D'Amico}},
  \bibinfo {author} {\bibfnamefont {Leonardo}\ \bibnamefont {Senatore}},
  \bibinfo {author} {\bibfnamefont {Cheng}\ \bibnamefont {Zhao}}, \ and\
  \bibinfo {author} {\bibfnamefont {Yifu}\ \bibnamefont {Cai}},\ }\bibfield
  {title} {\enquote {\bibinfo {title} {{BOSS Correlation Function analysis from
  the Effective Field Theory of Large-Scale Structure}},}\ }\href {\doibase
  10.1088/1475-7516/2022/02/036} {\bibfield  {journal} {\bibinfo  {journal}
  {JCAP}\ }\textbf {\bibinfo {volume} {02}},\ \bibinfo {pages} {036} (\bibinfo
  {year} {2022})},\ \Eprint {http://arxiv.org/abs/2110.07539} {arXiv:2110.07539
  [astro-ph.CO]} \BibitemShut {NoStop}%
\bibitem [{\citenamefont {Zhang}\ and\ \citenamefont
  {Cai}(2022)}]{Zhang:2021uyp}%
  \BibitemOpen
  \bibfield  {author} {\bibinfo {author} {\bibfnamefont {Pierre}\ \bibnamefont
  {Zhang}}\ and\ \bibinfo {author} {\bibfnamefont {Yifu}\ \bibnamefont {Cai}},\
  }\bibfield  {title} {\enquote {\bibinfo {title} {{BOSS full-shape analysis
  from the EFTofLSS with exact time dependence}},}\ }\href {\doibase
  10.1088/1475-7516/2022/01/031} {\bibfield  {journal} {\bibinfo  {journal}
  {JCAP}\ }\textbf {\bibinfo {volume} {01}},\ \bibinfo {pages} {031} (\bibinfo
  {year} {2022})},\ \Eprint {http://arxiv.org/abs/2111.05739} {arXiv:2111.05739
  [astro-ph.CO]} \BibitemShut {NoStop}%
\bibitem [{\citenamefont {Philcox}\ and\ \citenamefont
  {Ivanov}(2022)}]{Philcox:2021kcw}%
  \BibitemOpen
  \bibfield  {author} {\bibinfo {author} {\bibfnamefont {Oliver H.~E.}\
  \bibnamefont {Philcox}}\ and\ \bibinfo {author} {\bibfnamefont {Mikhail~M.}\
  \bibnamefont {Ivanov}},\ }\bibfield  {title} {\enquote {\bibinfo {title}
  {{BOSS DR12 full-shape cosmology: \ensuremath{\Lambda}CDM constraints from
  the large-scale galaxy power spectrum and bispectrum monopole}},}\ }\href
  {\doibase 10.1103/PhysRevD.105.043517} {\bibfield  {journal} {\bibinfo
  {journal} {Phys. Rev. D}\ }\textbf {\bibinfo {volume} {105}},\ \bibinfo
  {pages} {043517} (\bibinfo {year} {2022})},\ \Eprint
  {http://arxiv.org/abs/2112.04515} {arXiv:2112.04515 [astro-ph.CO]}
  \BibitemShut {NoStop}%
\bibitem [{\citenamefont {Simon}\ \emph
  {et~al.}(2022{\natexlab{a}})\citenamefont {Simon}, \citenamefont {Zhang},\
  and\ \citenamefont {Poulin}}]{Simon:2022csv}%
  \BibitemOpen
  \bibfield  {author} {\bibinfo {author} {\bibfnamefont {Th\'eo}\ \bibnamefont
  {Simon}}, \bibinfo {author} {\bibfnamefont {Pierre}\ \bibnamefont {Zhang}}, \
  and\ \bibinfo {author} {\bibfnamefont {Vivian}\ \bibnamefont {Poulin}},\
  }\bibfield  {title} {\enquote {\bibinfo {title} {{Cosmological inference from
  the EFTofLSS: the eBOSS QSO full-shape analysis}},}\ }\href@noop {} {\
  (\bibinfo {year} {2022}{\natexlab{a}})},\ \Eprint
  {http://arxiv.org/abs/2210.14931} {arXiv:2210.14931 [astro-ph.CO]}
  \BibitemShut {NoStop}%
\bibitem [{\citenamefont {Chudaykin}\ and\ \citenamefont
  {Ivanov}(2023)}]{Chudaykin:2022nru}%
  \BibitemOpen
  \bibfield  {author} {\bibinfo {author} {\bibfnamefont {Anton}\ \bibnamefont
  {Chudaykin}}\ and\ \bibinfo {author} {\bibfnamefont {Mikhail~M.}\
  \bibnamefont {Ivanov}},\ }\bibfield  {title} {\enquote {\bibinfo {title}
  {{Cosmological constraints from the power spectrum of eBOSS quasars}},}\
  }\href {\doibase 10.1103/PhysRevD.107.043518} {\bibfield  {journal} {\bibinfo
   {journal} {Phys. Rev. D}\ }\textbf {\bibinfo {volume} {107}},\ \bibinfo
  {pages} {043518} (\bibinfo {year} {2023})},\ \Eprint
  {http://arxiv.org/abs/2210.17044} {arXiv:2210.17044 [astro-ph.CO]}
  \BibitemShut {NoStop}%
\bibitem [{\citenamefont {Smith}\ \emph {et~al.}(2022)\citenamefont {Smith},
  \citenamefont {Poulin},\ and\ \citenamefont {Simon}}]{Smith:2022iax}%
  \BibitemOpen
  \bibfield  {author} {\bibinfo {author} {\bibfnamefont {Tristan~L.}\
  \bibnamefont {Smith}}, \bibinfo {author} {\bibfnamefont {Vivian}\
  \bibnamefont {Poulin}}, \ and\ \bibinfo {author} {\bibfnamefont {Th\'eo}\
  \bibnamefont {Simon}},\ }\bibfield  {title} {\enquote {\bibinfo {title}
  {{Assessing the robustness of sound horizon-free determinations of the Hubble
  constant}},}\ }\href@noop {} {\  (\bibinfo {year} {2022})},\ \Eprint
  {http://arxiv.org/abs/2208.12992} {arXiv:2208.12992 [astro-ph.CO]}
  \BibitemShut {NoStop}%
\bibitem [{\citenamefont {Simon}\ \emph
  {et~al.}(2022{\natexlab{b}})\citenamefont {Simon}, \citenamefont
  {Franco~Abell\'an}, \citenamefont {Du}, \citenamefont {Poulin},\ and\
  \citenamefont {Tsai}}]{Simon:2022ftd}%
  \BibitemOpen
  \bibfield  {author} {\bibinfo {author} {\bibfnamefont {Th\'eo}\ \bibnamefont
  {Simon}}, \bibinfo {author} {\bibfnamefont {Guillermo}\ \bibnamefont
  {Franco~Abell\'an}}, \bibinfo {author} {\bibfnamefont {Peizhi}\ \bibnamefont
  {Du}}, \bibinfo {author} {\bibfnamefont {Vivian}\ \bibnamefont {Poulin}}, \
  and\ \bibinfo {author} {\bibfnamefont {Yuhsin}\ \bibnamefont {Tsai}},\
  }\bibfield  {title} {\enquote {\bibinfo {title} {{Constraining decaying dark
  matter with BOSS data and the effective field theory of large-scale
  structures}},}\ }\href {\doibase 10.1103/PhysRevD.106.023516} {\bibfield
  {journal} {\bibinfo  {journal} {Phys. Rev. D}\ }\textbf {\bibinfo {volume}
  {106}},\ \bibinfo {pages} {023516} (\bibinfo {year} {2022}{\natexlab{b}})},\
  \Eprint {http://arxiv.org/abs/2203.07440} {arXiv:2203.07440 [astro-ph.CO]}
  \BibitemShut {NoStop}%
\bibitem [{\citenamefont {Kumar}\ \emph {et~al.}(2022)\citenamefont {Kumar},
  \citenamefont {Nunes},\ and\ \citenamefont {Yadav}}]{Kumar:2022vee}%
  \BibitemOpen
  \bibfield  {author} {\bibinfo {author} {\bibfnamefont {Suresh}\ \bibnamefont
  {Kumar}}, \bibinfo {author} {\bibfnamefont {Rafael~C.}\ \bibnamefont
  {Nunes}}, \ and\ \bibinfo {author} {\bibfnamefont {Priya}\ \bibnamefont
  {Yadav}},\ }\bibfield  {title} {\enquote {\bibinfo {title} {{Updating
  non-standard neutrinos properties with Planck-CMB data and full-shape
  analysis of BOSS and eBOSS galaxies}},}\ }\href@noop {} {\  (\bibinfo {year}
  {2022})},\ \Eprint {http://arxiv.org/abs/2205.04292} {arXiv:2205.04292
  [astro-ph.CO]} \BibitemShut {NoStop}%
\bibitem [{\citenamefont {Nunes}\ \emph {et~al.}(2022)\citenamefont {Nunes},
  \citenamefont {Vagnozzi}, \citenamefont {Kumar}, \citenamefont
  {Di~Valentino},\ and\ \citenamefont {Mena}}]{Nunes:2022bhn}%
  \BibitemOpen
  \bibfield  {author} {\bibinfo {author} {\bibfnamefont {Rafael~C.}\
  \bibnamefont {Nunes}}, \bibinfo {author} {\bibfnamefont {Sunny}\ \bibnamefont
  {Vagnozzi}}, \bibinfo {author} {\bibfnamefont {Suresh}\ \bibnamefont
  {Kumar}}, \bibinfo {author} {\bibfnamefont {Eleonora}\ \bibnamefont
  {Di~Valentino}}, \ and\ \bibinfo {author} {\bibfnamefont {Olga}\ \bibnamefont
  {Mena}},\ }\bibfield  {title} {\enquote {\bibinfo {title} {{New tests of dark
  sector interactions from the full-shape galaxy power spectrum}},}\ }\href
  {\doibase 10.1103/PhysRevD.105.123506} {\bibfield  {journal} {\bibinfo
  {journal} {Phys. Rev. D}\ }\textbf {\bibinfo {volume} {105}},\ \bibinfo
  {pages} {123506} (\bibinfo {year} {2022})},\ \Eprint
  {http://arxiv.org/abs/2203.08093} {arXiv:2203.08093 [astro-ph.CO]}
  \BibitemShut {NoStop}%
\bibitem [{\citenamefont {Niedermann}\ and\ \citenamefont
  {Sloth}(2021)}]{Niedermann:2020qbw}%
  \BibitemOpen
  \bibfield  {author} {\bibinfo {author} {\bibfnamefont {Florian}\ \bibnamefont
  {Niedermann}}\ and\ \bibinfo {author} {\bibfnamefont {Martin~S.}\
  \bibnamefont {Sloth}},\ }\bibfield  {title} {\enquote {\bibinfo {title} {{New
  Early Dark Energy is compatible with current LSS data}},}\ }\href {\doibase
  10.1103/PhysRevD.103.103537} {\bibfield  {journal} {\bibinfo  {journal}
  {Phys. Rev. D}\ }\textbf {\bibinfo {volume} {103}},\ \bibinfo {pages}
  {103537} (\bibinfo {year} {2021})},\ \Eprint
  {http://arxiv.org/abs/2009.00006} {arXiv:2009.00006 [astro-ph.CO]}
  \BibitemShut {NoStop}%
\bibitem [{\citenamefont {Lagu\"e}\ \emph {et~al.}(2021)\citenamefont
  {Lagu\"e}, \citenamefont {Bond}, \citenamefont {Hlo\v{z}ek}, \citenamefont
  {Rogers}, \citenamefont {Marsh},\ and\ \citenamefont {Grin}}]{Lague:2021frh}%
  \BibitemOpen
  \bibfield  {author} {\bibinfo {author} {\bibfnamefont {Alex}\ \bibnamefont
  {Lagu\"e}}, \bibinfo {author} {\bibfnamefont {J.~Richard}\ \bibnamefont
  {Bond}}, \bibinfo {author} {\bibfnamefont {Ren\'ee}\ \bibnamefont
  {Hlo\v{z}ek}}, \bibinfo {author} {\bibfnamefont {Keir~K.}\ \bibnamefont
  {Rogers}}, \bibinfo {author} {\bibfnamefont {David J.~E.}\ \bibnamefont
  {Marsh}}, \ and\ \bibinfo {author} {\bibfnamefont {Daniel}\ \bibnamefont
  {Grin}},\ }\bibfield  {title} {\enquote {\bibinfo {title} {{Constraining
  Ultralight Axions with Galaxy Surveys}},}\ }\href@noop {} {\  (\bibinfo
  {year} {2021})},\ \Eprint {http://arxiv.org/abs/2104.07802} {arXiv:2104.07802
  [astro-ph.CO]} \BibitemShut {NoStop}%
\bibitem [{\citenamefont {Carrilho}\ \emph {et~al.}(2022)\citenamefont
  {Carrilho}, \citenamefont {Moretti},\ and\ \citenamefont
  {Pourtsidou}}]{Carrilho:2022mon}%
  \BibitemOpen
  \bibfield  {author} {\bibinfo {author} {\bibfnamefont {Pedro}\ \bibnamefont
  {Carrilho}}, \bibinfo {author} {\bibfnamefont {Chiara}\ \bibnamefont
  {Moretti}}, \ and\ \bibinfo {author} {\bibfnamefont {Alkistis}\ \bibnamefont
  {Pourtsidou}},\ }\bibfield  {title} {\enquote {\bibinfo {title} {{Cosmology
  with the EFTofLSS and BOSS: dark energy constraints and a note on priors}},}\
  }\href@noop {} {\  (\bibinfo {year} {2022})},\ \Eprint
  {http://arxiv.org/abs/2207.14784} {arXiv:2207.14784 [astro-ph.CO]}
  \BibitemShut {NoStop}%
\bibitem [{\citenamefont {Simon}\ \emph
  {et~al.}(2022{\natexlab{c}})\citenamefont {Simon}, \citenamefont {Zhang},
  \citenamefont {Poulin},\ and\ \citenamefont {Smith}}]{Simon:2022adh}%
  \BibitemOpen
  \bibfield  {author} {\bibinfo {author} {\bibfnamefont {Th\'eo}\ \bibnamefont
  {Simon}}, \bibinfo {author} {\bibfnamefont {Pierre}\ \bibnamefont {Zhang}},
  \bibinfo {author} {\bibfnamefont {Vivian}\ \bibnamefont {Poulin}}, \ and\
  \bibinfo {author} {\bibfnamefont {Tristan~L.}\ \bibnamefont {Smith}},\
  }\bibfield  {title} {\enquote {\bibinfo {title} {{Updated constraints from
  the effective field theory analysis of BOSS power spectrum on Early Dark
  Energy}},}\ }\href@noop {} {\  (\bibinfo {year} {2022}{\natexlab{c}})},\
  \Eprint {http://arxiv.org/abs/2208.05930} {arXiv:2208.05930 [astro-ph.CO]}
  \BibitemShut {NoStop}%
\bibitem [{\citenamefont {Sch\"oneberg}\ \emph {et~al.}(2023)\citenamefont
  {Sch\"oneberg}, \citenamefont {Franco~Abell\'an}, \citenamefont {Simon},
  \citenamefont {Bartlett}, \citenamefont {Patel},\ and\ \citenamefont
  {Smith}}]{Schoneberg:2023rnx}%
  \BibitemOpen
  \bibfield  {author} {\bibinfo {author} {\bibfnamefont {Nils}\ \bibnamefont
  {Sch\"oneberg}}, \bibinfo {author} {\bibfnamefont {Guillermo}\ \bibnamefont
  {Franco~Abell\'an}}, \bibinfo {author} {\bibfnamefont {Th\'eo}\ \bibnamefont
  {Simon}}, \bibinfo {author} {\bibfnamefont {Alexa}\ \bibnamefont {Bartlett}},
  \bibinfo {author} {\bibfnamefont {Yashvi}\ \bibnamefont {Patel}}, \ and\
  \bibinfo {author} {\bibfnamefont {Tristan~L.}\ \bibnamefont {Smith}},\
  }\bibfield  {title} {\enquote {\bibinfo {title} {{The weak, the strong and
  the ugly -- A comparative analysis of interacting stepped dark radiation}},}\
  }\href@noop {} {\  (\bibinfo {year} {2023})},\ \Eprint
  {http://arxiv.org/abs/2306.12469} {arXiv:2306.12469 [astro-ph.CO]}
  \BibitemShut {NoStop}%
\bibitem [{\citenamefont {Allali}\ \emph {et~al.}(2023)\citenamefont {Allali},
  \citenamefont {Rompineve},\ and\ \citenamefont {Hertzberg}}]{Allali:2023zbi}%
  \BibitemOpen
  \bibfield  {author} {\bibinfo {author} {\bibfnamefont {Itamar~J.}\
  \bibnamefont {Allali}}, \bibinfo {author} {\bibfnamefont {Fabrizio}\
  \bibnamefont {Rompineve}}, \ and\ \bibinfo {author} {\bibfnamefont {Mark~P.}\
  \bibnamefont {Hertzberg}},\ }\bibfield  {title} {\enquote {\bibinfo {title}
  {{Dark Sectors with Mass Thresholds Face Cosmological Datasets}},}\
  }\href@noop {} {\  (\bibinfo {year} {2023})},\ \Eprint
  {http://arxiv.org/abs/2305.14166} {arXiv:2305.14166 [astro-ph.CO]}
  \BibitemShut {NoStop}%
\bibitem [{\citenamefont {D'Amico}\ \emph
  {et~al.}(2022{\natexlab{b}})\citenamefont {D'Amico}, \citenamefont
  {Lewandowski}, \citenamefont {Senatore},\ and\ \citenamefont
  {Zhang}}]{DAmico:2022gki}%
  \BibitemOpen
  \bibfield  {author} {\bibinfo {author} {\bibfnamefont {Guido}\ \bibnamefont
  {D'Amico}}, \bibinfo {author} {\bibfnamefont {Matthew}\ \bibnamefont
  {Lewandowski}}, \bibinfo {author} {\bibfnamefont {Leonardo}\ \bibnamefont
  {Senatore}}, \ and\ \bibinfo {author} {\bibfnamefont {Pierre}\ \bibnamefont
  {Zhang}},\ }\bibfield  {title} {\enquote {\bibinfo {title} {{Limits on
  primordial non-Gaussianities from BOSS galaxy-clustering data}},}\
  }\href@noop {} {\  (\bibinfo {year} {2022}{\natexlab{b}})},\ \Eprint
  {http://arxiv.org/abs/2201.11518} {arXiv:2201.11518 [astro-ph.CO]}
  \BibitemShut {NoStop}%
\bibitem [{\citenamefont {Chudaykin}\ \emph {et~al.}(2020)\citenamefont
  {Chudaykin}, \citenamefont {Ivanov}, \citenamefont {Philcox},\ and\
  \citenamefont {Simonovi\'c}}]{Chudaykin:2020aoj}%
  \BibitemOpen
  \bibfield  {author} {\bibinfo {author} {\bibfnamefont {Anton}\ \bibnamefont
  {Chudaykin}}, \bibinfo {author} {\bibfnamefont {Mikhail~M.}\ \bibnamefont
  {Ivanov}}, \bibinfo {author} {\bibfnamefont {Oliver H.~E.}\ \bibnamefont
  {Philcox}}, \ and\ \bibinfo {author} {\bibfnamefont {Marko}\ \bibnamefont
  {Simonovi\'c}},\ }\bibfield  {title} {\enquote {\bibinfo {title} {{Nonlinear
  perturbation theory extension of the Boltzmann code CLASS}},}\ }\href
  {\doibase 10.1103/PhysRevD.102.063533} {\bibfield  {journal} {\bibinfo
  {journal} {Phys. Rev. D}\ }\textbf {\bibinfo {volume} {102}},\ \bibinfo
  {pages} {063533} (\bibinfo {year} {2020})},\ \Eprint
  {http://arxiv.org/abs/2004.10607} {arXiv:2004.10607 [astro-ph.CO]}
  \BibitemShut {NoStop}%
\bibitem [{\citenamefont {Heymans}\ \emph {et~al.}(2021)\citenamefont {Heymans}
  \emph {et~al.}}]{Heymans:2020gsg}%
  \BibitemOpen
  \bibfield  {author} {\bibinfo {author} {\bibfnamefont {Catherine}\
  \bibnamefont {Heymans}} \emph {et~al.},\ }\bibfield  {title} {\enquote
  {\bibinfo {title} {{KiDS-1000 Cosmology: Multi-probe weak gravitational
  lensing and spectroscopic galaxy clustering constraints}},}\ }\href {\doibase
  10.1051/0004-6361/202039063} {\bibfield  {journal} {\bibinfo  {journal}
  {Astron. Astrophys.}\ }\textbf {\bibinfo {volume} {646}},\ \bibinfo {pages}
  {A140} (\bibinfo {year} {2021})},\ \Eprint {http://arxiv.org/abs/2007.15632}
  {arXiv:2007.15632 [astro-ph.CO]} \BibitemShut {NoStop}%
\bibitem [{\citenamefont {Abbott}\ \emph {et~al.}(2021)\citenamefont {Abbott}
  \emph {et~al.}}]{DES:2021wwk}%
  \BibitemOpen
  \bibfield  {author} {\bibinfo {author} {\bibfnamefont {T.~M.~C.}\
  \bibnamefont {Abbott}} \emph {et~al.} (\bibinfo {collaboration} {DES}),\
  }\bibfield  {title} {\enquote {\bibinfo {title} {{Dark Energy Survey Year 3
  Results: Cosmological Constraints from Galaxy Clustering and Weak
  Lensing}},}\ }\href@noop {} {\  (\bibinfo {year} {2021})},\ \Eprint
  {http://arxiv.org/abs/2105.13549} {arXiv:2105.13549 [astro-ph.CO]}
  \BibitemShut {NoStop}%
\bibitem [{\citenamefont {Lange}\ \emph {et~al.}(2021)\citenamefont {Lange},
  \citenamefont {Leauthaud}, \citenamefont {Singh}, \citenamefont {Guo},
  \citenamefont {Zhou}, \citenamefont {Smith},\ and\ \citenamefont
  {Cyr-Racine}}]{Lange:2020mnl}%
  \BibitemOpen
  \bibfield  {author} {\bibinfo {author} {\bibfnamefont {Johannes~U.}\
  \bibnamefont {Lange}}, \bibinfo {author} {\bibfnamefont {Alexie}\
  \bibnamefont {Leauthaud}}, \bibinfo {author} {\bibfnamefont {Sukhdeep}\
  \bibnamefont {Singh}}, \bibinfo {author} {\bibfnamefont {Hong}\ \bibnamefont
  {Guo}}, \bibinfo {author} {\bibfnamefont {Rongpu}\ \bibnamefont {Zhou}},
  \bibinfo {author} {\bibfnamefont {Tristan~L.}\ \bibnamefont {Smith}}, \ and\
  \bibinfo {author} {\bibfnamefont {Francis-Yan}\ \bibnamefont {Cyr-Racine}},\
  }\bibfield  {title} {\enquote {\bibinfo {title} {{On the halo-mass and radial
  scale dependence of the lensing is low effect}},}\ }\href {\doibase
  10.1093/mnras/stab189} {\bibfield  {journal} {\bibinfo  {journal} {Mon. Not.
  Roy. Astron. Soc.}\ }\textbf {\bibinfo {volume} {502}},\ \bibinfo {pages}
  {2074--2086} (\bibinfo {year} {2021})},\ \Eprint
  {http://arxiv.org/abs/2011.02377} {arXiv:2011.02377 [astro-ph.CO]}
  \BibitemShut {NoStop}%
\bibitem [{\citenamefont {Amon}\ \emph {et~al.}(2022)\citenamefont {Amon} \emph
  {et~al.}}]{Amon:2022ycy}%
  \BibitemOpen
  \bibfield  {author} {\bibinfo {author} {\bibfnamefont {A.}~\bibnamefont
  {Amon}} \emph {et~al.},\ }\bibfield  {title} {\enquote {\bibinfo {title}
  {{Consistent lensing and clustering in a low-$S_8$ Universe with BOSS, DES
  Year 3, HSC Year 1 and KiDS-1000}},}\ }\href@noop {} {\  (\bibinfo {year}
  {2022})},\ \Eprint {http://arxiv.org/abs/2202.07440} {arXiv:2202.07440
  [astro-ph.CO]} \BibitemShut {NoStop}%
\bibitem [{\citenamefont {Abdalla}\ \emph {et~al.}(2022)\citenamefont {Abdalla}
  \emph {et~al.}}]{Abdalla:2022yfr}%
  \BibitemOpen
  \bibfield  {author} {\bibinfo {author} {\bibfnamefont {Elcio}\ \bibnamefont
  {Abdalla}} \emph {et~al.},\ }\bibfield  {title} {\enquote {\bibinfo {title}
  {{Cosmology intertwined: A review of the particle physics, astrophysics, and
  cosmology associated with the cosmological tensions and anomalies}},}\ }\href
  {\doibase 10.1016/j.jheap.2022.04.002} {\bibfield  {journal} {\bibinfo
  {journal} {JHEAp}\ }\textbf {\bibinfo {volume} {34}},\ \bibinfo {pages}
  {49--211} (\bibinfo {year} {2022})},\ \Eprint
  {http://arxiv.org/abs/2203.06142} {arXiv:2203.06142 [astro-ph.CO]}
  \BibitemShut {NoStop}%
\bibitem [{\citenamefont {Sch\"oneberg}\ \emph {et~al.}(2021)\citenamefont
  {Sch\"oneberg}, \citenamefont {Franco~Abell\'an}, \citenamefont
  {P\'erez~S\'anchez}, \citenamefont {Witte}, \citenamefont {Poulin},\ and\
  \citenamefont {Lesgourgues}}]{Schoneberg:2021qvd}%
  \BibitemOpen
  \bibfield  {author} {\bibinfo {author} {\bibfnamefont {Nils}\ \bibnamefont
  {Sch\"oneberg}}, \bibinfo {author} {\bibfnamefont {Guillermo}\ \bibnamefont
  {Franco~Abell\'an}}, \bibinfo {author} {\bibfnamefont {Andrea}\ \bibnamefont
  {P\'erez~S\'anchez}}, \bibinfo {author} {\bibfnamefont {Samuel~J.}\
  \bibnamefont {Witte}}, \bibinfo {author} {\bibfnamefont {Vivian}\
  \bibnamefont {Poulin}}, \ and\ \bibinfo {author} {\bibfnamefont {Julien}\
  \bibnamefont {Lesgourgues}},\ }\bibfield  {title} {\enquote {\bibinfo {title}
  {{The $H_0$ Olympics: A fair ranking of proposed models}},}\ }\href@noop {}
  {\  (\bibinfo {year} {2021})},\ \Eprint {http://arxiv.org/abs/2107.10291}
  {arXiv:2107.10291 [astro-ph.CO]} \BibitemShut {NoStop}%
\bibitem [{\citenamefont {Riess}\ \emph {et~al.}(2021)\citenamefont {Riess}
  \emph {et~al.}}]{Riess:2021jrx}%
  \BibitemOpen
  \bibfield  {author} {\bibinfo {author} {\bibfnamefont {Adam~G.}\ \bibnamefont
  {Riess}} \emph {et~al.},\ }\bibfield  {title} {\enquote {\bibinfo {title} {{A
  Comprehensive Measurement of the Local Value of the Hubble Constant with 1
  km/s/Mpc Uncertainty from the Hubble Space Telescope and the SH0ES Team}},}\
  }\href@noop {} {\  (\bibinfo {year} {2021})},\ \Eprint
  {http://arxiv.org/abs/2112.04510} {arXiv:2112.04510 [astro-ph.CO]}
  \BibitemShut {NoStop}%
\bibitem [{\citenamefont {Poulin}\ \emph {et~al.}(2019)\citenamefont {Poulin},
  \citenamefont {Smith}, \citenamefont {Karwal},\ and\ \citenamefont
  {Kamionkowski}}]{Poulin:2018cxd}%
  \BibitemOpen
  \bibfield  {author} {\bibinfo {author} {\bibfnamefont {Vivian}\ \bibnamefont
  {Poulin}}, \bibinfo {author} {\bibfnamefont {Tristan~L.}\ \bibnamefont
  {Smith}}, \bibinfo {author} {\bibfnamefont {Tanvi}\ \bibnamefont {Karwal}}, \
  and\ \bibinfo {author} {\bibfnamefont {Marc}\ \bibnamefont {Kamionkowski}},\
  }\bibfield  {title} {\enquote {\bibinfo {title} {{Early Dark Energy Can
  Resolve The Hubble Tension}},}\ }\href {\doibase
  10.1103/PhysRevLett.122.221301} {\bibfield  {journal} {\bibinfo  {journal}
  {Phys. Rev. Lett.}\ }\textbf {\bibinfo {volume} {122}},\ \bibinfo {pages}
  {221301} (\bibinfo {year} {2019})},\ \Eprint
  {http://arxiv.org/abs/1811.04083} {arXiv:1811.04083 [astro-ph.CO]}
  \BibitemShut {NoStop}%
\bibitem [{\citenamefont {Aghanim}\ \emph
  {et~al.}(2020{\natexlab{a}})\citenamefont {Aghanim} \emph
  {et~al.}}]{Planck:2018lbu}%
  \BibitemOpen
  \bibfield  {author} {\bibinfo {author} {\bibfnamefont {N.}~\bibnamefont
  {Aghanim}} \emph {et~al.} (\bibinfo {collaboration} {Planck}),\ }\bibfield
  {title} {\enquote {\bibinfo {title} {{Planck 2018 results. VIII.
  Gravitational lensing}},}\ }\href {\doibase 10.1051/0004-6361/201833886}
  {\bibfield  {journal} {\bibinfo  {journal} {Astron. Astrophys.}\ }\textbf
  {\bibinfo {volume} {641}},\ \bibinfo {pages} {A8} (\bibinfo {year}
  {2020}{\natexlab{a}})},\ \Eprint {http://arxiv.org/abs/1807.06210}
  {arXiv:1807.06210 [astro-ph.CO]} \BibitemShut {NoStop}%
\bibitem [{\citenamefont {Schöneberg}\ \emph {et~al.}(2019)\citenamefont
  {Schöneberg}, \citenamefont {Lesgourgues},\ and\ \citenamefont
  {Hooper}}]{Schoneberg_2019}%
  \BibitemOpen
  \bibfield  {author} {\bibinfo {author} {\bibfnamefont {Nils}\ \bibnamefont
  {Schöneberg}}, \bibinfo {author} {\bibfnamefont {Julien}\ \bibnamefont
  {Lesgourgues}}, \ and\ \bibinfo {author} {\bibfnamefont {Deanna~C.}\
  \bibnamefont {Hooper}},\ }\bibfield  {title} {\enquote {\bibinfo {title} {The
  {BAO}+{BBN} take on the hubble tension},}\ }\href {\doibase
  10.1088/1475-7516/2019/10/029} {\bibfield  {journal} {\bibinfo  {journal}
  {Journal of Cosmology and Astroparticle Physics}\ }\textbf {\bibinfo {volume}
  {2019}},\ \bibinfo {pages} {029--029} (\bibinfo {year} {2019})}\BibitemShut
  {NoStop}%
\bibitem [{\citenamefont {Consiglio}\ \emph {et~al.}(2018)\citenamefont
  {Consiglio}, \citenamefont {de~Salas}, \citenamefont {Mangano}, \citenamefont
  {Miele}, \citenamefont {Pastor},\ and\ \citenamefont
  {Pisanti}}]{Consiglio_2018}%
  \BibitemOpen
  \bibfield  {author} {\bibinfo {author} {\bibfnamefont {R.}~\bibnamefont
  {Consiglio}}, \bibinfo {author} {\bibfnamefont {P.F.}\ \bibnamefont
  {de~Salas}}, \bibinfo {author} {\bibfnamefont {G.}~\bibnamefont {Mangano}},
  \bibinfo {author} {\bibfnamefont {G.}~\bibnamefont {Miele}}, \bibinfo
  {author} {\bibfnamefont {S.}~\bibnamefont {Pastor}}, \ and\ \bibinfo {author}
  {\bibfnamefont {O.}~\bibnamefont {Pisanti}},\ }\bibfield  {title} {\enquote
  {\bibinfo {title} {{PArthENoPE} reloaded},}\ }\href {\doibase
  10.1016/j.cpc.2018.06.022} {\bibfield  {journal} {\bibinfo  {journal}
  {Computer Physics Communications}\ }\textbf {\bibinfo {volume} {233}},\
  \bibinfo {pages} {237--242} (\bibinfo {year} {2018})}\BibitemShut {NoStop}%
\bibitem [{\citenamefont {Cooke}\ \emph {et~al.}(2018)\citenamefont {Cooke},
  \citenamefont {Pettini},\ and\ \citenamefont {Steidel}}]{Cooke_2018}%
  \BibitemOpen
  \bibfield  {author} {\bibinfo {author} {\bibfnamefont {Ryan~J.}\ \bibnamefont
  {Cooke}}, \bibinfo {author} {\bibfnamefont {Max}\ \bibnamefont {Pettini}}, \
  and\ \bibinfo {author} {\bibfnamefont {Charles~C.}\ \bibnamefont {Steidel}},\
  }\bibfield  {title} {\enquote {\bibinfo {title} {One percent determination of
  the primordial deuterium abundance},}\ }\href {\doibase
  10.3847/1538-4357/aaab53} {\bibfield  {journal} {\bibinfo  {journal} {The
  Astrophysical Journal}\ }\textbf {\bibinfo {volume} {855}},\ \bibinfo {pages}
  {102} (\bibinfo {year} {2018})}\BibitemShut {NoStop}%
\bibitem [{\citenamefont {Aver}\ \emph {et~al.}(2015)\citenamefont {Aver},
  \citenamefont {Olive},\ and\ \citenamefont {Skillman}}]{Aver_2015}%
  \BibitemOpen
  \bibfield  {author} {\bibinfo {author} {\bibfnamefont {Erik}\ \bibnamefont
  {Aver}}, \bibinfo {author} {\bibfnamefont {Keith~A.}\ \bibnamefont {Olive}},
  \ and\ \bibinfo {author} {\bibfnamefont {Evan~D.}\ \bibnamefont {Skillman}},\
  }\bibfield  {title} {\enquote {\bibinfo {title} {The effects of he i
  $\lambda$10830 on helium abundance determinations},}\ }\href {\doibase
  10.1088/1475-7516/2015/07/011} {\bibfield  {journal} {\bibinfo  {journal}
  {Journal of Cosmology and Astroparticle Physics}\ }\textbf {\bibinfo {volume}
  {2015}},\ \bibinfo {pages} {011--011} (\bibinfo {year} {2015})}\BibitemShut
  {NoStop}%
\bibitem [{\citenamefont {Brinckmann}\ and\ \citenamefont
  {Lesgourgues}(2018)}]{Brinckmann:2018cvx}%
  \BibitemOpen
  \bibfield  {author} {\bibinfo {author} {\bibfnamefont {Thejs}\ \bibnamefont
  {Brinckmann}}\ and\ \bibinfo {author} {\bibfnamefont {Julien}\ \bibnamefont
  {Lesgourgues}},\ }\bibfield  {title} {\enquote {\bibinfo {title}
  {{MontePython 3: boosted MCMC sampler and other features}},}\ }\href@noop {}
  {\  (\bibinfo {year} {2018})},\ \Eprint {http://arxiv.org/abs/1804.07261}
  {arXiv:1804.07261} \BibitemShut {NoStop}%
\bibitem [{\citenamefont {Lewis}(2019)}]{Lewis:2019xzd}%
  \BibitemOpen
  \bibfield  {author} {\bibinfo {author} {\bibfnamefont {Antony}\ \bibnamefont
  {Lewis}},\ }\bibfield  {title} {\enquote {\bibinfo {title} {{GetDist: a
  Python package for analysing Monte Carlo samples}},}\ }\href@noop {} {\
  (\bibinfo {year} {2019})},\ \Eprint {http://arxiv.org/abs/1910.13970}
  {arXiv:1910.13970 [astro-ph.IM]} \BibitemShut {NoStop}%
\bibitem [{\citenamefont {Nishimichi}\ \emph {et~al.}(2020)\citenamefont
  {Nishimichi}, \citenamefont {D'Amico}, \citenamefont {Ivanov}, \citenamefont
  {Senatore}, \citenamefont {Simonovi\'c}, \citenamefont {Takada},
  \citenamefont {Zaldarriaga},\ and\ \citenamefont
  {Zhang}}]{Nishimichi:2020tvu}%
  \BibitemOpen
  \bibfield  {author} {\bibinfo {author} {\bibfnamefont {Takahiro}\
  \bibnamefont {Nishimichi}}, \bibinfo {author} {\bibfnamefont {Guido}\
  \bibnamefont {D'Amico}}, \bibinfo {author} {\bibfnamefont {Mikhail~M.}\
  \bibnamefont {Ivanov}}, \bibinfo {author} {\bibfnamefont {Leonardo}\
  \bibnamefont {Senatore}}, \bibinfo {author} {\bibfnamefont {Marko}\
  \bibnamefont {Simonovi\'c}}, \bibinfo {author} {\bibfnamefont {Masahiro}\
  \bibnamefont {Takada}}, \bibinfo {author} {\bibfnamefont {Matias}\
  \bibnamefont {Zaldarriaga}}, \ and\ \bibinfo {author} {\bibfnamefont
  {Pierre}\ \bibnamefont {Zhang}},\ }\bibfield  {title} {\enquote {\bibinfo
  {title} {{Blinded challenge for precision cosmology with large-scale
  structure: results from effective field theory for the redshift-space galaxy
  power spectrum}},}\ }\href {\doibase 10.1103/PhysRevD.102.123541} {\bibfield
  {journal} {\bibinfo  {journal} {Phys. Rev. D}\ }\textbf {\bibinfo {volume}
  {102}},\ \bibinfo {pages} {123541} (\bibinfo {year} {2020})},\ \Eprint
  {http://arxiv.org/abs/2003.08277} {arXiv:2003.08277 [astro-ph.CO]}
  \BibitemShut {NoStop}%
\bibitem [{\citenamefont {Angulo}\ \emph {et~al.}(2015)\citenamefont {Angulo},
  \citenamefont {Fasiello}, \citenamefont {Senatore},\ and\ \citenamefont
  {Vlah}}]{Angulo:2015eqa}%
  \BibitemOpen
  \bibfield  {author} {\bibinfo {author} {\bibfnamefont {Raul}\ \bibnamefont
  {Angulo}}, \bibinfo {author} {\bibfnamefont {Matteo}\ \bibnamefont
  {Fasiello}}, \bibinfo {author} {\bibfnamefont {Leonardo}\ \bibnamefont
  {Senatore}}, \ and\ \bibinfo {author} {\bibfnamefont {Zvonimir}\ \bibnamefont
  {Vlah}},\ }\bibfield  {title} {\enquote {\bibinfo {title} {{On the Statistics
  of Biased Tracers in the Effective Field Theory of Large Scale
  Structures}},}\ }\href {\doibase 10.1088/1475-7516/2015/09/029,
  10.1088/1475-7516/2015/9/029} {\bibfield  {journal} {\bibinfo  {journal}
  {JCAP}\ }\textbf {\bibinfo {volume} {1509}},\ \bibinfo {pages} {029}
  (\bibinfo {year} {2015})},\ \Eprint {http://arxiv.org/abs/1503.08826}
  {arXiv:1503.08826 [astro-ph.CO]} \BibitemShut {NoStop}%
\bibitem [{\citenamefont {Fujita}\ \emph {et~al.}(2016)\citenamefont {Fujita},
  \citenamefont {Mauerhofer}, \citenamefont {Senatore}, \citenamefont {Vlah},\
  and\ \citenamefont {Angulo}}]{Fujita:2016dne}%
  \BibitemOpen
  \bibfield  {author} {\bibinfo {author} {\bibfnamefont {Tomohiro}\
  \bibnamefont {Fujita}}, \bibinfo {author} {\bibfnamefont {Valentin}\
  \bibnamefont {Mauerhofer}}, \bibinfo {author} {\bibfnamefont {Leonardo}\
  \bibnamefont {Senatore}}, \bibinfo {author} {\bibfnamefont {Zvonimir}\
  \bibnamefont {Vlah}}, \ and\ \bibinfo {author} {\bibfnamefont {Raul}\
  \bibnamefont {Angulo}},\ }\bibfield  {title} {\enquote {\bibinfo {title}
  {{Very Massive Tracers and Higher Derivative Biases}},}\ }\href@noop {} {\
  (\bibinfo {year} {2016})},\ \Eprint {http://arxiv.org/abs/1609.00717}
  {arXiv:1609.00717 [astro-ph.CO]} \BibitemShut {NoStop}%
\bibitem [{\citenamefont {D'Amico}\ \emph
  {et~al.}(2021{\natexlab{b}})\citenamefont {D'Amico}, \citenamefont
  {Senatore}, \citenamefont {Zhang},\ and\ \citenamefont
  {Nishimichi}}]{DAmico:2021ymi}%
  \BibitemOpen
  \bibfield  {author} {\bibinfo {author} {\bibfnamefont {Guido}\ \bibnamefont
  {D'Amico}}, \bibinfo {author} {\bibfnamefont {Leonardo}\ \bibnamefont
  {Senatore}}, \bibinfo {author} {\bibfnamefont {Pierre}\ \bibnamefont
  {Zhang}}, \ and\ \bibinfo {author} {\bibfnamefont {Takahiro}\ \bibnamefont
  {Nishimichi}},\ }\bibfield  {title} {\enquote {\bibinfo {title} {{Taming
  redshift-space distortion effects in the EFTofLSS and its application to
  data}},}\ }\href@noop {} {\  (\bibinfo {year} {2021}{\natexlab{b}})},\
  \Eprint {http://arxiv.org/abs/2110.00016} {arXiv:2110.00016 [astro-ph.CO]}
  \BibitemShut {NoStop}%
\bibitem [{\citenamefont {Ivanov}\ \emph {et~al.}(2022)\citenamefont {Ivanov},
  \citenamefont {Philcox}, \citenamefont {Nishimichi}, \citenamefont
  {Simonovi\'c}, \citenamefont {Takada},\ and\ \citenamefont
  {Zaldarriaga}}]{Ivanov:2021kcd}%
  \BibitemOpen
  \bibfield  {author} {\bibinfo {author} {\bibfnamefont {Mikhail~M.}\
  \bibnamefont {Ivanov}}, \bibinfo {author} {\bibfnamefont {Oliver H.~E.}\
  \bibnamefont {Philcox}}, \bibinfo {author} {\bibfnamefont {Takahiro}\
  \bibnamefont {Nishimichi}}, \bibinfo {author} {\bibfnamefont {Marko}\
  \bibnamefont {Simonovi\'c}}, \bibinfo {author} {\bibfnamefont {Masahiro}\
  \bibnamefont {Takada}}, \ and\ \bibinfo {author} {\bibfnamefont {Matias}\
  \bibnamefont {Zaldarriaga}},\ }\bibfield  {title} {\enquote {\bibinfo {title}
  {{Precision analysis of the redshift-space galaxy bispectrum}},}\ }\href
  {\doibase 10.1103/PhysRevD.105.063512} {\bibfield  {journal} {\bibinfo
  {journal} {Phys. Rev. D}\ }\textbf {\bibinfo {volume} {105}},\ \bibinfo
  {pages} {063512} (\bibinfo {year} {2022})},\ \Eprint
  {http://arxiv.org/abs/2110.10161} {arXiv:2110.10161 [astro-ph.CO]}
  \BibitemShut {NoStop}%
\bibitem [{\citenamefont {Mirbabayi}\ \emph {et~al.}(2015)\citenamefont
  {Mirbabayi}, \citenamefont {Schmidt},\ and\ \citenamefont
  {Zaldarriaga}}]{Mirbabayi:2014zca}%
  \BibitemOpen
  \bibfield  {author} {\bibinfo {author} {\bibfnamefont {Mehrdad}\ \bibnamefont
  {Mirbabayi}}, \bibinfo {author} {\bibfnamefont {Fabian}\ \bibnamefont
  {Schmidt}}, \ and\ \bibinfo {author} {\bibfnamefont {Matias}\ \bibnamefont
  {Zaldarriaga}},\ }\bibfield  {title} {\enquote {\bibinfo {title} {{Biased
  Tracers and Time Evolution}},}\ }\href {\doibase
  10.1088/1475-7516/2015/07/030} {\bibfield  {journal} {\bibinfo  {journal}
  {JCAP}\ }\textbf {\bibinfo {volume} {07}},\ \bibinfo {pages} {030} (\bibinfo
  {year} {2015})},\ \Eprint {http://arxiv.org/abs/1412.5169} {arXiv:1412.5169
  [astro-ph.CO]} \BibitemShut {NoStop}%
\bibitem [{\citenamefont {Fujita}\ and\ \citenamefont
  {Vlah}(2020)}]{Fujita:2020xtd}%
  \BibitemOpen
  \bibfield  {author} {\bibinfo {author} {\bibfnamefont {Tomohiro}\
  \bibnamefont {Fujita}}\ and\ \bibinfo {author} {\bibfnamefont {Zvonimir}\
  \bibnamefont {Vlah}},\ }\bibfield  {title} {\enquote {\bibinfo {title}
  {{Perturbative description of biased tracers using consistency relations of
  LSS}},}\ }\href {\doibase 10.1088/1475-7516/2020/10/059} {\bibfield
  {journal} {\bibinfo  {journal} {JCAP}\ }\textbf {\bibinfo {volume} {10}},\
  \bibinfo {pages} {059} (\bibinfo {year} {2020})},\ \Eprint
  {http://arxiv.org/abs/2003.10114} {arXiv:2003.10114 [astro-ph.CO]}
  \BibitemShut {NoStop}%
\bibitem [{\citenamefont {Lewandowski}\ \emph {et~al.}(2018)\citenamefont
  {Lewandowski}, \citenamefont {Senatore}, \citenamefont {Prada}, \citenamefont
  {Zhao},\ and\ \citenamefont {Chuang}}]{Lewandowski:2015ziq}%
  \BibitemOpen
  \bibfield  {author} {\bibinfo {author} {\bibfnamefont {Matthew}\ \bibnamefont
  {Lewandowski}}, \bibinfo {author} {\bibfnamefont {Leonardo}\ \bibnamefont
  {Senatore}}, \bibinfo {author} {\bibfnamefont {Francisco}\ \bibnamefont
  {Prada}}, \bibinfo {author} {\bibfnamefont {Cheng}\ \bibnamefont {Zhao}}, \
  and\ \bibinfo {author} {\bibfnamefont {Chia-Hsun}\ \bibnamefont {Chuang}},\
  }\bibfield  {title} {\enquote {\bibinfo {title} {{EFT of large scale
  structures in redshift space}},}\ }\href {\doibase
  10.1103/PhysRevD.97.063526} {\bibfield  {journal} {\bibinfo  {journal} {Phys.
  Rev.}\ }\textbf {\bibinfo {volume} {D97}},\ \bibinfo {pages} {063526}
  (\bibinfo {year} {2018})},\ \Eprint {http://arxiv.org/abs/1512.06831}
  {arXiv:1512.06831 [astro-ph.CO]} \BibitemShut {NoStop}%
\bibitem [{\citenamefont {Senatore}\ and\ \citenamefont
  {Trevisan}(2018)}]{Senatore:2017pbn}%
  \BibitemOpen
  \bibfield  {author} {\bibinfo {author} {\bibfnamefont {Leonardo}\
  \bibnamefont {Senatore}}\ and\ \bibinfo {author} {\bibfnamefont {Gabriele}\
  \bibnamefont {Trevisan}},\ }\bibfield  {title} {\enquote {\bibinfo {title}
  {{On the IR-Resummation in the EFTofLSS}},}\ }\href {\doibase
  10.1088/1475-7516/2018/05/019} {\bibfield  {journal} {\bibinfo  {journal}
  {JCAP}\ }\textbf {\bibinfo {volume} {1805}},\ \bibinfo {pages} {019}
  (\bibinfo {year} {2018})},\ \Eprint {http://arxiv.org/abs/1710.02178}
  {arXiv:1710.02178 [astro-ph.CO]} \BibitemShut {NoStop}%
\bibitem [{\citenamefont {Blas}\ \emph {et~al.}(2016)\citenamefont {Blas},
  \citenamefont {Garny}, \citenamefont {Ivanov},\ and\ \citenamefont
  {Sibiryakov}}]{Blas:2016sfa}%
  \BibitemOpen
  \bibfield  {author} {\bibinfo {author} {\bibfnamefont {Diego}\ \bibnamefont
  {Blas}}, \bibinfo {author} {\bibfnamefont {Mathias}\ \bibnamefont {Garny}},
  \bibinfo {author} {\bibfnamefont {Mikhail~M.}\ \bibnamefont {Ivanov}}, \ and\
  \bibinfo {author} {\bibfnamefont {Sergey}\ \bibnamefont {Sibiryakov}},\
  }\bibfield  {title} {\enquote {\bibinfo {title} {{Time-Sliced Perturbation
  Theory II: Baryon Acoustic Oscillations and Infrared Resummation}},}\ }\href
  {\doibase 10.1088/1475-7516/2016/07/028} {\bibfield  {journal} {\bibinfo
  {journal} {JCAP}\ }\textbf {\bibinfo {volume} {1607}},\ \bibinfo {pages}
  {028} (\bibinfo {year} {2016})},\ \Eprint {http://arxiv.org/abs/1605.02149}
  {arXiv:1605.02149 [astro-ph.CO]} \BibitemShut {NoStop}%
\bibitem [{\citenamefont {Ivanov}\ and\ \citenamefont
  {Sibiryakov}(2018)}]{Ivanov:2018gjr}%
  \BibitemOpen
  \bibfield  {author} {\bibinfo {author} {\bibfnamefont {Mikhail~M.}\
  \bibnamefont {Ivanov}}\ and\ \bibinfo {author} {\bibfnamefont {Sergey}\
  \bibnamefont {Sibiryakov}},\ }\bibfield  {title} {\enquote {\bibinfo {title}
  {{Infrared Resummation for Biased Tracers in Redshift Space}},}\ }\href
  {\doibase 10.1088/1475-7516/2018/07/053} {\bibfield  {journal} {\bibinfo
  {journal} {JCAP}\ }\textbf {\bibinfo {volume} {1807}},\ \bibinfo {pages}
  {053} (\bibinfo {year} {2018})},\ \Eprint {http://arxiv.org/abs/1804.05080}
  {arXiv:1804.05080 [astro-ph.CO]} \BibitemShut {NoStop}%
\bibitem [{\citenamefont {Lewandowski}\ and\ \citenamefont
  {Senatore}(2020)}]{Lewandowski:2018ywf}%
  \BibitemOpen
  \bibfield  {author} {\bibinfo {author} {\bibfnamefont {Matthew}\ \bibnamefont
  {Lewandowski}}\ and\ \bibinfo {author} {\bibfnamefont {Leonardo}\
  \bibnamefont {Senatore}},\ }\bibfield  {title} {\enquote {\bibinfo {title}
  {{An analytic implementation of the IR-resummation for the BAO peak}},}\
  }\href {\doibase 10.1088/1475-7516/2020/03/018} {\bibfield  {journal}
  {\bibinfo  {journal} {JCAP}\ }\textbf {\bibinfo {volume} {03}},\ \bibinfo
  {pages} {018} (\bibinfo {year} {2020})},\ \Eprint
  {http://arxiv.org/abs/1810.11855} {arXiv:1810.11855 [astro-ph.CO]}
  \BibitemShut {NoStop}%
\bibitem [{\citenamefont {Aghanim}\ \emph
  {et~al.}(2020{\natexlab{b}})\citenamefont {Aghanim} \emph
  {et~al.}}]{Planck:2018vyg}%
  \BibitemOpen
  \bibfield  {author} {\bibinfo {author} {\bibfnamefont {N.}~\bibnamefont
  {Aghanim}} \emph {et~al.} (\bibinfo {collaboration} {Planck}),\ }\bibfield
  {title} {\enquote {\bibinfo {title} {{Planck 2018 results. VI. Cosmological
  parameters}},}\ }\href {\doibase 10.1051/0004-6361/201833910} {\bibfield
  {journal} {\bibinfo  {journal} {Astron. Astrophys.}\ }\textbf {\bibinfo
  {volume} {641}},\ \bibinfo {pages} {A6} (\bibinfo {year}
  {2020}{\natexlab{b}})},\ \bibinfo {note} {[Erratum: Astron.Astrophys. 652, C4
  (2021)]},\ \Eprint {http://arxiv.org/abs/1807.06209} {arXiv:1807.06209
  [astro-ph.CO]} \BibitemShut {NoStop}%
\bibitem [{\citenamefont {{Aghamousa, Amir and
  others}}({2016})}]{Aghamousa:2016zmz}%
  \BibitemOpen
  \bibfield  {author} {\bibinfo {author} {\bibnamefont {{Aghamousa, Amir and
  others}}} (\bibinfo {collaboration} {{DESI}}),\ }\bibfield  {title} {\enquote
  {\bibinfo {title} {{The DESI Experiment Part I: Science,Targeting, and Survey
  Design}},}\ }\href@noop {} {\  (\bibinfo {year} {{2016}})},\ \Eprint
  {http://arxiv.org/abs/{1611.00036}} {{1611.00036}} \BibitemShut {NoStop}%
\bibitem [{\citenamefont {Amendola}\ \emph {et~al.}(2018)\citenamefont
  {Amendola} \emph {et~al.}}]{Amendola:2016saw}%
  \BibitemOpen
  \bibfield  {author} {\bibinfo {author} {\bibfnamefont {Luca}\ \bibnamefont
  {Amendola}} \emph {et~al.},\ }\bibfield  {title} {\enquote {\bibinfo {title}
  {{Cosmology and fundamental physics with the Euclid satellite}},}\ }\href
  {\doibase 10.1007/s41114-017-0010-3} {\bibfield  {journal} {\bibinfo
  {journal} {Living Rev. Rel.}\ }\textbf {\bibinfo {volume} {21}},\ \bibinfo
  {pages} {2} (\bibinfo {year} {2018})},\ \Eprint
  {http://arxiv.org/abs/1606.00180} {1606.00180} \BibitemShut {NoStop}%
\bibitem [{\citenamefont {Ade}\ \emph {et~al.}(2014)\citenamefont {Ade} \emph
  {et~al.}}]{Planck:2013nga}%
  \BibitemOpen
  \bibfield  {author} {\bibinfo {author} {\bibfnamefont {P.~A.~R.}\
  \bibnamefont {Ade}} \emph {et~al.} (\bibinfo {collaboration} {Planck}),\
  }\bibfield  {title} {\enquote {\bibinfo {title} {{Planck intermediate
  results. XVI. Profile likelihoods for cosmological parameters}},}\ }\href
  {\doibase 10.1051/0004-6361/201323003} {\bibfield  {journal} {\bibinfo
  {journal} {Astron. Astrophys.}\ }\textbf {\bibinfo {volume} {566}},\ \bibinfo
  {pages} {A54} (\bibinfo {year} {2014})},\ \Eprint
  {http://arxiv.org/abs/1311.1657} {arXiv:1311.1657 [astro-ph.CO]} \BibitemShut
  {NoStop}%
\bibitem [{\citenamefont {Herold}\ \emph {et~al.}(2021)\citenamefont {Herold},
  \citenamefont {Ferreira},\ and\ \citenamefont {Komatsu}}]{Herold:2021ksg}%
  \BibitemOpen
  \bibfield  {author} {\bibinfo {author} {\bibfnamefont {Laura}\ \bibnamefont
  {Herold}}, \bibinfo {author} {\bibfnamefont {Elisa G.~M.}\ \bibnamefont
  {Ferreira}}, \ and\ \bibinfo {author} {\bibfnamefont {Eiichiro}\ \bibnamefont
  {Komatsu}},\ }\bibfield  {title} {\enquote {\bibinfo {title} {{New constraint
  on Early Dark Energy from Planck and BOSS data using the profile
  likelihood}},}\ }\href@noop {} {\  (\bibinfo {year} {2021})},\ \Eprint
  {http://arxiv.org/abs/2112.12140} {arXiv:2112.12140 [astro-ph.CO]}
  \BibitemShut {NoStop}%
\bibitem [{\citenamefont {Reeves}\ \emph {et~al.}(2022)\citenamefont {Reeves},
  \citenamefont {Herold}, \citenamefont {Vagnozzi}, \citenamefont {Sherwin},\
  and\ \citenamefont {Ferreira}}]{Reeves:2022aoi}%
  \BibitemOpen
  \bibfield  {author} {\bibinfo {author} {\bibfnamefont {Alexander}\
  \bibnamefont {Reeves}}, \bibinfo {author} {\bibfnamefont {Laura}\
  \bibnamefont {Herold}}, \bibinfo {author} {\bibfnamefont {Sunny}\
  \bibnamefont {Vagnozzi}}, \bibinfo {author} {\bibfnamefont {Blake~D.}\
  \bibnamefont {Sherwin}}, \ and\ \bibinfo {author} {\bibfnamefont {Elisa
  G.~M.}\ \bibnamefont {Ferreira}},\ }\bibfield  {title} {\enquote {\bibinfo
  {title} {{Restoring cosmological concordance with early dark energy and
  massive neutrinos?}}}\ }\href@noop {} {\  (\bibinfo {year} {2022})},\ \Eprint
  {http://arxiv.org/abs/2207.01501} {arXiv:2207.01501 [astro-ph.CO]}
  \BibitemShut {NoStop}%
\bibitem [{\citenamefont {G\'omez-Valent}(2022)}]{Gomez-Valent:2022hkb}%
  \BibitemOpen
  \bibfield  {author} {\bibinfo {author} {\bibfnamefont {Adri\`a}\ \bibnamefont
  {G\'omez-Valent}},\ }\bibfield  {title} {\enquote {\bibinfo {title} {{A fast
  test to assess the impact of marginalization in Monte Carlo analyses, and its
  application to cosmology}},}\ }\href@noop {} {\  (\bibinfo {year} {2022})},\
  \Eprint {http://arxiv.org/abs/2203.16285} {arXiv:2203.16285 [astro-ph.CO]}
  \BibitemShut {NoStop}%
\bibitem [{\citenamefont {Gil-Marín}\ \emph
  {et~al.}(2016{\natexlab{a}})\citenamefont {Gil-Marín} \emph
  {et~al.}}]{Gil-Marin:2015sqa}%
  \BibitemOpen
  \bibfield  {author} {\bibinfo {author} {\bibfnamefont {Héctor}\ \bibnamefont
  {Gil-Marín}} \emph {et~al.},\ }\bibfield  {title} {\enquote {\bibinfo
  {title} {{The clustering of galaxies in the SDSS-III Baryon Oscillation
  Spectroscopic Survey: RSD measurement from the LOS-dependent power spectrum
  of DR12 BOSS galaxies}},}\ }\href {\doibase 10.1093/mnras/stw1096} {\bibfield
   {journal} {\bibinfo  {journal} {Mon. Not. Roy. Astron. Soc.}\ }\textbf
  {\bibinfo {volume} {460}},\ \bibinfo {pages} {4188--4209} (\bibinfo {year}
  {2016}{\natexlab{a}})},\ \Eprint {http://arxiv.org/abs/1509.06386}
  {arXiv:1509.06386 [astro-ph.CO]} \BibitemShut {NoStop}%
\bibitem [{\citenamefont {Zhao}\ \emph {et~al.}(2021)\citenamefont {Zhao} \emph
  {et~al.}}]{Zhao:2020bib}%
  \BibitemOpen
  \bibfield  {author} {\bibinfo {author} {\bibfnamefont {Cheng}\ \bibnamefont
  {Zhao}} \emph {et~al.},\ }\bibfield  {title} {\enquote {\bibinfo {title}
  {{The completed SDSS-IV extended Baryon Oscillation Spectroscopic Survey:
  1000 multi-tracer mock catalogues with redshift evolution and systematics for
  galaxies and quasars of the final data release}},}\ }\href {\doibase
  10.1093/mnras/stab510} {\bibfield  {journal} {\bibinfo  {journal} {Mon. Not.
  Roy. Astron. Soc.}\ }\textbf {\bibinfo {volume} {503}},\ \bibinfo {pages}
  {1149--1173} (\bibinfo {year} {2021})},\ \Eprint
  {http://arxiv.org/abs/2007.08997} {arXiv:2007.08997 [astro-ph.CO]}
  \BibitemShut {NoStop}%
\bibitem [{\citenamefont {Hand}\ \emph {et~al.}(2018)\citenamefont {Hand},
  \citenamefont {Feng}, \citenamefont {Beutler}, \citenamefont {Li},
  \citenamefont {Modi}, \citenamefont {Seljak},\ and\ \citenamefont
  {Slepian}}]{Hand:2017pqn}%
  \BibitemOpen
  \bibfield  {author} {\bibinfo {author} {\bibfnamefont {Nick}\ \bibnamefont
  {Hand}}, \bibinfo {author} {\bibfnamefont {Yu}~\bibnamefont {Feng}}, \bibinfo
  {author} {\bibfnamefont {Florian}\ \bibnamefont {Beutler}}, \bibinfo {author}
  {\bibfnamefont {Yin}\ \bibnamefont {Li}}, \bibinfo {author} {\bibfnamefont
  {Chirag}\ \bibnamefont {Modi}}, \bibinfo {author} {\bibfnamefont {Uros}\
  \bibnamefont {Seljak}}, \ and\ \bibinfo {author} {\bibfnamefont {Zachary}\
  \bibnamefont {Slepian}},\ }\bibfield  {title} {\enquote {\bibinfo {title}
  {{nbodykit: an open-source, massively parallel toolkit for large-scale
  structure}},}\ }\href {\doibase 10.3847/1538-3881/aadae0} {\bibfield
  {journal} {\bibinfo  {journal} {Astron. J.}\ }\textbf {\bibinfo {volume}
  {156}},\ \bibinfo {pages} {160} (\bibinfo {year} {2018})},\ \Eprint
  {http://arxiv.org/abs/1712.05834} {arXiv:1712.05834 [astro-ph.IM]}
  \BibitemShut {NoStop}%
\bibitem [{\citenamefont {Zhao}(2023)}]{Zhao:2023iwb}%
  \BibitemOpen
  \bibfield  {author} {\bibinfo {author} {\bibfnamefont {Cheng}\ \bibnamefont
  {Zhao}},\ }\bibfield  {title} {\enquote {\bibinfo {title} {{Fast Correlation
  Function Calculator -- A high-performance pair counting toolkit}},}\
  }\href@noop {} {\  (\bibinfo {year} {2023})},\ \Eprint
  {http://arxiv.org/abs/2301.12557} {arXiv:2301.12557 [astro-ph.IM]}
  \BibitemShut {NoStop}%
\bibitem [{\citenamefont {Beutler}\ and\ \citenamefont
  {McDonald}(2021)}]{Beutler:2021eqq}%
  \BibitemOpen
  \bibfield  {author} {\bibinfo {author} {\bibfnamefont {Florian}\ \bibnamefont
  {Beutler}}\ and\ \bibinfo {author} {\bibfnamefont {Patrick}\ \bibnamefont
  {McDonald}},\ }\bibfield  {title} {\enquote {\bibinfo {title} {{Unified
  galaxy power spectrum measurements from 6dFGS, BOSS, and eBOSS}},}\ }\href
  {\doibase 10.1088/1475-7516/2021/11/031} {\bibfield  {journal} {\bibinfo
  {journal} {JCAP}\ }\textbf {\bibinfo {volume} {11}},\ \bibinfo {pages} {031}
  (\bibinfo {year} {2021})},\ \Eprint {http://arxiv.org/abs/2106.06324}
  {arXiv:2106.06324 [astro-ph.CO]} \BibitemShut {NoStop}%
\bibitem [{\citenamefont {Philcox}(2021)}]{Philcox:2020vbm}%
  \BibitemOpen
  \bibfield  {author} {\bibinfo {author} {\bibfnamefont {Oliver H.~E.}\
  \bibnamefont {Philcox}},\ }\bibfield  {title} {\enquote {\bibinfo {title}
  {{Cosmology without window functions: Quadratic estimators for the galaxy
  power spectrum}},}\ }\href {\doibase 10.1103/PhysRevD.103.103504} {\bibfield
  {journal} {\bibinfo  {journal} {Phys. Rev. D}\ }\textbf {\bibinfo {volume}
  {103}},\ \bibinfo {pages} {103504} (\bibinfo {year} {2021})},\ \Eprint
  {http://arxiv.org/abs/2012.09389} {arXiv:2012.09389 [astro-ph.CO]}
  \BibitemShut {NoStop}%
\bibitem [{\citenamefont {Gil-Marín}\ \emph
  {et~al.}(2016{\natexlab{b}})\citenamefont {Gil-Marín} \emph
  {et~al.}}]{Gil-Marin:2015nqa}%
  \BibitemOpen
  \bibfield  {author} {\bibinfo {author} {\bibfnamefont {Héctor}\ \bibnamefont
  {Gil-Marín}} \emph {et~al.},\ }\bibfield  {title} {\enquote {\bibinfo
  {title} {{The clustering of galaxies in the SDSS-III Baryon Oscillation
  Spectroscopic Survey: BAO measurement from the LOS-dependent power spectrum
  of DR12 BOSS galaxies}},}\ }\href {\doibase 10.1093/mnras/stw1264} {\bibfield
   {journal} {\bibinfo  {journal} {Mon. Not. Roy. Astron. Soc.}\ }\textbf
  {\bibinfo {volume} {460}},\ \bibinfo {pages} {4210--4219} (\bibinfo {year}
  {2016}{\natexlab{b}})},\ \Eprint {http://arxiv.org/abs/1509.06373}
  {arXiv:1509.06373 [astro-ph.CO]} \BibitemShut {NoStop}%
\bibitem [{\citenamefont {Beutler}\ \emph {et~al.}(2017)\citenamefont {Beutler}
  \emph {et~al.}}]{BOSS:2016hvq}%
  \BibitemOpen
  \bibfield  {author} {\bibinfo {author} {\bibfnamefont {Florian}\ \bibnamefont
  {Beutler}} \emph {et~al.} (\bibinfo {collaboration} {BOSS}),\ }\bibfield
  {title} {\enquote {\bibinfo {title} {{The clustering of galaxies in the
  completed SDSS-III Baryon Oscillation Spectroscopic Survey: baryon acoustic
  oscillations in the Fourier space}},}\ }\href {\doibase
  10.1093/mnras/stw2373} {\bibfield  {journal} {\bibinfo  {journal} {Mon. Not.
  Roy. Astron. Soc.}\ }\textbf {\bibinfo {volume} {464}},\ \bibinfo {pages}
  {3409--3430} (\bibinfo {year} {2017})},\ \Eprint
  {http://arxiv.org/abs/1607.03149} {arXiv:1607.03149 [astro-ph.CO]}
  \BibitemShut {NoStop}%
\bibitem [{\citenamefont {Philcox}\ \emph {et~al.}(2020)\citenamefont
  {Philcox}, \citenamefont {Ivanov}, \citenamefont {Simonović},\ and\
  \citenamefont {Zaldarriaga}}]{Philcox:2020vvt}%
  \BibitemOpen
  \bibfield  {author} {\bibinfo {author} {\bibfnamefont {Oliver H.~E.}\
  \bibnamefont {Philcox}}, \bibinfo {author} {\bibfnamefont {Mikhail~M.}\
  \bibnamefont {Ivanov}}, \bibinfo {author} {\bibfnamefont {Marko}\
  \bibnamefont {Simonović}}, \ and\ \bibinfo {author} {\bibfnamefont {Matias}\
  \bibnamefont {Zaldarriaga}},\ }\bibfield  {title} {\enquote {\bibinfo {title}
  {{Combining Full-Shape and BAO Analyses of Galaxy Power Spectra: A 1.6\%
  CMB-independent constraint on H0}},}\ }\href@noop {} {\  (\bibinfo {year}
  {2020})},\ \Eprint {http://arxiv.org/abs/2002.04035} {arXiv:2002.04035
  [astro-ph.CO]} \BibitemShut {NoStop}%
\bibitem [{\citenamefont {Kitaura}\ \emph {et~al.}(2016)\citenamefont {Kitaura}
  \emph {et~al.}}]{Kitaura:2015uqa}%
  \BibitemOpen
  \bibfield  {author} {\bibinfo {author} {\bibfnamefont {Francisco-Shu}\
  \bibnamefont {Kitaura}} \emph {et~al.},\ }\bibfield  {title} {\enquote
  {\bibinfo {title} {{The clustering of galaxies in the SDSS-III Baryon
  Oscillation Spectroscopic Survey: mock galaxy catalogues for the BOSS Final
  Data Release}},}\ }\href {\doibase 10.1093/mnras/stv2826} {\bibfield
  {journal} {\bibinfo  {journal} {Mon. Not. Roy. Astron. Soc.}\ }\textbf
  {\bibinfo {volume} {456}},\ \bibinfo {pages} {4156--4173} (\bibinfo {year}
  {2016})},\ \Eprint {http://arxiv.org/abs/1509.06400} {arXiv:1509.06400
  [astro-ph.CO]} \BibitemShut {NoStop}%
\bibitem [{\citenamefont {Reid}\ \emph {et~al.}(2016)\citenamefont {Reid} \emph
  {et~al.}}]{Reid:2015gra}%
  \BibitemOpen
  \bibfield  {author} {\bibinfo {author} {\bibfnamefont {Beth}\ \bibnamefont
  {Reid}} \emph {et~al.},\ }\bibfield  {title} {\enquote {\bibinfo {title}
  {{SDSS-III Baryon Oscillation Spectroscopic Survey Data Release 12: galaxy
  target selection and large scale structure catalogues}},}\ }\href {\doibase
  10.1093/mnras/stv2382} {\bibfield  {journal} {\bibinfo  {journal} {Mon. Not.
  Roy. Astron. Soc.}\ }\textbf {\bibinfo {volume} {455}},\ \bibinfo {pages}
  {1553--1573} (\bibinfo {year} {2016})},\ \Eprint
  {http://arxiv.org/abs/1509.06529} {arXiv:1509.06529 [astro-ph.CO]}
  \BibitemShut {NoStop}%
\bibitem [{\citenamefont {de~Mattia}\ and\ \citenamefont
  {Ruhlmann-Kleider}(2019)}]{deMattia:2019vdg}%
  \BibitemOpen
  \bibfield  {author} {\bibinfo {author} {\bibfnamefont {Arnaud}\ \bibnamefont
  {de~Mattia}}\ and\ \bibinfo {author} {\bibfnamefont {Vanina}\ \bibnamefont
  {Ruhlmann-Kleider}},\ }\bibfield  {title} {\enquote {\bibinfo {title}
  {{Integral constraints in spectroscopic surveys}},}\ }\href {\doibase
  10.1088/1475-7516/2019/08/036} {\bibfield  {journal} {\bibinfo  {journal}
  {JCAP}\ }\textbf {\bibinfo {volume} {08}},\ \bibinfo {pages} {036} (\bibinfo
  {year} {2019})},\ \Eprint {http://arxiv.org/abs/1904.08851} {arXiv:1904.08851
  [astro-ph.CO]} \BibitemShut {NoStop}%
\bibitem [{\citenamefont {Chen}\ \emph {et~al.}(2020)\citenamefont {Chen},
  \citenamefont {Vlah},\ and\ \citenamefont {White}}]{Chen:2020fxs}%
  \BibitemOpen
  \bibfield  {author} {\bibinfo {author} {\bibfnamefont {Shi-Fan}\ \bibnamefont
  {Chen}}, \bibinfo {author} {\bibfnamefont {Zvonimir}\ \bibnamefont {Vlah}}, \
  and\ \bibinfo {author} {\bibfnamefont {Martin}\ \bibnamefont {White}},\
  }\bibfield  {title} {\enquote {\bibinfo {title} {{Consistent Modeling of
  Velocity Statistics and Redshift-Space Distortions in One-Loop Perturbation
  Theory}},}\ }\href {\doibase 10.1088/1475-7516/2020/07/062} {\bibfield
  {journal} {\bibinfo  {journal} {JCAP}\ }\textbf {\bibinfo {volume} {07}},\
  \bibinfo {pages} {062} (\bibinfo {year} {2020})},\ \Eprint
  {http://arxiv.org/abs/2005.00523} {arXiv:2005.00523 [astro-ph.CO]}
  \BibitemShut {NoStop}%
\bibitem [{\citenamefont {Chen}\ \emph {et~al.}(2021)\citenamefont {Chen},
  \citenamefont {Vlah}, \citenamefont {Castorina},\ and\ \citenamefont
  {White}}]{Chen:2020zjt}%
  \BibitemOpen
  \bibfield  {author} {\bibinfo {author} {\bibfnamefont {Shi-Fan}\ \bibnamefont
  {Chen}}, \bibinfo {author} {\bibfnamefont {Zvonimir}\ \bibnamefont {Vlah}},
  \bibinfo {author} {\bibfnamefont {Emanuele}\ \bibnamefont {Castorina}}, \
  and\ \bibinfo {author} {\bibfnamefont {Martin}\ \bibnamefont {White}},\
  }\bibfield  {title} {\enquote {\bibinfo {title} {{Redshift-Space Distortions
  in Lagrangian Perturbation Theory}},}\ }\href {\doibase
  10.1088/1475-7516/2021/03/100} {\bibfield  {journal} {\bibinfo  {journal}
  {JCAP}\ }\textbf {\bibinfo {volume} {03}},\ \bibinfo {pages} {100} (\bibinfo
  {year} {2021})},\ \Eprint {http://arxiv.org/abs/2012.04636} {arXiv:2012.04636
  [astro-ph.CO]} \BibitemShut {NoStop}%
\bibitem [{\citenamefont {Maus}\ \emph {et~al.}(2023)\citenamefont {Maus},
  \citenamefont {Chen},\ and\ \citenamefont {White}}]{Maus:2023rtr}%
  \BibitemOpen
  \bibfield  {author} {\bibinfo {author} {\bibfnamefont {Mark}\ \bibnamefont
  {Maus}}, \bibinfo {author} {\bibfnamefont {Shi-Fan}\ \bibnamefont {Chen}}, \
  and\ \bibinfo {author} {\bibfnamefont {Martin}\ \bibnamefont {White}},\
  }\bibfield  {title} {\enquote {\bibinfo {title} {{A comparison of template
  vs. direct model fitting for redshift-space distortions in BOSS}},}\
  }\href@noop {} {\  (\bibinfo {year} {2023})},\ \Eprint
  {http://arxiv.org/abs/2302.07430} {arXiv:2302.07430 [astro-ph.CO]}
  \BibitemShut {NoStop}%
\bibitem [{\citenamefont {Amon}\ and\ \citenamefont
  {Efstathiou}(2022)}]{Amon:2022azi}%
  \BibitemOpen
  \bibfield  {author} {\bibinfo {author} {\bibfnamefont {Alexandra}\
  \bibnamefont {Amon}}\ and\ \bibinfo {author} {\bibfnamefont {George}\
  \bibnamefont {Efstathiou}},\ }\bibfield  {title} {\enquote {\bibinfo {title}
  {{A non-linear solution to the $S_8$ tension?}}}\ }\href@noop {} {\
  (\bibinfo {year} {2022})},\ \Eprint {http://arxiv.org/abs/2206.11794}
  {arXiv:2206.11794 [astro-ph.CO]} \BibitemShut {NoStop}%
\bibitem [{\citenamefont {Heymans}\ \emph {et~al.}(2012)\citenamefont {Heymans}
  \emph {et~al.}}]{Heymans:2012gg}%
  \BibitemOpen
  \bibfield  {author} {\bibinfo {author} {\bibfnamefont {Catherine}\
  \bibnamefont {Heymans}} \emph {et~al.},\ }\bibfield  {title} {\enquote
  {\bibinfo {title} {{CFHTLenS: The Canada-France-Hawaii Telescope Lensing
  Survey}},}\ }\href {\doibase 10.1111/j.1365-2966.2012.21952.x} {\bibfield
  {journal} {\bibinfo  {journal} {Mon. Not. Roy. Astron. Soc.}\ }\textbf
  {\bibinfo {volume} {427}},\ \bibinfo {pages} {146} (\bibinfo {year}
  {2012})},\ \Eprint {http://arxiv.org/abs/1210.0032} {arXiv:1210.0032}
  \BibitemShut {NoStop}%
\bibitem [{\citenamefont {Abbott}\ \emph {et~al.}(2022)\citenamefont {Abbott}
  \emph {et~al.}}]{DES:2022urg}%
  \BibitemOpen
  \bibfield  {author} {\bibinfo {author} {\bibfnamefont {T.~M.~C.}\
  \bibnamefont {Abbott}} \emph {et~al.} (\bibinfo {collaboration} {DES, SPT}),\
  }\bibfield  {title} {\enquote {\bibinfo {title} {{Joint analysis of DES Year
  3 data and CMB lensing from SPT and Planck III: Combined cosmological
  constraints}},}\ }\href@noop {} {\  (\bibinfo {year} {2022})},\ \Eprint
  {http://arxiv.org/abs/2206.10824} {arXiv:2206.10824 [astro-ph.CO]}
  \BibitemShut {NoStop}%
\bibitem [{\citenamefont {Alam}\ \emph {et~al.}(2021)\citenamefont {Alam} \emph
  {et~al.}}]{eBOSS:2020yzd}%
  \BibitemOpen
  \bibfield  {author} {\bibinfo {author} {\bibfnamefont {Shadab}\ \bibnamefont
  {Alam}} \emph {et~al.} (\bibinfo {collaboration} {eBOSS}),\ }\bibfield
  {title} {\enquote {\bibinfo {title} {{Completed SDSS-IV extended Baryon
  Oscillation Spectroscopic Survey: Cosmological implications from two decades
  of spectroscopic surveys at the Apache Point Observatory}},}\ }\href
  {\doibase 10.1103/PhysRevD.103.083533} {\bibfield  {journal} {\bibinfo
  {journal} {Phys. Rev. D}\ }\textbf {\bibinfo {volume} {103}},\ \bibinfo
  {pages} {083533} (\bibinfo {year} {2021})},\ \Eprint
  {http://arxiv.org/abs/2007.08991} {arXiv:2007.08991 [astro-ph.CO]}
  \BibitemShut {NoStop}%
\bibitem [{\citenamefont {Vargas-Maga\~na}\ \emph {et~al.}(2018)\citenamefont
  {Vargas-Maga\~na} \emph {et~al.}}]{Vargas-Magana:2016imr}%
  \BibitemOpen
  \bibfield  {author} {\bibinfo {author} {\bibfnamefont {Mariana}\ \bibnamefont
  {Vargas-Maga\~na}} \emph {et~al.},\ }\bibfield  {title} {\enquote {\bibinfo
  {title} {{The clustering of galaxies in the completed SDSS-III Baryon
  Oscillation Spectroscopic Survey: theoretical systematics and Baryon Acoustic
  Oscillations in the galaxy correlation function}},}\ }\href {\doibase
  10.1093/mnras/sty571} {\bibfield  {journal} {\bibinfo  {journal} {Mon. Not.
  Roy. Astron. Soc.}\ }\textbf {\bibinfo {volume} {477}},\ \bibinfo {pages}
  {1153--1188} (\bibinfo {year} {2018})},\ \Eprint
  {http://arxiv.org/abs/1610.03506} {arXiv:1610.03506 [astro-ph.CO]}
  \BibitemShut {NoStop}%
\bibitem [{\citenamefont {Brieden}\ \emph {et~al.}(2022)\citenamefont
  {Brieden}, \citenamefont {Gil-Mar\'\i{}n},\ and\ \citenamefont
  {Verde}}]{Brieden:2022lsd}%
  \BibitemOpen
  \bibfield  {author} {\bibinfo {author} {\bibfnamefont {Samuel}\ \bibnamefont
  {Brieden}}, \bibinfo {author} {\bibfnamefont {H\'ector}\ \bibnamefont
  {Gil-Mar\'\i{}n}}, \ and\ \bibinfo {author} {\bibfnamefont {Licia}\
  \bibnamefont {Verde}},\ }\bibfield  {title} {\enquote {\bibinfo {title}
  {{Model-agnostic interpretation of 10 billion years of cosmic evolution
  traced by BOSS and eBOSS data}},}\ }\href@noop {} {\  (\bibinfo {year}
  {2022})},\ \Eprint {http://arxiv.org/abs/2204.11868} {arXiv:2204.11868
  [astro-ph.CO]} \BibitemShut {NoStop}%
\end{thebibliography}%

\end{document}